\documentclass[structabstract]{aa}    
\usepackage{amssymb,amsmath}
\usepackage{graphicx}  
\usepackage{txfonts}
\usepackage{natbib}
\usepackage{color}   
\bibpunct{(}{)}{;}{a}{}{,} 
\begin{document}  
\title{Radiation thermo-chemical models of protoplanetary discs}
\subtitle{IV Modelling CO ro-vibrational emission from Herbig~Ae
  discs} \author{W. F. Thi\inst{1}, I. Kamp\inst{2},
  P. Woitke\inst{3}, G. van der Plas\inst{4}, R. Bertelsen\inst{2},
  L. Wiesenfeld\inst{1}}
\titlerunning{Modelling CO ro-vibrational emission from Herbig~Ae discs}
\institute{UJF-Grenoble 1 / CNRS-INSU, Institut de Plan\'{e}tologie et
  d'Astrophysique (IPAG) UMR 5274, Grenoble, F-38041, France
  \email{thiw@obs.ujf-grenoble.fr} 
  \and Kapteyn Astronomical
  Institute, P.O. Box 800, 9700 AV Groningen, The Netherlands 
  \and
  SUPA, School of Physics \& Astronomy, University of St. Andrews,
  North Haugh, St. Andrews KY16 9SS, UK 
  \and Departamento de
  Astronom\'{i}a, Facultad de Ciencias F\'{i}sicas y Matem\'{a}ticas,
  Universidad de Chile, Chile
}
\authorrunning{W.-F. Thi}

\date{Received January 1, 2012; accepted october 16, 2012}

 
  \abstract
  {The carbon monoxide (CO) ro-vibrational emission from discs around
    Herbig~Ae stars and T~Tauri stars with strong ultraviolet
    emissions suggests that fluorescence pumping from the ground
    $X^1\Sigma^+$ to the electronic $A^1\Pi$ state of CO should be
    taken into account in disc models.}
  {We wish to understand the excitation mechanism of CO ro-vibrational
    emission seen in Herbig~Ae discs, in particular in transitions
    involving highly excited rotational and vibrational levels.}
  {We implemented a CO model molecule that includes up to 50
    rotational levels within nine vibrational levels for the ground
    and $A$-excited states in the radiative-photochemical code {\sc
      ProDiMo}. We took CO collisions with hydrogen molecules (H$_2$),
    hydrogen atoms (H), helium (He), and electrons into account. We
    estimated the missing collision rates using standard scaling laws
    and discussed their limitations. We tested the effectiveness of UV
    fluorescence pumping for the population of high-vibrational levels
    ($v$=1--9, $J$=1--50) for four Herbig~Ae disc models (disc mass
    $M_{\mathrm{disc}}$=10$^{-2}$, 10$^{-4}$ and inner radius
    $R_{\mathrm{disc}}$=1, 20~AU). We tested the effect of infrared
    (IR) pumping on the CO vibrational temperature and the rotational
    population in the ground vibrational level.}
  {UV fluorescence and IR pumping impact on the population of
    ro-vibrational $v>1$ levels. The $v=1$ rotational levels are
    populated at rotational temperatures between the radiation
    temperature around 4.6~$\mu$m and the gas kinetic temperature.
    The UV pumping efficiency increases with decreasing disc mass. The
    consequence is that the vibrational temperatures
    $T_{\mathrm{vib}}$, which measure the relative populations between
    the vibrational levels, are higher than the disc gas kinetic
    temperatures (suprathermal population of the vibrational
    levels). The effect is more important for low-density gases
    because of lower collisional de-excitations.The UV pumping is more
    efficient for low-mass ($M_{\mathrm{disc}}<$~10$^{-3}$
    M$_{\odot}$) than high-mass ($M_{\mathrm{disc}}>$~10$^{-3}$
    M$_{\odot}$) discs. Rotational temperatures from fundamental
    transitions derived using optically thick $^{12}$CO $v=1-0$ lines
    do not reflect the gas kinetic temperature.  Uncertainties in the
    rate coefficients within an order of magnitude result in
    variations in the CO line fluxes up to 20\%. CO pure rotational
    levels with energies lower than 1000~K are populated in LTE but
    are sensitive to a number of vibrational levels included in the
    model. The $^{12}$CO pure rotational lines are highly optically
    thick for transition from levels up to
    $E_{\mathrm{upper}}$=2000~K. The model line fluxes are comparable
    with the observed line fluxes from typical HerbigAe low- and
    high-mass discs.}  {}

   \keywords{ Stars: protoplanetary disks, molecular processes,
     radiation mechanisms, radiative transfer.}

   \maketitle
%
\section{Introduction}

Terrestrial planets form in the inner region of protoplanetary discs
(typically $R<$3-5 AU), where the gas reaches temperatures of a few
hundred to a few thousand degrees Kelvin and the dust grains are warm
enough such that water is not frozen onto their surfaces
\citep{Kamp2010A&A...510A..18K}. The growth of solid cores in the
inner disc is slower than in the outer disc, where grains are coated
with an icy mantle. As a consequence, the solid bodies never reach the
critical mass to attract gravitationally the disc gas before the gas
disc has dissipated \citep{Armitage2010apf..book.....A}.
   
At a few hundred to a few thousand degrees, molecules are excited to
their ro-vibrational levels and emit in the near- and
mid-infrared. Probing the warm gas emissions requires high
angular-resolution observations in the infrared. Carbon monoxide is
one of the most abundant species in protoplanetary discs, and its
ro-vibrational transitions around 4.6 micron ($M$-band) are the most
commonly detected lines with high spectral resolution adaptive-optics
(AO) assisted spectrometers around T~Tauri
\citep{Najita2003ApJ...589..931N,Rettig2004ApJ...616L.163R,Salyk2011ApJ...743..112S}
and Herbig~Ae stars
\citep{Blake2004ApJ...606L..73B,Carmona2005A&A...436..977C,Brittain2009ApJ...702...85B,vanderPlas2009A&A...500.1137V}.  
The CO peak emission is also seen displaced from the source centre,
suggesting the presence of an inner hole
\citep{Brittain2003ApJ...588..535B,Goto2006ApJ...652..758G,Brown2012ApJ...744..116B}. The
spectro-astrometric technique further improves our understanding of
the emission by reaching spatial resolutions of the order of
0.1--0.5~AU for nearby (100-140~pc) discs
\citep{Pontoppidan2008ApJ...684.1323P}. \citet{Goto2012A&A...539A..81G}
observed the CO ro-vibrational emission from the \object{HD~100546}
disc with a spatial resolution of 10~AU at 100~pc. They found warm CO
gas (400-500) out to a distance of 50~AU, which is evidence for the
presence of a warm molecular disc atmosphere. Very dense and hot gas
located within a few tenths of AU from the stars can be detected by
the CO overtone emissions \citep{
  Tatulli2008A&A...489.1151T,Berthoud2007ApJ...660..461B,Thi2005A&A...430L..61T}.

Line profiles are diverse, from broad single peaked to clear double
peaked; the latter is a signature of Keplerian rotation of the gas
between 1 and 5~AU
\citep{Salyk2009ApJ...699..330S}. \citet{Bast2011A&A...527A.119B}
focused on 8 out of 50 discs that show broad single-peaked CO
profiles. They proposed that those 8 discs have either highly
turbulent inner disc gas and/or that a slowly moving disc wind
contributes to the emission.

The $M$-band observations show lines emitted from a large range of
excitation levels (from fundamental $v=1 \rightarrow 0$ $P$(1) and
$R$(0) lines to ``hot'' lines ($v'>1$, $\Delta v=1$) which make it
possible to probe disc regions with different excitation conditions.
CO vibrational diagrams of Herbig~Ae discs indicate vibrational gas
temperatures higher than the rotational ones (a few thousand K instead
of a few hundred K). This suggests that the population of high
vibrational levels ($v>$1) may be dominated by IR and UV fluorescence
excitation
\citep{Brittain2007ApJ...659..685B,Brown2012ApJ...744..116B}. The
$v>$~1 ro-vibrational levels can also be efficiently populated by UV
fluorescence for T~Tauri discs, especially for strongly accreting
systems with excess UV emission
\citep{Bast2011A&A...527A.119B}. Fluorescence excitation in T~Tauri
discs is supported by {\it Hubble Space Telescope} observations of
UV-pumped CO emission \citep{France2011ApJ...734...31F}.
  
Many models of CO ro-vibrational emission made the assumption of local
thermodynamical equilibrium (LTE) population
\citep{Hugelmeyer2009A&A...498..793H,Regaly2010A&A...523A..69R,Salyk2011ApJ...743..112S}. Nevertheless,
a detailed understanding of the CO ro-vibrational lines, especially
the spatial location of the lines and the efficiency of the IR/UV
fluorescence, requires non local thermodynamic equilibrium (NLTE)
modelling and a chemo-physical code that includes detailed continuum
and line radiative transfer.  The accuracy of NLTE modelling depends
on many factors, including the availability of accurate collision
rates. The difficulties associated with the measurement and
computation of accurate rates result in sparse trustworthy
data. Completeness in the rate coefficients can be attained only by
using approximate extrapolation rules derived from physical
considerations.  \citet{Krotkov1980ApJ...240..940K} and
\citet{Scoville1980ApJ...240..929S} studied the efficiency of UV and
IR pumping but considered only pure-vibrational levels.

In this paper, we describe the implementation of a large CO model
molecule, which includes many ro-vibrational levels in the ground
electronic state $X$ and the electronic state $A$, into the code {\sc
  ProDiMo}. The radiative chemo-physical code {\sc ProDiMo} solves the
continuum radiative transfer, the gas heating and cooling balance, the
gas chemistry, and the disc vertical hydrostatic structure
self-consistently \citep[][paper I]{Woitke2009A&A...501..383W}. This
paper is the fourth in a series dealing with aspects in protoplanetary
disc modelling \citep[][paper II \&
III]{Kamp2010A&A...510A..18K,Thi2011MNRAS.412..711T}.

We will use the code to model two typical discs around an A-type star:
a solar-nebula-mass disc ($M_{\mathrm{disc}}=0.01$ M$_\odot$), and a
low-mass disc ($M_{\mathrm{disc}}=10^{-4}$ M$_\odot$). The choice of
these models isq justified by the expectation that the strength of
UV/IR fluorescence will depend on the density and optical depth in the
discs.

The paper is organized as follows: we present and discuss the CO
spectroscopic and collisional data implemented in the {\sc ProDiMo}
code in Sec.~\ref{molecular_data}. The main feature of the {\sc
  ProDiMo} code and the disc model parameters are described in
Sec.~\ref{code_description} and \ref{model_description}. In
Sec.~\ref{results_discussion}, we discuss the model outputs. Finally,
we present our conclusions and recommendations on the interpretations
of CO ro-vibrational observations in discs in Sec.~\ref{conclusions}.

\section{CO molecular data}\label{molecular_data}

\subsection{Spectroscopic data}

\citet{Tashkun2010JQSRT.111.1106T} compiled and evaluated existing
experimental frequencies for transitions within the ground electronic
state. Their work is based on accurate experimental frequency
measurements
\citep[e.g.][]{George1994ApPhB..59..159G,Gendriesch2009A&A...497..927G}.

\begin{figure*}[ht]
  \centering  
 \includegraphics[scale=0.5]{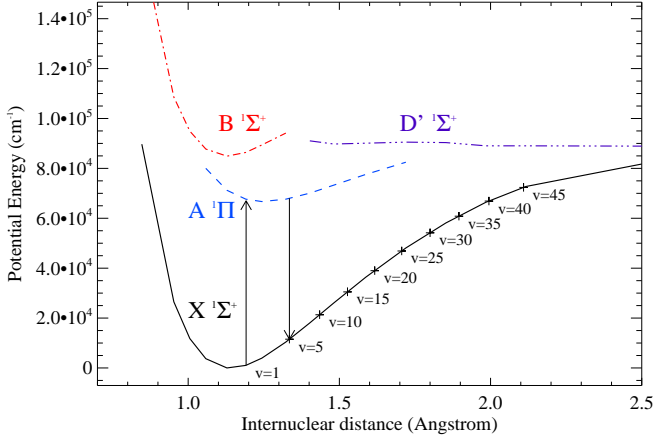}
 \includegraphics[scale=0.5]{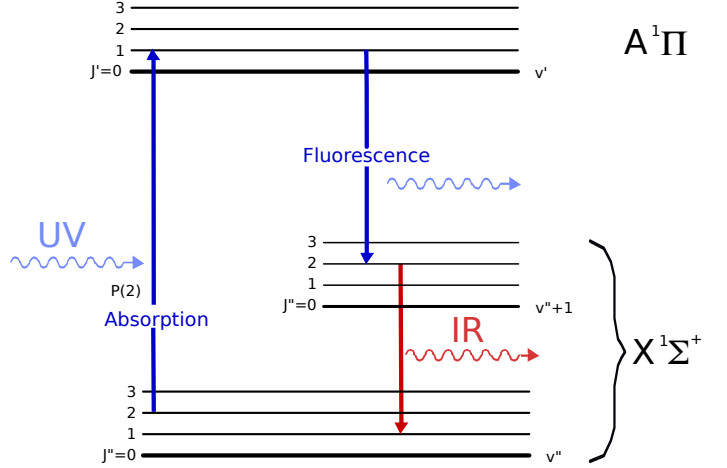}
 \caption{CO energy potential curves on the left and a schematic of
   the UV fluorescence mechanism on the right. We are concerned with
   transitions between the ground electronic state $X^1\Sigma^+$ and
   the electronic excited state $A^1\Pi$.  The arrows show a possible
   UV-fluorescence pumping from a ground electronic level
   ($X^1\Sigma^+$) at $v''=1$ to a level in the electronic state
   $A^1\Pi$, followed by a de-excitation to the level $v''=2$ in the
   ground electronic state. The value for the curves are taken from
   \citet{Cooper1987JChPh..87..424C}.}
\label{CO_potential_energy_curves}
\end{figure*}

We compared their energy levels and frequencies with those from
\citet{Chandra1996A&AS..117..557C} and \citet{Hure1993JMoSp.160..335H}
and found no significant differences. In addition,
\citet{Chandra1996A&AS..117..557C} computed Einstein spontaneous
probabilities $A^1\Pi$ for all ro-vibrational transitions between 140
rotational levels and 10 vibrational levels in the ground electronic
state. We adopted their transition probabilities, which compared well
with other studies \citep[e.g.][]{Okada2002JQSRT..72..813O}. The
transitions between the electronic excited ro-vibrational levels $A$
and the ground ro-vibrational levels $X^1\Sigma^+$ (4th positive
system), which occur around $\sim$~1600 \AA, were derived from the
band-averaged transition probabilities of
\citet{Beegle1999A&A...347..375B} and
\citet{Borges2001JMoSp.209...24B} using the formula and
H\"{o}ln-London factors in \citet{Morton1994ApJS...95..301M}, which
are given in
Table~\ref{HolnLondonTable}. Figure~\ref{CO_potential_energy_curves}
shows a sketch of the fluorescence mechanism involving a
ro-vibrational level in the ground electronic state $X^1\Sigma^+$ and
a ro-vibrational level in the excited state $A^1\Pi$.
    
\begin{table}[th]
  \caption{H\"{o}ln-London factors taken from \citet{Morton1994ApJS...95..301M}.}             
  \label{HolnLondonTable}      
\centering          
\begin{tabular}{llll}     
\hline\hline       
        & \multicolumn{1}{c}{$P$} & \multicolumn{1}{c}{$Q$} & \multicolumn{1}{c}{$R$}\\
        & $J'=J''-1$ & $J'=J''$ & $J'=J''+1$ \\
\hline
$^1\Sigma$-$^1\Sigma$ & $J''$ & ... & $J''+1$\\
                     & $J'+1$ & ... & $J'$\\
$^1\Pi$-$^1\Sigma$    & $J''-1$ & $2J''+1$ & $J''+2$ \\
                     & $J'$    & $2J'+1$ & $J'+1$\\
\hline                    
\end{tabular}
\end{table}

\subsection{Collisional data}

Although CO is one of the most studied interstellar molecules with a
large amount of experimental and theoretical collision rates
available, the lack of completeness of any individual study obliged us
to use scaling laws to fill the gaps. Since protoplanetary discs show
a large range of physical and chemical properties, we need to consider
collisions of CO with H, He, H$_2$, and electrons. In the rest of the
section, we will first review existing theoretical and experimental
collision rates before discussing the strengths and limitations of
scaling laws.

\subsubsection{CO-H collision rates }

The $v$=1$\rightarrow$0 CO-H de-excitation rate coefficients were
first measured by \citet{Millikan1963JChPh..39.3209M} in shock-tube
experiments. Subsequent works include the shock-tube experiments by
\citet{vonRosenberg1971JChPh..54.1974V}, \citet{Glass1982}, and more
recently \citet{Kozlov2000} (between 2000 and 3000~K). Shock-tube
experimental measurements are de-excitation timescales,which can be
fitted in terms of the Landau-Teller rate coefficients in cm$^{3}$
s$^{-1}$ \citep{Ayres1989ApJ...338.1033A}:
\begin{equation}
k_{v'\rightarrow v''}=4.2\times 10^{-19} \frac{\exp{[B-0.0069A\beta^{1/3}]}}{\beta(1-e^{-\beta})} \label{eq_LandauTeller}
\end{equation}
with $\beta = E_{\rm v=1}/T$ and the values for $A$ and $B$ given in
Table~\ref{LandauTellerTable}. \citet{Neufeld1994ApJ...428..170N} gave
a simple formula that fits the shock-tube data for $T>$~300~K:
\begin{equation}
k_v(1\rightarrow0) = 7.9 \times 10^{-13} T \exp{(-1208 T)}.
\end{equation}
This formula supersedes an older one given by
\citet{Hollenbach1979ApJS...41..555H}:
\begin{equation}
k_v(1\rightarrow0) = 3\times10^{-12} \sqrt{T} e^{-(2\times 10^3/T)^{3.43}}  e^{-3080/T}.
\end{equation}
Experimental values for transitions other than the $v$=1$\rightarrow$0
are not available.
\begin{table}[th]
  \caption{Landau-Teller coefficients for CO-partner collisions derived from shock-tube experiments.}             
\label{LandauTellerTable}      
\centering          
\begin{tabular}{llll}     
\hline\hline       
 Partner & $A$ & $B$ & Ref.\\
\hline
H     &  3    &  18 &  1\\
H     &  53   &  19 &  2\\
H     & 188.9 &  36 & 3, 4\\  
H$_2$ &  64   &  19.1 & 1\\
He    &  87   &  19.1 & 4\\ 
\hline                    
\end{tabular}
\tablefoot{(1) \citet{Glass1982}.(2) \citet{Millikan1963JChPh..39.3209M}. (3) \citet{vonRosenberg1971JChPh..54.1974V,Kozlov2000}. (4) \citet{Millikan1964JChPh..40.2594M}.}
\end{table}
\citet{Balakrishnan2002ApJ...568..443B} performed quantal scattering
computations for rotational and band-averaged vibrational collision
rates between CO and hydrogen atoms. They used the close-coupling
method for the rotational transitions and the infinite order sudden
approximation (IOS, \citealt{Flower2007mcim.book.....F}) for the
vibrational transitions for a large range of gas temperatures (5 $<$
$T$ $<$ 3000~K). The IOS approximation reduces the computational
effort but is strictly valid only for energetic collisions,
i.e. $\Delta E \ll kT$, where $\Delta E$ is the energy difference between
two levels. The close-coupling computations were done for rotational
levels up to $J$=7. The vibrational transition rates were computed up
to $v=$~4.

\begin{figure}
  \centering  
 \includegraphics[scale=0.35,angle=90]{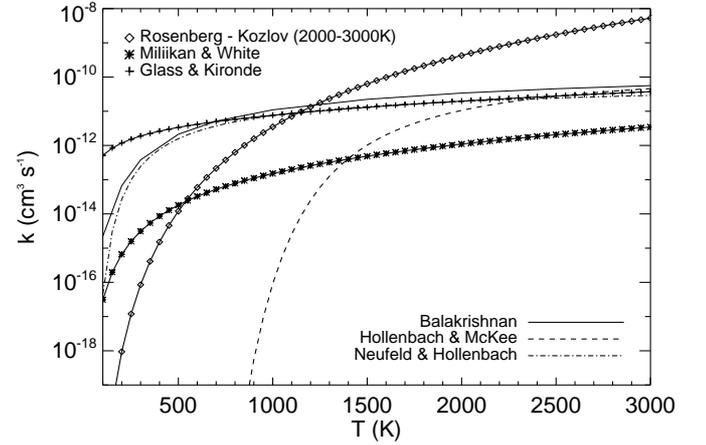}
 \caption{Vibrational de-excitation ($v$=1$\rightarrow$0) rate
   coefficients of CO by collisions with H (in cm$^3$ s$^{-1}$). The
   large discrepancies between the experimental values may stem from
   the reactivity of H-atoms with CO. We adopted the theoretical rates
   from \citet{Balakrishnan2002ApJ...568..443B}, which match the
   experimental values of \citet{Glass1982} (plus signs).}
\label{H_rate_comparison}
\end{figure}

Figure~\ref{H_rate_comparison} shows the different rate
coefficients. The most striking feature is the large dispersion (a few
orders of magnitude) between the
sources. \citet{Ayres1989ApJ...338.1033A} chose the values of
\citet{Glass1982}, which are much larger than the older values of
\citet{vonRosenberg1971JChPh..54.1974V}. However, recent measurements
by \cite{Kozlov2000} reproduced the values of
\citet{vonRosenberg1971JChPh..54.1974V}. Interestingly, the
theoretical values calculated by
\citet{Balakrishnan2002ApJ...568..443B} are much closer to the values
of \citet{Glass1982} and are best approximated by the formula of
\citet{Neufeld1994ApJ...428..170N}. One possible reason for the large
discrepancies is the reactivity of CO with atomic hydrogen (an open
shell species), which can falsify the measurements. We chose to use
the theoretical rates by \citet{Balakrishnan2002ApJ...568..443B}.

\subsubsection{CO-He collision rates }

\begin{figure}
  \centering
 \includegraphics[scale=0.35,angle=90]{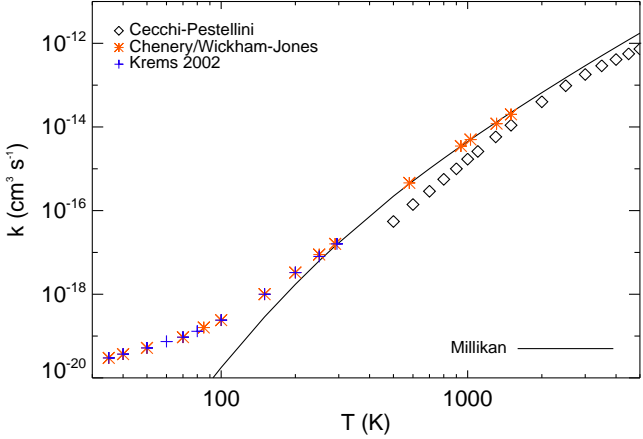}
 \caption{Vibrational de-excitation ($v$=1$\rightarrow$0) rate
   coefficients of CO by collisions with He (in cm$^3$ s$^{-1}$). We
   adopted in this study the rates of \citet{Krems2002JChPh.116.4525K}
   at temperatures $<$ 1500 K (plus signs) and those of
   \citet{Cecchi-Pestellini2002ApJ...571.1015C} at very high
   temperatures ($>$ 1500 K, diamonds).}
\label{He_rate_comparison}
\end{figure}

\citet{Cecchi-Pestellini2002ApJ...571.1015C} computed rotational and
vibrational rate de-excitation coefficients of CO in collision with
He.  The rotation rate coefficients in the ground vibrational level
are computed by the quantal close-coupling method. The vibrational
rates are computed for 500$\leq$ $T$ $\leq$ 1300~K using the IOS
approximation. \citet{Krems2002JChPh.116.4525K} calculated
quantum-mechanically ro-vibrational rate coefficients for 35
$\leq$~$T$~$\leq$~1500~K. Rates were computed by
\citet{Flower2012MNRAS.tmp.3481F} and \citet{Reid1995JChPh.103.2528R}.
The experimental fit coefficients of
\citet{Millikan1964JChPh..40.2594M} are given in
Table~\ref{LandauTellerTable}.
\citet{Wickham-Jones1987JChPh..87.5294W} reported experimental data
from the PhD thesis of Chenery (1984), who measured rates down to
35~K. The experimental rate coefficients are compared to the
theoretical values in Fig.~\ref{He_rate_comparison}. The theoretical
values of \citet{Cecchi-Pestellini2002ApJ...571.1015C} are a factor
$\sim$~2 smaller than the experimental values, whereas the values
computed by \citet{Krems2002JChPh.116.4525K} match the shock-tube data
down to $T$=~300~K. The formula \ref{eq_LandauTeller} that fits the
experimental shock-tube data under-predicts the values for
$T\leq$~300~K. We adopted the values of
\citet{Krems2002JChPh.116.4525K} up to $T$=~1500~K and those of
\citet{Cecchi-Pestellini2002ApJ...571.1015C} above $T$=~1500~K. The
rate coefficients for CO in collisions with He are around three orders
of magnitude smaller than the collision rates with H. This behaviour
can be explained by the fact that He is a close-shell species.

\subsubsection{CO-H$_2$ collision rates }

In the case of collisional excitation of CO by H$_2$, we need to
consider the collisions with ortho-H$_2$ and para-H$_2$
separately. \citet{Yang2010ApJ...718.1062Y} computed with the
close-coupling method rates for temperatures between 1 and 3000~K
involving ground vibrational CO transitions and with rotational levels
up to $J$=40.  \\
Rates for vibrational transitions are scarcer and are given averaged
over the rotational levels. Between 30 and 70~K,
\citet{Reid1997JChPh.106.4931R} showed experimental and theoretical
rates for the CO v=1$\rightarrow$0 transition. Between 70 and 290~K,
rates derived from shock-tube experiments are available
\citep{Hooker1963JChPh..38..214H,Andrews1975CPL....36..271A,Andrews1976CPL....41..565A,Glass1982}. At
temperatures higher than 300~K, we can assume that the Landau-Teller
law is valid and expand the shock-tube rates to a few thousand K
\citep{Ayres1989ApJ...338.1033A}. \citet{Neufeld1994ApJ...428..170N}
provided a convenient analytical expression of the shock-tube data
($T>$290 K):
\begin{equation} 
k_v(1\rightarrow0) = 4.3 \times 10^{-14} T \exp{(-68 T^{1/3})}.
\end{equation}
Alternatively, one can use the Landau-Teller formula with the values
in Table~\ref{LandauTellerTable}.

\begin{figure}
  \centering
 \includegraphics[scale=0.35,angle=90]{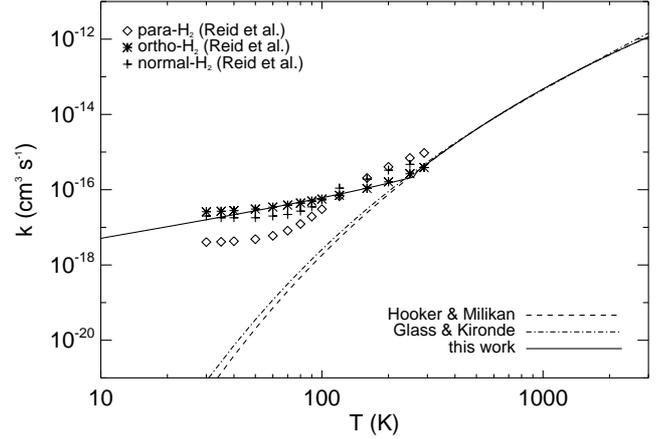}
 \caption{Vibrational de-excitation ($v$=1$\rightarrow$0) rate
   coefficients of CO by collisions with H$_2$ (in cm$^3$
   s$^{-1}$). Normal H$_2$ is a mix of o-H$_2$ and p-H$_2$ with an
   ortho-to-para ratio at LTE.}
\label{H2_rate_comparison}
\end{figure}

In Fig.~\ref{H2_rate_comparison}, we compare the two shock-tube
experimental rate coefficients, the experimental and theoretical
values of \citet{Reid1997JChPh.106.4931R}, and our adopted fit
\begin{equation}
k_v(1\rightarrow0) = 5\times 10^{-19} T \exp{(2 \times 10^{-3} T)},
\end{equation}
for $T\leq$~300~K and the fit to the \citet{Hooker1963JChPh..38..214H}
data by \citet{Neufeld1994ApJ...428..170N} above 300~K. We notice a
large discrepancy between the experimental and theoretical rates at
low temperatures. The shock-tube data are consistent with each
other. It is known that the rate extrapolations to lower temperatures
with the Landau-Teller law may not be correct. Collisions with H are
around two orders of magnitude more effective than collisions with
H$_2$ at all temperatures.

\subsubsection{CO-electron collision rates }

\begin{figure}
  \centering
 \includegraphics[scale=0.35,angle=90]{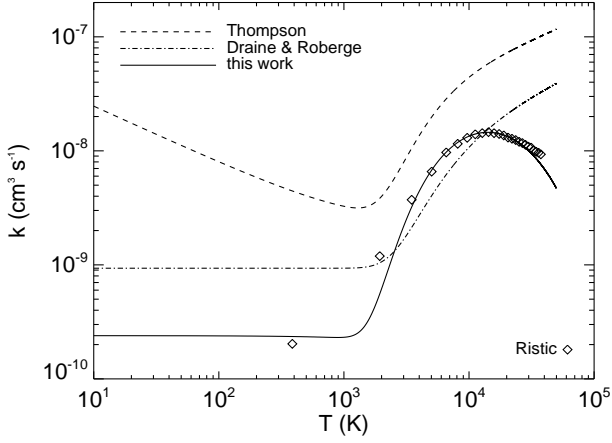}
 \caption{Vibrational de-excitation ($v$=1$\rightarrow$0) rate
   coefficients of CO by collisions with electrons (in cm$^3$ s$^{-1}$).}
\label{elec_rate_comparison}
\end{figure}

The collision excitation and de-excitation rates of CO by electrons
for rotational transitions are computed using the theory of
\citet{Dickinson1977A&A....54..645D} based on the Born approximation
for $\Delta J=\pm$~1 (see also
\citealt{Randell1996JPhB...29.2049R}). We adopted a rotational
constant $B_{{\rm CO}}=$~1.922529~cm$^{-1}$, a centrifugal distortion
$D_{{\rm CO}}=$~6.1210779 $\times$ 10$^{-6}$~cm$^{-1}$, and a dipole
moment of 0.112 Debye.  \citet{Ristic2007CP....336...58R} measured the
vibrational rates of CO by collisions with electrons for vibrational
levels up to ten. \citet{Morgan1993JPhB...26.2429M} also computed
cross sections for vibrational transitions.
\cite{Thompson1973ApJ...181.1039T} gave a simple formula for the
v=1$\rightarrow$0 rate
\begin{equation}
k_v(1\rightarrow 0) = 1.4 \times 10^{-9}{\beta}^{-1/2}
\left(1+\beta+19 e^{-3.22\beta}(1+4.22\beta)\right),
\end{equation}
where again $\beta = E_{\rm v=1}/T$. An alternative formula is given by
\citet{Draine1984ApJ...282..491D}
\begin{multline}
  k_v({\rm v}=1\rightarrow 0) = 1.9 \times 10^{-11}\sqrt{T}\\
  \left(\sqrt{\frac{2420}{T}}+10\sqrt{1+\frac{10500}{T}}\exp{\left(-\frac{10210}{T}\right)}\right).
\end{multline}
Finally, we propose our own analytical formula, which fits the values of
\citet{Ristic2007CP....336...58R}
\begin{multline}
k_v({\rm v}=1\rightarrow0) =1.9 \times 10^{-11} T^{1/3}\\
\left(\left(\frac{2000}{T}\right)^{1/3}+50 \sqrt{1+\frac{70000}{T}}\exp{\left(-\frac{10210}{T}\right)}\right)\\
\exp{(-4.5\times 10^{-5} T)}. \label{eq_CO_e_ProdiMo}
\end{multline}
The measurements for the $v$=1$\rightarrow$0 transition are shown in
Fig.~\ref{elec_rate_comparison} and compared to the three analytical
fitting formulae. The \citet{Thompson1973ApJ...181.1039T} formula
deviates significantly from the experimental data. We used a
combination of formula (\ref{eq_CO_e_ProdiMo}), which fits the
experimental data relatively well down to 400~K.Below 400~K, we
adopted a flat rate coefficient.

\subsection{Rate coefficient scaling laws}

The current experimental and theoretical works do not provide all the
rate coefficients needed to model the CO statistical ro-vibrational
population of the ground $X^1\Sigma^+$ and excited $A^1\Pi$ electronic
levels. Therefore, we have to resort to extrapolation and scaling laws
to fill the gaps.

\subsubsection{Extrapolating rotational transition rates}

A few extrapolating laws attempt to estimate rates from existing
ones. We used the energy-corrected sudden (ECS) scaling law with a
hybrid exponential power-law for the base ($k_{{\rm J''}\rightarrow
  0}$) rates
\citep{Depristo1979JChPh..71..850D,Depristo1979CP.....44..171D,Green1993ApJ...412..436G}.
\citet{Goldflam1977JChPh..67.5661G} showed that rates between any
levels can be derived knowing the rates to the ground level within the
IOS approximation. To overcome the limitation of collision energies
greater than the level energy differences,
\citet{Depristo1979JChPh..71..850D} proposed a correction for the
inelasticity and finite collision duration effects using second-order
perturbation theory for low-collision energies (i.e. for collisions at
low temperatures). A de-excitation ($J'>J''$) rate is given by the
ESC-EP (energy-corrected sudden scaling law with a hybrid exponential
power-law for the base rates) formula:
\begin{multline}
k_{0,J' \rightarrow  0,J''}=(2 J''+1)\exp{\left[\frac{(E_{J'}-E_{J''})}{kT}\right]}\Omega^2_{J'}\\
\times \sum_{L} (2{L}+1)\Omega^{-2}_{L}\left(\begin{array}{ccc} {J'} & {L} & {J''} \\ 0 & 0 & 0 \end{array}\right)^2 k_{{0,J'}\rightarrow{0,0}},
\end{multline}
where 
\begin{equation}
\left(\begin{array}{ccc} {J'} & {L} & {J''} \\ 0 & 0 & 0 \end{array}\right)
\end{equation}
is the Wigner 3-$j$ symbol and $E_{J'}$ and $E_{J''=J'-1}$ are the
upper and lower level energy, respectively. The correction factor
introduced by \citet{Depristo1979JChPh..71..850D} is defined by
\begin{equation}
\Omega_{J'}=\frac{1}{1+\tau_{J'}/6},
\end{equation}
with
\begin{equation}
\tau_{J'} = 0.065\omega_{J'}l_c(\mu/T)^{1/2},
\end{equation}
where $\mu$ is the reduced mass in atomic mass units, $\omega_{J'}$ is
the inelasticity to the next lowest level ($E_{J'}-E_{J''=J'-1}$) in
cm$^{-1}$, and $l_c$ is an adjustable scaling length in \AA. We
adopted $l_c$=3 \AA \ \citep{Schoier2005A&A...432..369S}. The ESC
formula is recovered for $l_c$=0 \AA.

\begin{figure}
  \centering
 \includegraphics[scale=0.35,angle=90]{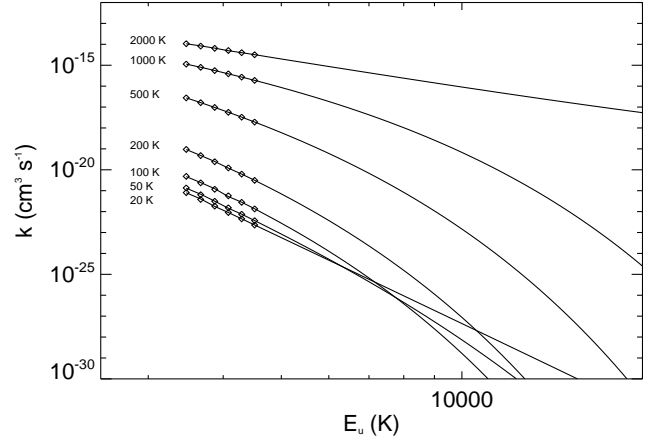}
 \caption{Extrapolated CO -- o-H$_2$ pure rotational rate coefficients. The
   x-axis corresponds to the upper level energy in Kelvin.The diamonds
   are rates from \citet{Yang2010ApJ...718.1062Y}.}
\label{CO_rate_extrapolation}
\end{figure}

The scaling law requires the knowledge of the base rates
($J\rightarrow 0$). Collision rates with hydrogen atoms to the ground
level are known for $J$ up to 7, with H$_2$ up to 40, and only for
$J$=1$\rightarrow$0 for electrons.  Further estimates of the rates to
the ground level are needed. The ESC-EP model provides an
approximation for rates to the ground level
\citep{Green1993ApJ...412..436G,Millot1990JChPh..93.8001M}:
\begin{equation}\label{eq_esc_cp2}
  k_{{0,J'}\rightarrow{0,0}}=A(T)\left[J'(J'+1)\right]^{-\gamma}\exp{(-\beta E_{J'}/kT)},
\end{equation}
where $A(T)$, $\gamma$, and $\beta$ are free parameters, which are
estimated by fitting equation \ref{eq_esc_cp2} to the existing data
using a least-square solver for over-constrained systems. We performed
the extrapolation for collisions with o-H$_2$, p-H$_2$, He, and H but
not with electrons. An example of extrapolated rate coefficients is
shown in Fig.~\ref{CO_rate_extrapolation}.

\subsubsection{Extrapolating vibrational transition rates}

The extrapolation of an existing rate to low temperature was performed
by fitting the available rates to the formula
\begin{equation}
k_v = a \times \exp{(T+b)},
\end{equation}
where $a$ and $b$ are the two fitting parameters.

The extrapolation of rates between two vibrational levels without any
known rate follows physically motivated scaling laws.  Based on the
initial work of \citet{Procaccia1975JChPh..63.4261P},
\citet{Chandra2001A&A...376..356C} proposed using the Landau-Teller
relationship for transitions between adjacent states ($\Delta {\rm
  v}=1$)
\begin{equation}
  k_{v}(v' \rightarrow v'-1)= v' k_{v}(1 \rightarrow 0),\label{eq_chandra1}
\end{equation}
and for $\Delta {\rm v}>1$:
\begin{equation}
k_{v}(v' \rightarrow v'') = \frac{2(v''-1)+1}{2(v'-2)+1}v' k_{v}(1 \rightarrow 0)\label{eq_chandra2}
\end{equation}
\citet{Elitzur1983ApJ...266..609E} argued for a different formulation
\begin{multline}
  k_{v}(v' \rightarrow v'') = (v'-v'') k_{v}(1 \rightarrow 0) \\
  \exp{\left(-(v'-v''-1)\frac{1.5E_{v}/kT}{1+1.5E_v/kT}\right)}.\label{eq_elitzur1}
\end{multline}
For $\Delta v=1$ ($v''=v'-1$), this last formula reduces to
\begin{equation}
k_{v}(v' \rightarrow v'-1) = k_{v}(1 \rightarrow 0).\label{eq_elitzur2}
\end{equation}

\citet{Scoville1980ApJ...240..929S} argued that the probabilities of
collision-induced vibrational transitions are proportional to the
corresponding radiative transition matrix elements (Born-Coulomb long
range interaction). Therefore, the vibrational de-excitation rate
coefficients from $v'$ to $v''$ can be derived from the
$v=1\rightarrow 0$ rate coefficients:
\begin{equation}
  k_{v}(v' \rightarrow v'') = \frac{A_{v'v''}}{A_{10}} k_{v}(1 \rightarrow 0), \\
\end{equation}
where $A_{v'v''}$ is the vibrational band transition probabilities.
The three formulations give quite different rate estimates. In order
to choose which one is the most appropriate, we compared the
predictions with detailed calculations and/or experimental values.

\begin{figure}
  \centering
 \includegraphics[scale=0.35,angle=90]{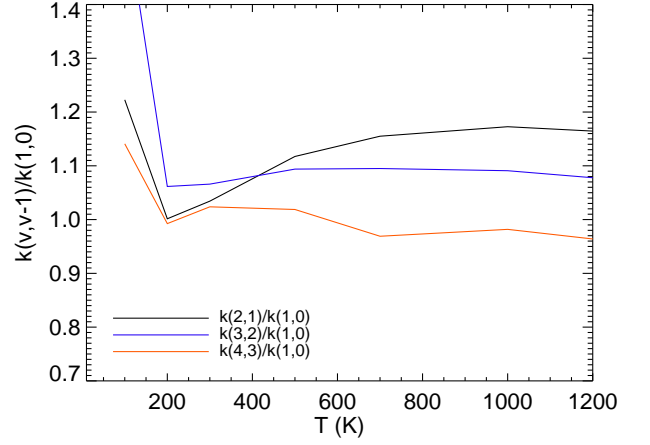}
 \caption{Vibrational de-excitation rate coefficient ratios
   $k$(v'$\rightarrow$v'-1)/$k$(1$\rightarrow$0) of CO by
   H. Rate coefficients are taken from
   \citet{Balakrishnan2002ApJ...568..443B}.}
\label{H_rate_ratios}
\end{figure}

Figure~\ref{H_rate_ratios} shows the rate coefficient ratios for
transitions between adjacent levels in collisions with atomic hydrogen
using the detailed calculation of
\citet{Balakrishnan2002ApJ...568..443B}. The ratios are grouped around
one with a small spread. The rate coefficient ratios suggest that the
expansion rules proposed by \citet{Elitzur1983ApJ...266..609E} are
more appropriate to expand the detailed rate coefficients.

\begin{figure}
  \centering
 \includegraphics[scale=0.35,angle=90]{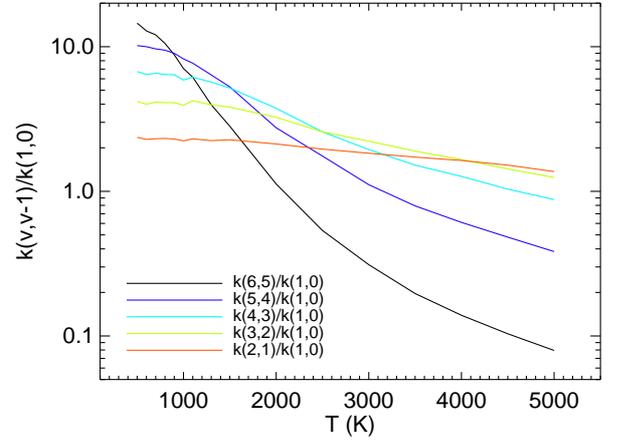}
 \caption{Vibrational de-excitation rate coefficient ratios
   $k$(v'$\rightarrow$v'-1)/$k$(1$\rightarrow$0) of CO by
   He. Rate coefficients are taken from
   \citet{Krems2002JChPh.116.4525K}.}
\label{He_rate_ratios}
\end{figure}

The rate coefficient ratios for CO-He collisions vary dramatically
with temperature (Fig.~\ref{He_rate_ratios}). For temperatures below
1000~K, the ratios are close to $2\times v'$, a value that corresponds
neither to the \citet{Chandra2001A&A...376..356C} prescription
($=v'$) nor to that championed by
\citet{Elitzur1983ApJ...266..609E} (=1). We chose to use the \cite{Chandra2001A&A...376..356C} for $\Delta v>$1 and the factor $2\times v'$ for $\Delta v=$1.

The choice is open concerning the rules to apply to estimate the rate
coefficients between CO and H$_2$ and between CO and electrons other
than $v''$=1$\rightarrow$0. We chose to use the same combination of
scaling rules as for CO-He collisions. We estimate the scaling rules
to provide rates correct within an order of magnitude.

\subsubsection{Estimating ro-vibrational transition rates}

The vibrational rates are averaged over all ro-vibrational
transitions.  To obtain individual rates between ro-vibrational
states, we apply the method described in
\citet{Faure2008A&A...492..257F}. This method assumes a complete
decoupling of rotational and vibrational levels. The state-to-state
ro-vibrational rates are related to the corresponding rates for the
pure rotational transitions in the ground vibrational state:
\begin{equation}
k_{v',J' \rightarrow v'',J''} = P_{v'v''}k_{0,J' \rightarrow 0,J''}.
\end{equation}
The factor $P_{{\rm v'v''}}$ is defined as
\begin{equation}
P_{v'v''}(T) = \frac{k_{v'\rightarrow v''}\sum_{J'} g_{J'}\exp{(-E_{v',J'}/kT)}}{\sum_{J'} \left(g_{J'}\exp{(-E_{v',J'}/kT)}\sum_{J''} k_{0,J'\rightarrow 0,J''}\right)}.
\end{equation}
This formula ensures detailed balance.

\subsubsection{Approximating electronic transition collision rates}

In the absence of collision rates connecting the ground $X^1\Sigma^+$
to the $A^1\Pi$ electronic levels, we adopted the rough values for
CO-electron collision rates provided by the $\bar{g}$ approximation.
The $\bar{g}$ \citep{vanRegemorter1962ApJ...136..906V} approximation
relies on the Born-dipole (also called Born-Coulomb) long-range
interaction approximation and relates the collision strength to the
transition probability and wavelength for allowed transitions
\citep{Itikawa2007mppc.book.....I}. The Born-dipole approximation is
valid for species with a permanent dipole, which is the case for CO
($\mu$=0.122 Debye). The collision rate for allowed transitions is
given by
\begin{equation}
k_{ul} = \frac{8.6291 \times 10^{-6}}{g_{u}\sqrt{T}}\Gamma
\end{equation}
For neutrals, the g-bar value can be approximated by
\begin{equation}
\bar{g}\simeq (kT/h\nu)
\end{equation}
The collision strength $\Gamma$ is related to g-bar by
\begin{equation}
\Gamma = 2.388 \times 10^{-6} \lambda_{\mu \rm m}^3g_u A_{ul} \bar{g}
\end{equation}
The approximation provides rate coefficients correct within an
order of magnitude.

\subsubsection{Combining theoretical, experimental, and estimated rates}

The resulting CO rate matrix after combining all existing and
estimated rates is complete in the sense that all ro-vibrational
levels in the ground electronic state connected by a radiative
transition are also connected by collisions. Therefore all levels in
the ground electronic levels can be populated in LTE at very high
densities and temperatures. We have neglected the radiative and
collisional transitions between levels in excited electronic states
since radiative transitions to the ground electronic state are very
fast. The ''completeness'' is required to ensure that we don't
artificially create sub-thermally populated levels.  The methodology
described above was implemented in the radiative photochemical code
{\sc ProDiMo}. The main collisional partner in ro-vibrational
transitions is the atomic hydrogen by a few orders of magnitude.

\section{CO fundamental and hot-band emission from Herbig~Ae discs}\label{models}

\subsection{The {\sc ProDiMo} code}\label{code_description}

A detailed description of the code can be found in
\citep[][]{Woitke2009A&A...501..383W}. Subsequent additions to the
code are documented in \citet[][]{Kamp2010A&A...510A..18K},
\citet[][]{Thi2011MNRAS.412..711T}, and
\citet{Woitke2011A&A...534A..44W}. Although not used in this study
X-ray heating and X-ray chemistry are included in {\sc ProDiMo}
\citep{Aresu2011A&A...526A.163A}. We deal in this paper with the
excitation mechanisms of the CO ro-vibrational levels and do not
attempt to fit actual observations.

\subsection{Disc model description}~\label{model_description}

We chose to model two representative discs, which can be found around
Herbig~Ae stars. Our generic star has an effective temperature of
8600~K, a mass of 2.2~M$_{\odot}$, and luminosity of
32~L$_{\odot}$. We have no intention to fit a particular object in
this study. The input stellar spectrum is taken from the {\sc PHOENIX}
database of stellar spectra \citep{Brott2005ESASP.576..565B}. The disc
model parameters are summarized in Table~\ref{tab_DiscParameters}. The
disc geometry is described by a single zone power-law surface-density
profile with index $\epsilon$, an inner radius $R_{\mathrm{in}}$, an
outer radius $R_{\mathrm{out}}$, a disc gas mass $M_{\mathrm{disc}}$,
and a standard gas-to-dust mass ratio $\Delta$ of 100 for all the
discs. The inner radius was set to two values to model the presence of
an inner disc hole and its effects on the CO ro-vibrational emissions.

\begin{center}
\begin{table}
  \caption{Disc and CO parameters. The parameters are defined in \citet{Woitke2009A&A...501..383W}}\label{tab_DiscParameters}
		\begin{tabular}{lll}
                  \hline
                  \noalign{\smallskip}   
                  stellar mass & $M_*$ & 2.2~M$_\odot$ \\ 
                  stellar luminosity &$L_*$ & 32~L$_\odot$ \\ 
                  effective temperature & $T_{\mathrm {eff}}$ & 8600~K\\
                  distance        & $d$ & 140~pc\\
                  disc inclination & $i$ & 0~\degr\ (face-on)\\
                  total disc mass & $M_{\mathrm{disc}}$ & 10$^{-2}$, 10$^{-4}$ M$_\odot$ \\ 
                  disc inner radius & $R_{\mathrm {in}}$  & 1, 20~AU \\
                  disc outer radius & $R_{\mathrm {out}}$  & 300~AU\\
                  vert. column density index & $\epsilon$ & 1 \\
                  inner rim soft edge  & & on\\
                  gas to dust mass ratio           &  $\delta$ & 100 \\
                  dust grain material mass density & $\rho_{\mathrm{dust}}$ & 3.5 g cm$^{-3}$ \\
                  minimum dust particle size       & $a_{\mathrm{min}}$ & 0.05 $\mu$m\\
                  maximum dust particle size       & $a_{\mathrm{max}}$ & 1000 $\mu$m\\
                  composition                      & & ISM silicate\\
                  dust size distribution power-law & $p$            & 3.5\\
                  H$_2$ cosmic ray ionization rate       & $\zeta_{\mathrm{CR}}$  & 1.7 $\times$ 10 $^{-17}$ s$^{-1}$\\
                  ISM UV field w.r.t. Draine field  & $\chi$          & 1.0 \\
                  abundance of PAHs relative to ISM & $f_{\rm PAH}$      & 0.1\\
                  $\alpha$ viscosity parameter      & $\alpha$           &  0.0 \\
                  turbulence width                  & $\delta v$         & 0.15~km~s$^{-1}$\\
                  \noalign{\smallskip}   
                  \hline
                  \noalign{\smallskip}   
                  reference scale height & $H_0$ & 15~AU\\
                  reference radius       & $R_0$ & 100~AU\\
                  flaring index          & $p$   & 1.2   \\
                  \noalign{\smallskip}   
                  \hline
                  \noalign{\smallskip}   
                  CO $X^1\Sigma^+$&  rot. levels & 50\\
                  CO $A^1\Pi$ & rot. levels & 50\\
                  CO $X^1\Sigma^+$ & vib. levels & up to 9 \\
                  CO $A^1\Pi$ & vib. levels & up to 9 \\
                  \hline
\end{tabular}  
\ \\ 
\end{table}
\end{center}
The dust grains are defined in both discs by their composition
(astronomical silicates, \citealt{Laor1993ApJ...402..441L}) and their
size distribution (a power-law ranging from $a_{\mathrm{min}}$ to
$a_{\mathrm{max}}$). We did not iterate on the vertical hydrostatic
profile but used instead a power-law $H=H_0(R/R_0)^{-p}$ to model the
gas scale-height $H$. We chose to model two flaring discs (flaring
index $p$=1.2) because observations suggest that the CO levels
populated by UV-fluorescence pumping seem to occur in flaring
discs. 

The two discs differ from each other only in the total (gas+dust) disc
mass $M_{\mathrm{disc}}$. The first disc has a mass consistent with
the typical solar nebula mass of 0.01 M$_\odot$, while the second disc
is more akin to an object in transition from a primordial gas-rich
disc to a secondary debris disc with a total mass of 10$^{-4}$
M$_\odot$. The choice of the two masses is motivated by the
possibility to assess the effects of infrared and UV fluorescence on
the CO population in discs with vastly different density structures.
The discs are assumed passively heated only (i.e. without viscous
heating: $\alpha=$~0), and we do not consider X-ray heating because
the X-ray luminosity is generally low compared to the UV luminosity in
Herbig~Ae stars. X-ray heating and chemistry can play an important
role in discs around T~Tauri stars
\citep{Aresu2011A&A...526A.163A,Aresu2012arXiv1209.0591A,Meijerink2012arXiv1208.4959M}.

The main source of heating for the gas is the stellar UV via
photoelectric effects on polycyclic aromatic hydrocarbons (PAHs)
assumed here to be circumcoronenes (C$_{54}$H$_{18}$). The abundance of
PAHs is assumed at 10\% of the standard interstellar abundance
($f_{\mathrm{PAH}}=0.1$).

The gas and dust were assumed well-mixed with no dust settling. A
detailed discussion on the heating and cooling in HerbigAe discs can
be found in \citet{Kamp2010A&A...510A..18K}.

The continuum radiative transfer and dust thermal balance are
calculated in two dimensions using the long-characteristic method
\citep{Woitke2009A&A...501..383W, Thi2011MNRAS.412..711T}. The
radiative transfer numerical implementation was benchmarked against
other radiative transfer codes \citep{Pinte2009A&A...498..967P}. The
gas temperatures were computed by balancing the heating and cooling
(mostly radiative) processes. The CO self-shielding against
photodissociation uses factors interpolated from values in look-up
tables \citep{Visser2009A&A...503..323V}.
\begin{figure}[ht]
  \centering
 \includegraphics[scale=0.5,angle=0]{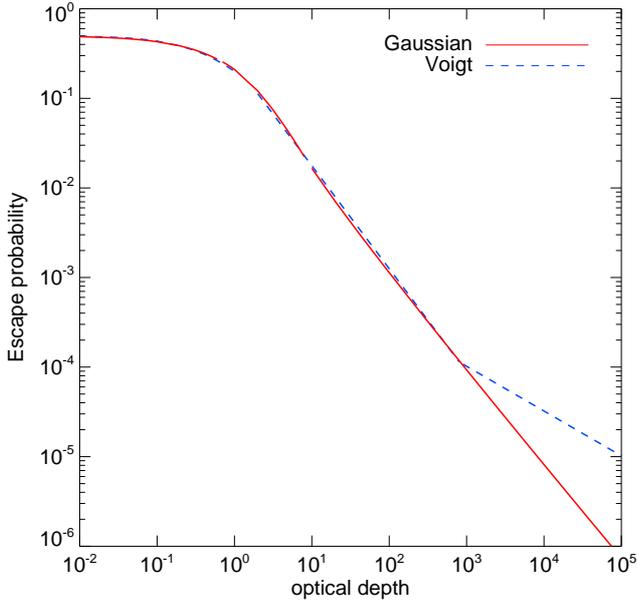}
 \caption{Escape probability for a semi-infinite slab. We adopted
   $a=10^{-3}$ for illustration.}
\label{voigt_escape_proba}
\end{figure}

\begin{table*}[th]
  \caption{CO LTE rotational temperature and vibrational temperature for the UV/no-UV models derived from analyzing the CO fundamental emission rotational diagrams. max($T_{\mathrm{dust}}$) is the maximum dust temperature attained in the disc. n.a. means not available.}             
\label{ModelTable}      
\centering          
\begin{tabular}{lllllllll}     
  \hline\hline 
  \noalign{\smallskip}   
  Model & $M_{\mathrm{disc}}$ & $R_{\mathrm {in}}$ & $A^1\Pi$ & $X^1\Sigma^+$ & UV& $T_{\mathrm{rot}}^{\mathrm{LTE}}$ & $T_{\mathrm{vib}}$ & max($T_{\mathrm{dust}}$)\\
  &                &   & number of & number of  & pumping? & $v=1\rightarrow 0$        & v=0,...,5    &  \\
  &    (M$_\odot$)   &    (AU)        & vib. levels & vib. levels  &  & (K)   &(K)      & (K)    \\   
  \noalign{\smallskip}   
  \hline
  \noalign{\smallskip}         
  1a     &  10$^{-2}$  &  1   & 9 & 9  & yes & 777 & 1413     & 857 \\ 
  1b     &  10$^{-2}$  &  1   & 0 & 9  & no  & 777 & 941-1413 & 857 \\
  1c     &  10$^{-2}$  &  1   & 0 & 5  & no  & n.a. & 1413     & 857 \\
  1d     &  10$^{-2}$  &  1   & 0 & 2  & no  & n.a. & n.a.     & 857 \\ 
  1e     &  10$^{-2}$  &  1   & 0 & 1  & no  & n.a. & n.a.     & 857 \\
  \noalign{\smallskip}   
  \hline  
  \noalign{\smallskip}         
  2a     &  10$^{-2}$  &  20   & 9 & 9 & yes & 655 & 2025 & 184 \\  
  2b     &  10$^{-2}$  &  20   & 0 & 9 & no  &655 & 576  & 184 \\  
  \noalign{\smallskip}   
  \hline
  \noalign{\smallskip}         
  3a     &  10$^{-4}$  &  1   & 9 &  9 & yes & 907 & 1740     & 815 \\  
  3b     &  10$^{-4}$  &  1   & 0 &  9 & no & 907 & 858-1192 & 815 \\ 
  3c     &  10$^{-4}$  &  1   & 0 &  5 & no & n.a. & 858      & 815 \\ 
  3d     &  10$^{-4}$  &  1   & 0 &  2 & no & n.a. & n.a.     & 815 \\ 
  3e     &  10$^{-4}$  &  1   & 0 &  1 & no & n.a. & n.a.     & 815 \\ 
  \noalign{\smallskip}   
  \hline
  \noalign{\smallskip}         
  4a     &  10$^{-4}$  &  20   & 9 & 9 & yes & 658 & 3177 & 172 \\  
  4b     &  10$^{-4}$  &  20   & 0 & 9 & no & 658 & 560  & 172 \\
  \noalign{\smallskip}           
  \hline                    
  \noalign{\smallskip}   
  5a     &  10$^{-3}$  &  1   & 9 & 9 & yes & 750  & 1438  & 838 \\
  5b     &  10$^{-3}$  &  1   & 0 & 9 & no & 750  & 1096  & 838 \\
  6a     &  10$^{-3}$  &  5   & 9 & 9 & yes & 772  & 1731  & 352 \\  
  6b     &  10$^{-3}$  &  5   & 0 & 9 & no & 772  & 1000  & 352 \\
  7a     &  2.5 $\times$ 10$^{-3}$  &  5   & 9 & 9 & yes & 734  & 1628 & 357 \\
  7b     &  2.5 $\times$ 10$^{-3}$  &  5   & 0 & 9 & no & 734  & 915  & 357 \\
  8a     &  5 $\times$ 10$^{-3}$  &  5   & 9 & 9 & yes & 726  & 1575 & 360 \\
  8b     &  5 $\times$ 10$^{-3}$  &  5   & 0 & 9 & no & 726  &  872 & 360 \\
  \noalign{\smallskip}         
  \hline                    
\end{tabular}  
\end{table*}
CO ro-vibrational transitions have moderate Einstein probabilities
($A$=1-10 s$^{-1}$) so that the line transfer occurs in the Doppler
core of the line profile.  On the other hand, UV-fluorescence pumping
occurs via electronic transitions ($A^1\Pi$--$X^1\Sigma^+$) with high
Einstein probabilities ($A$=10$^4$-10$^7$ s$^{-1}$) and thus with
large natural line width.  The line core becomes quickly optically
thick (line self-shielding) with the UV line transfer happening in the
Lorentzian wings of the line. Therefore the use of the Voigt profile
in the line transfer is warranted.

The CO levels are computed using a 1+1D escape probability method for
Voigt line profiles. We adapted the analytical formulation of
\citet{Apruzese1985JQSRT..34..447A} to match the formulation in
\citet{Woitke2009A&A...501..383W} at low optical depths and small
intrinsic line widths. At high optical depth, the escape probability
$\beta$ varies as $\tau^{-1/2}$ (Voigt profile) instead of $\tau^{-1}$
(Gaussian profile):
\begin{equation}
\beta= \left\{
\begin{array}{ll}
0.5(1+1.5\tau)^{-1} & \mathrm{if\ } \tau \le 1\\
0.25\tau^{-1.15}  &  \mathrm{if\ } 1<\tau<\tau_c\\
0.25\tau^{-0.5}\tau_c^{-0.65} & \mathrm{if\ } \tau>\tau_c,\\
\end{array}\right.
\end{equation}
where $a$ is defined as $\Gamma/(4\pi\Delta\nu_D)$ and $\tau_c$ is the
critical optical depth and is defined as $0.83/(a(1+\sqrt{a}))$. The
sum of the natural and collisional width is $\Gamma$ while the
effective Doppler width is $\Delta\nu_D$
\citep{Rybicki1986rpa..book.....R}.  The escape probability function
for a Voigt line profile is illustrated in
Fig.~\ref{voigt_escape_proba} for the case $a=10^{-3}$.  In addition
to the UV line shielding, dust grains contribute strongly to the UV
flux attenuation. The main limitation in using the escape probability
technique is that it does not take overlapping line effects into
account.
\begin{figure*}[ht] 
  \centering 
  \includegraphics[scale=0.3]{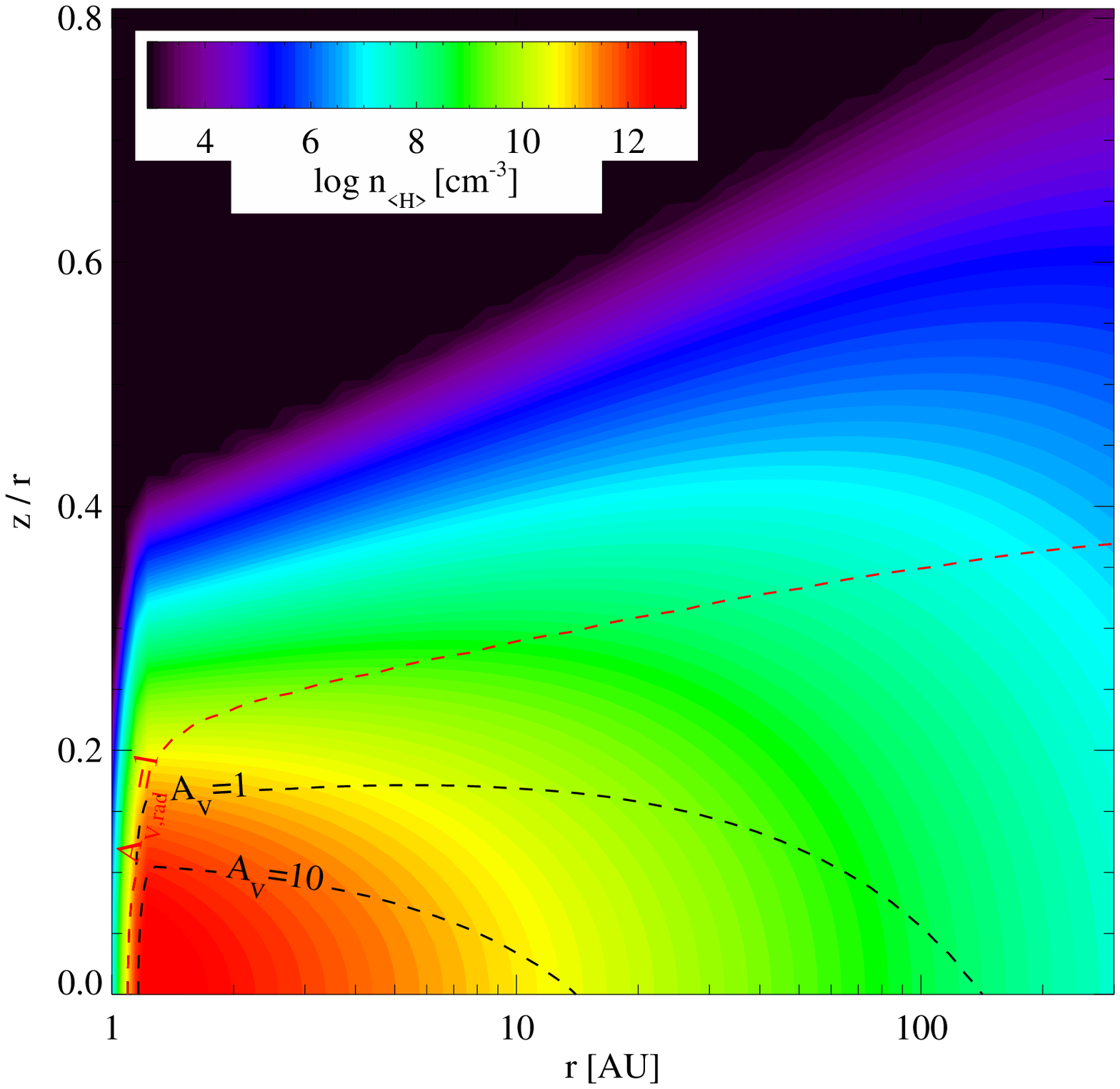}
  \includegraphics[scale=0.3]{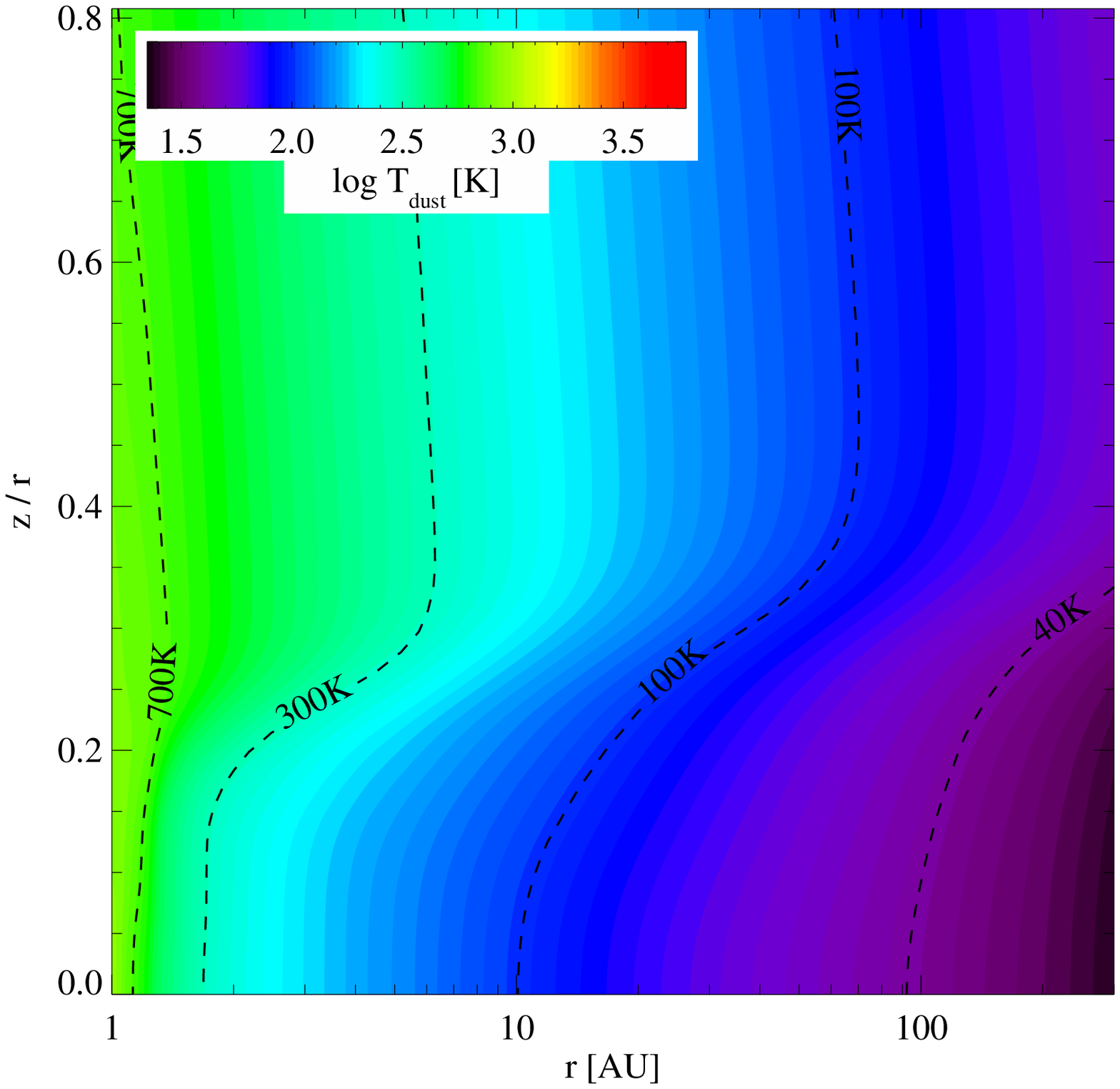}
  \includegraphics[scale=0.3]{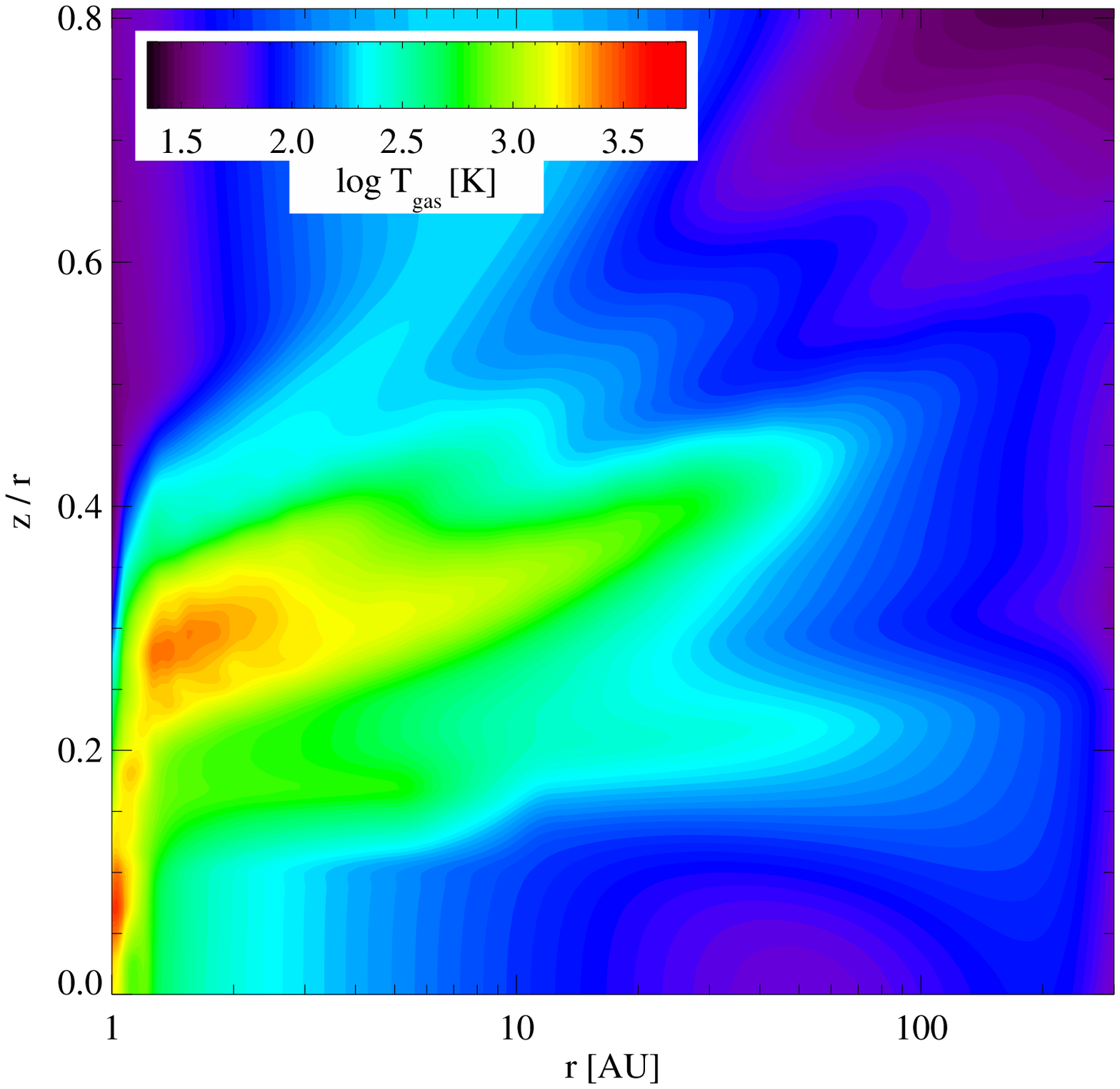}
\centering  
\caption{\label{disc_structur_3panels} Density, $T_{\mathrm{dust}}$,
  and $T_{\mathrm{gas}}$ structures for model 1a. The contours in
  black dashed lines in the density panel show the location of
  $A_{\mathrm{V}}$=1 and 10 from $z=\infty$. The red dashed lines mark
  the location of the radial $A_{\mathrm{V}}$=1 calculated from
  $R_{\mathrm{in}}$. In the middle panel various dust temperature
  contours are shown in black dashed lines. In our models,
  $T_{\mathrm{dust}}$ and $T_{\mathrm{gas}}$ are not equal.}

\end{figure*} 
\begin{figure*}[ht] 
  \centering 
  \includegraphics[scale=0.3]{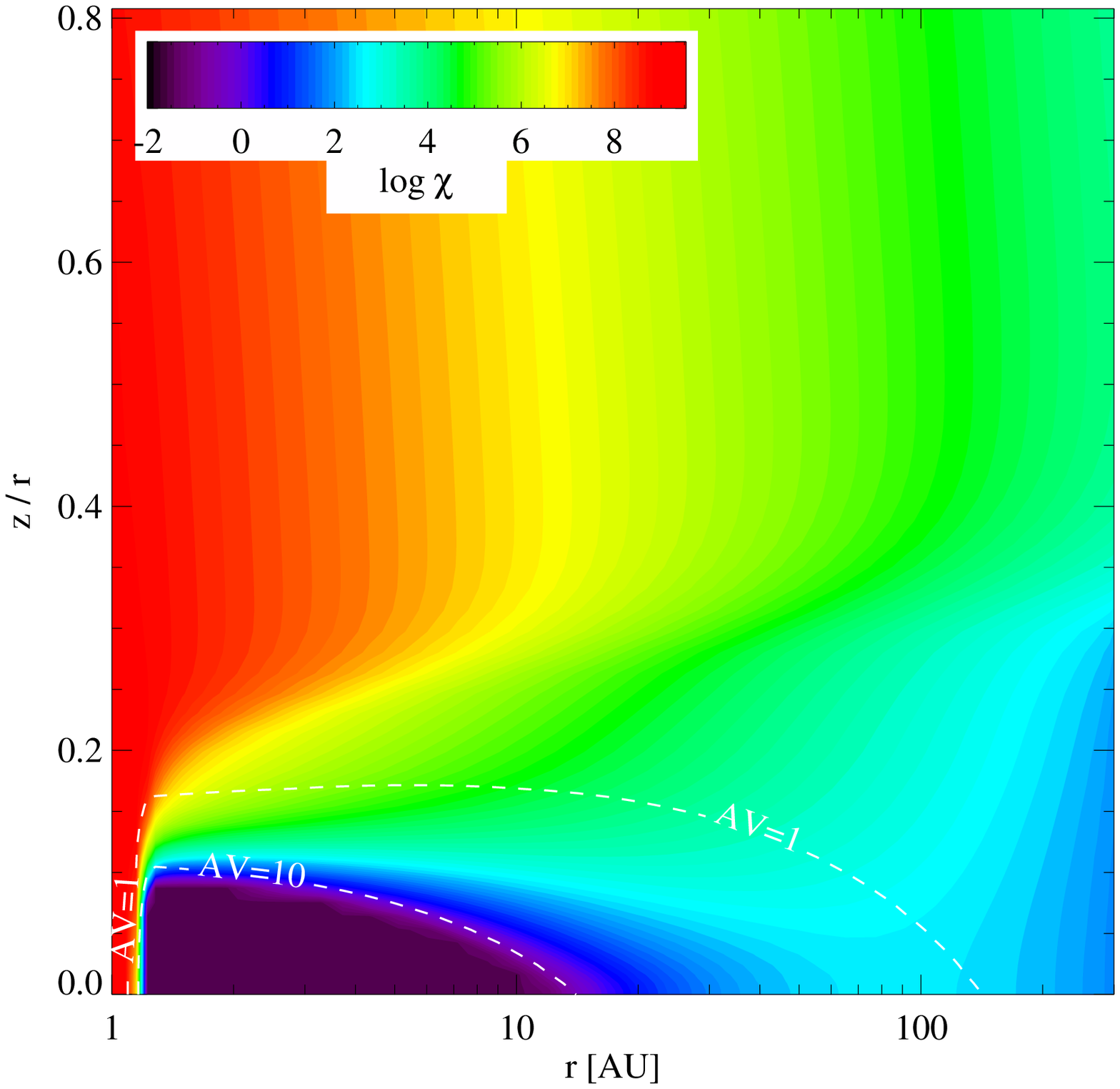}
  \includegraphics[scale=0.3]{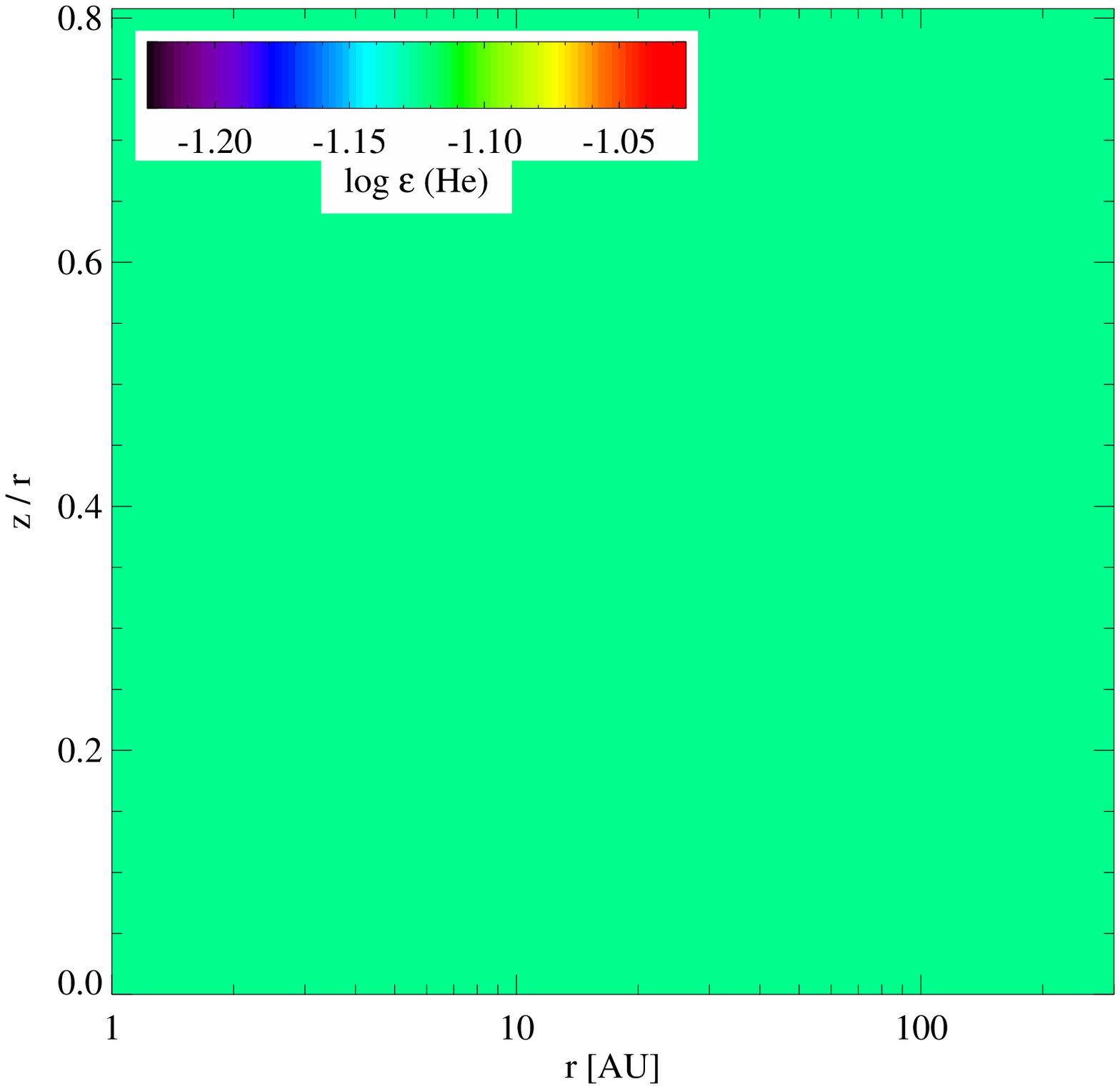}
  \includegraphics[scale=0.3]{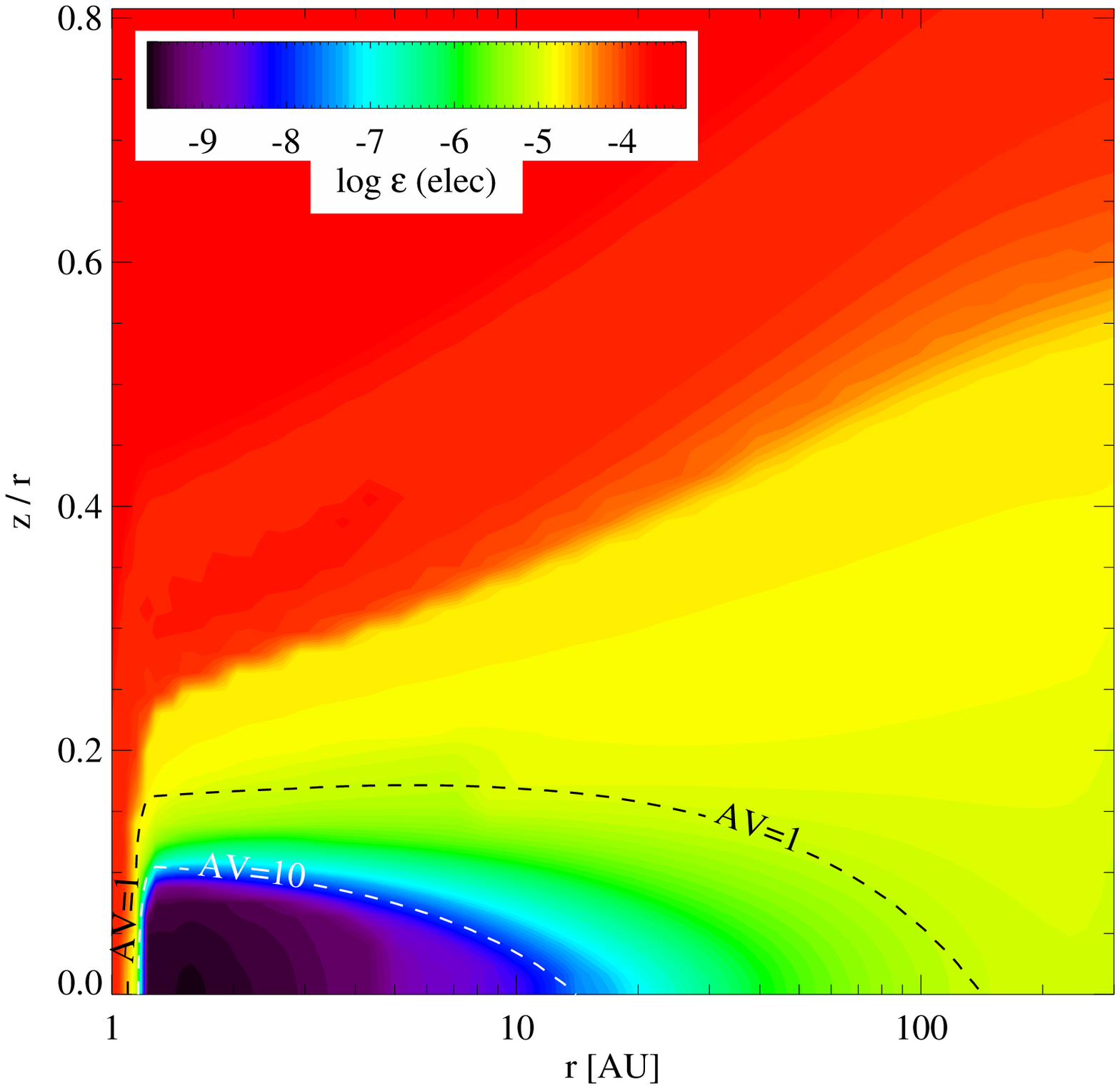}
  \includegraphics[scale=0.3]{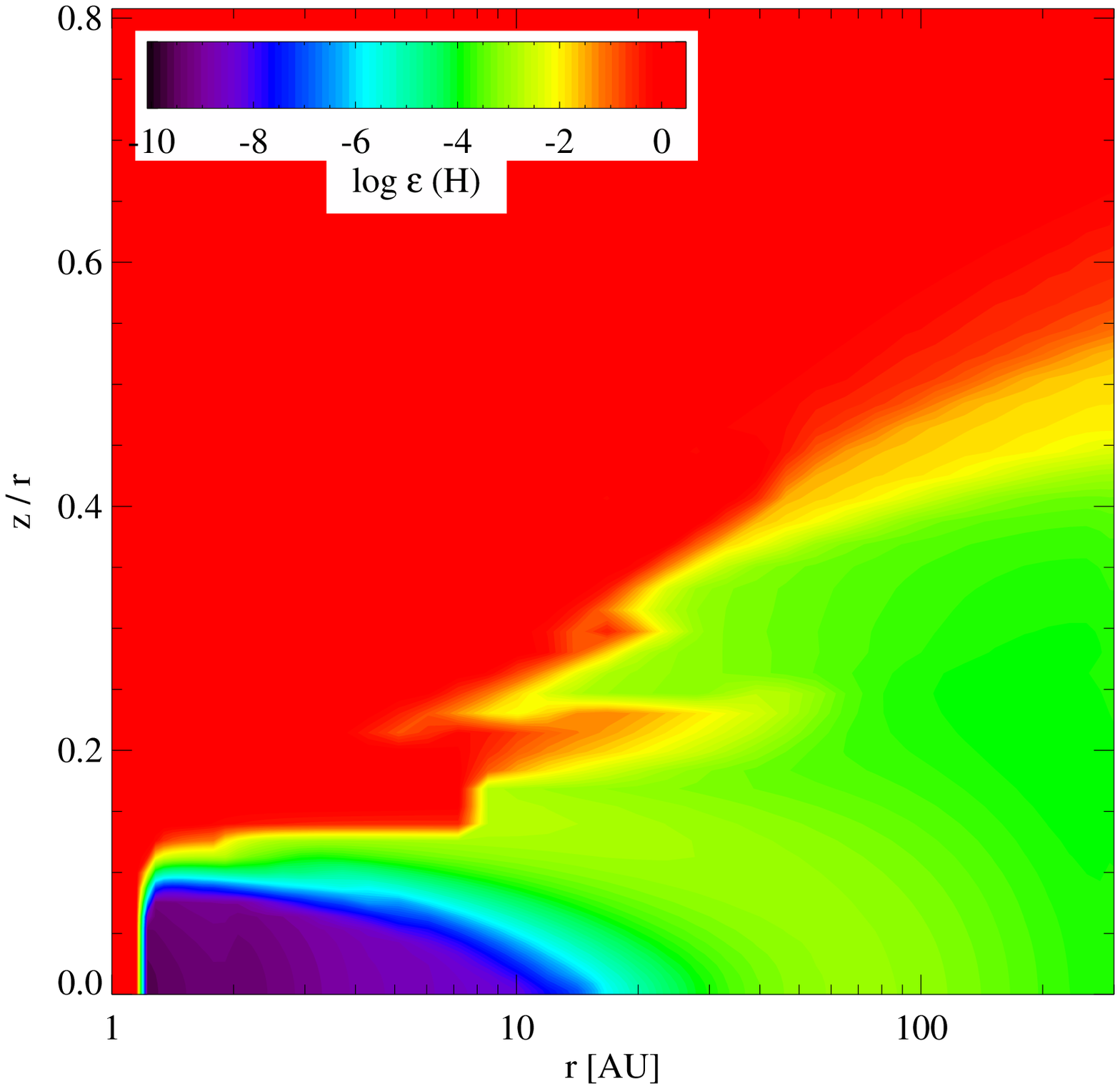}
  \includegraphics[scale=0.3]{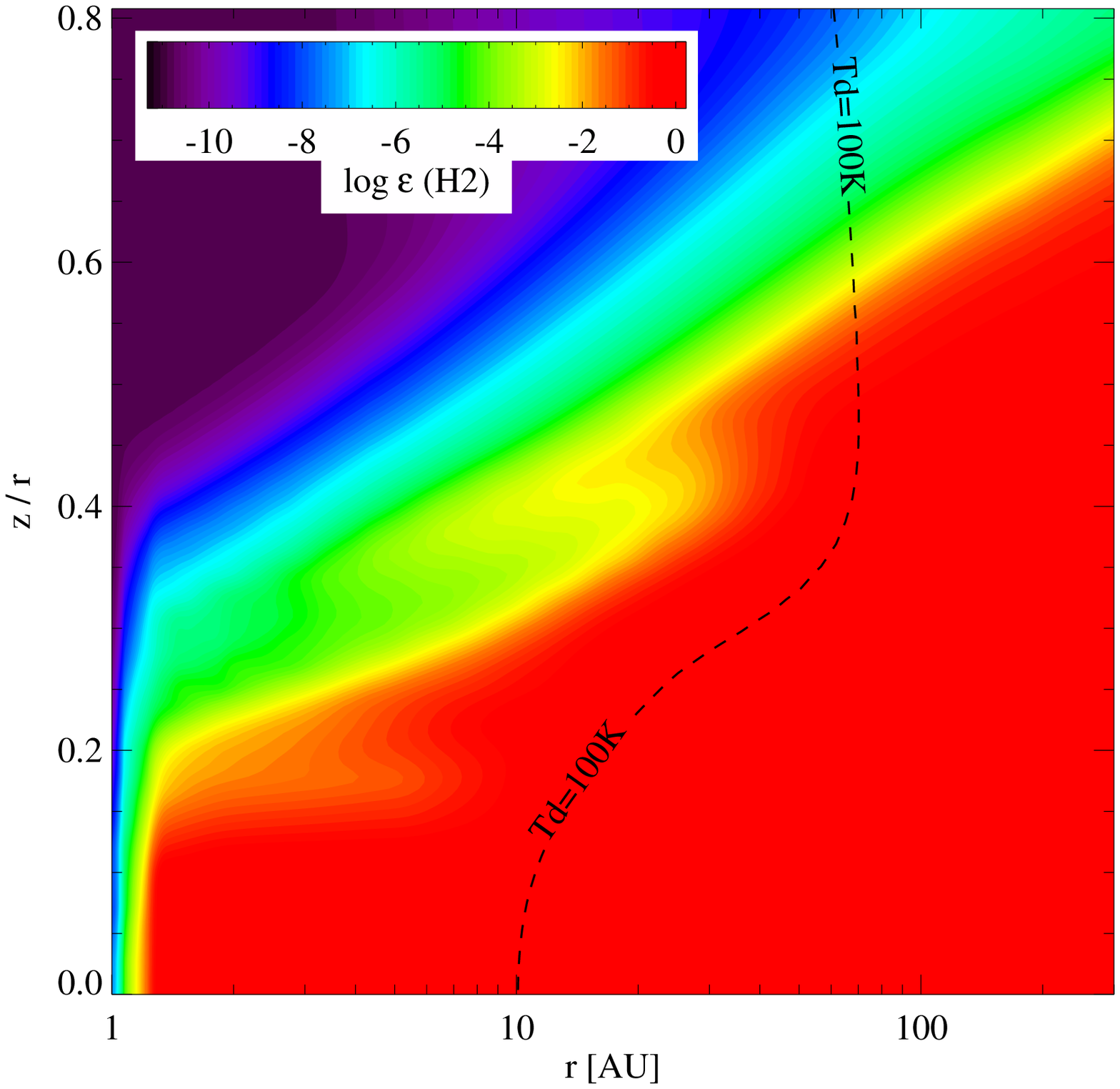}
  \includegraphics[scale=0.3]{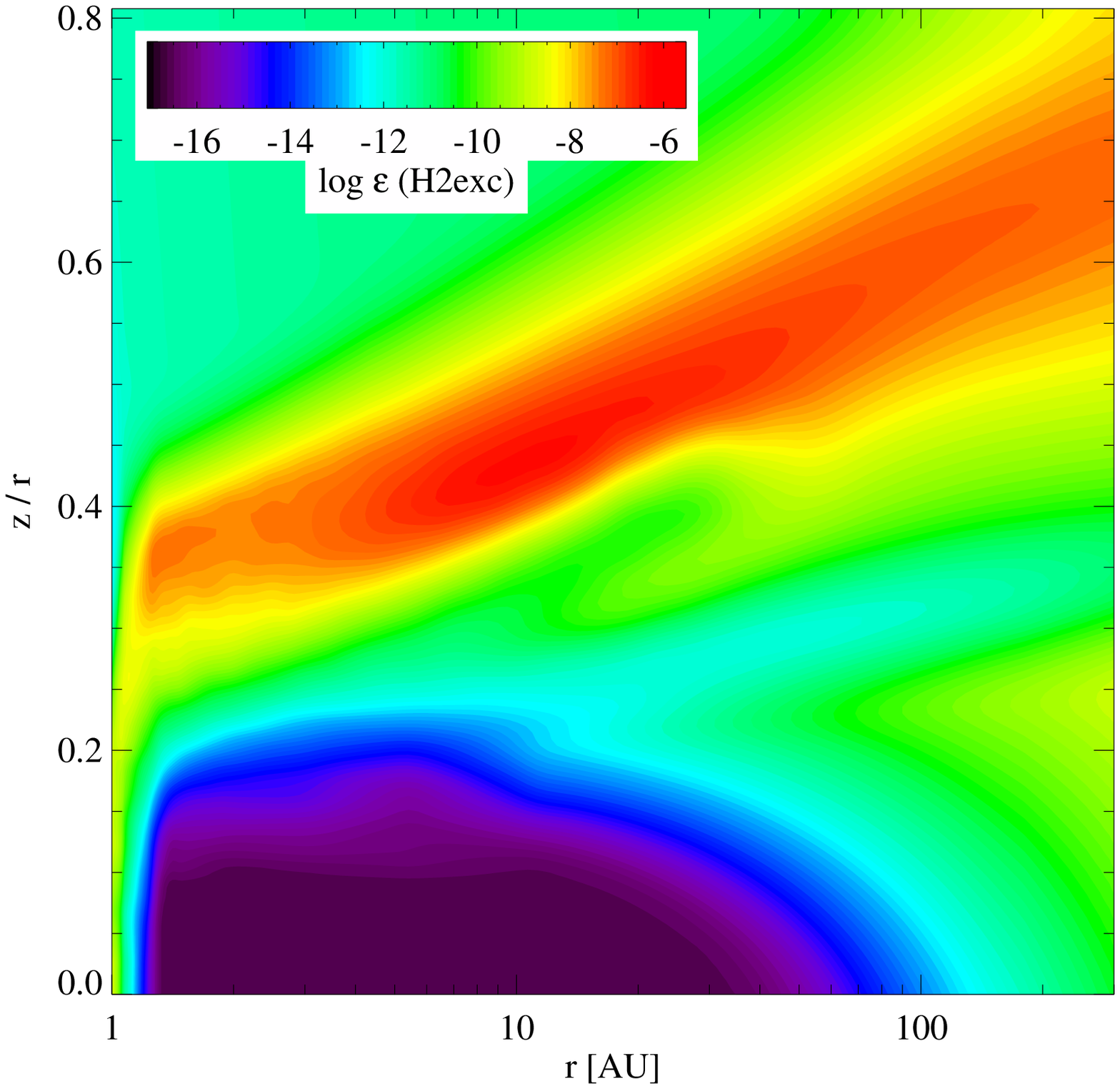}
  \includegraphics[scale=0.3]{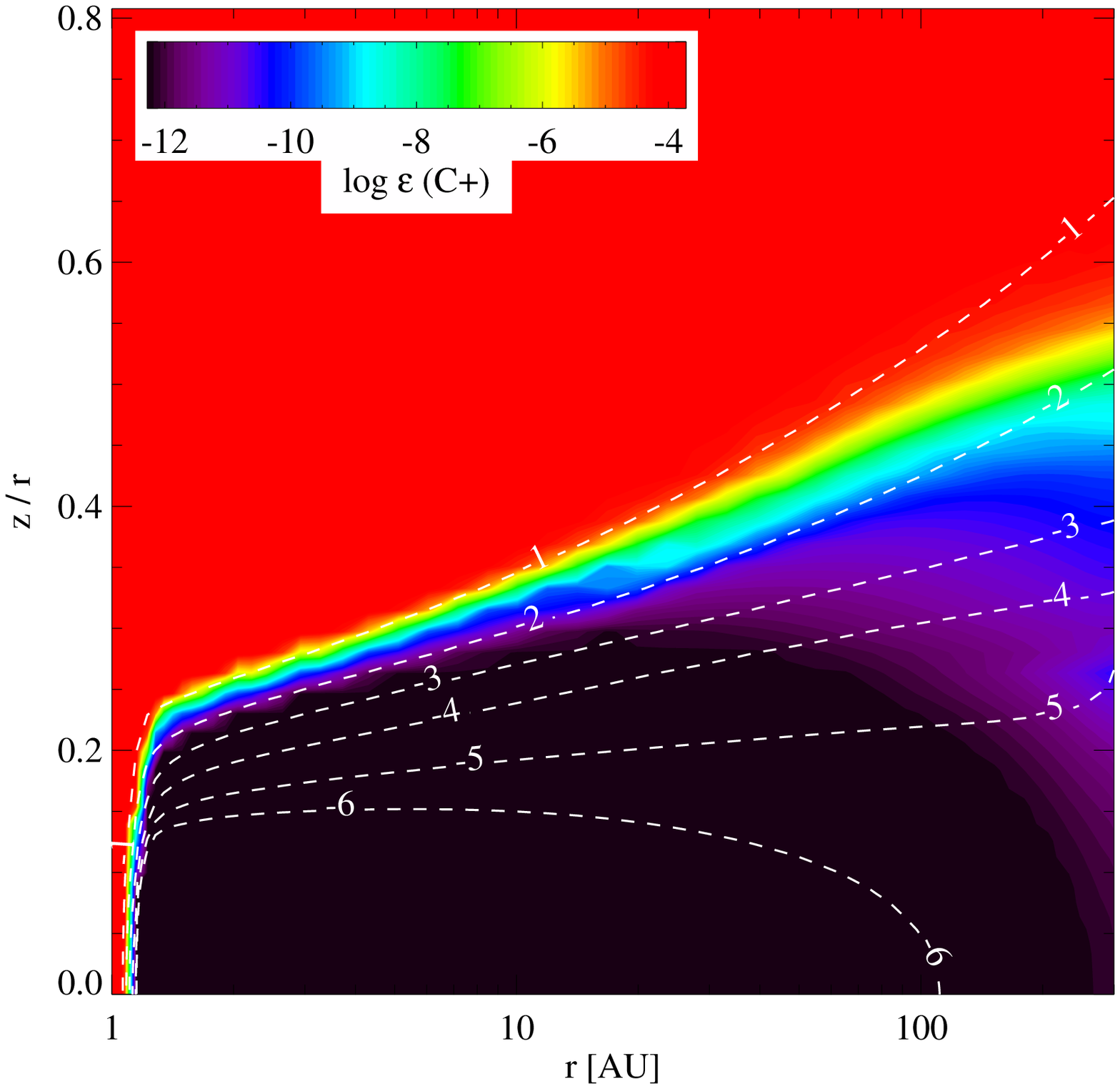}
  \includegraphics[scale=0.3]{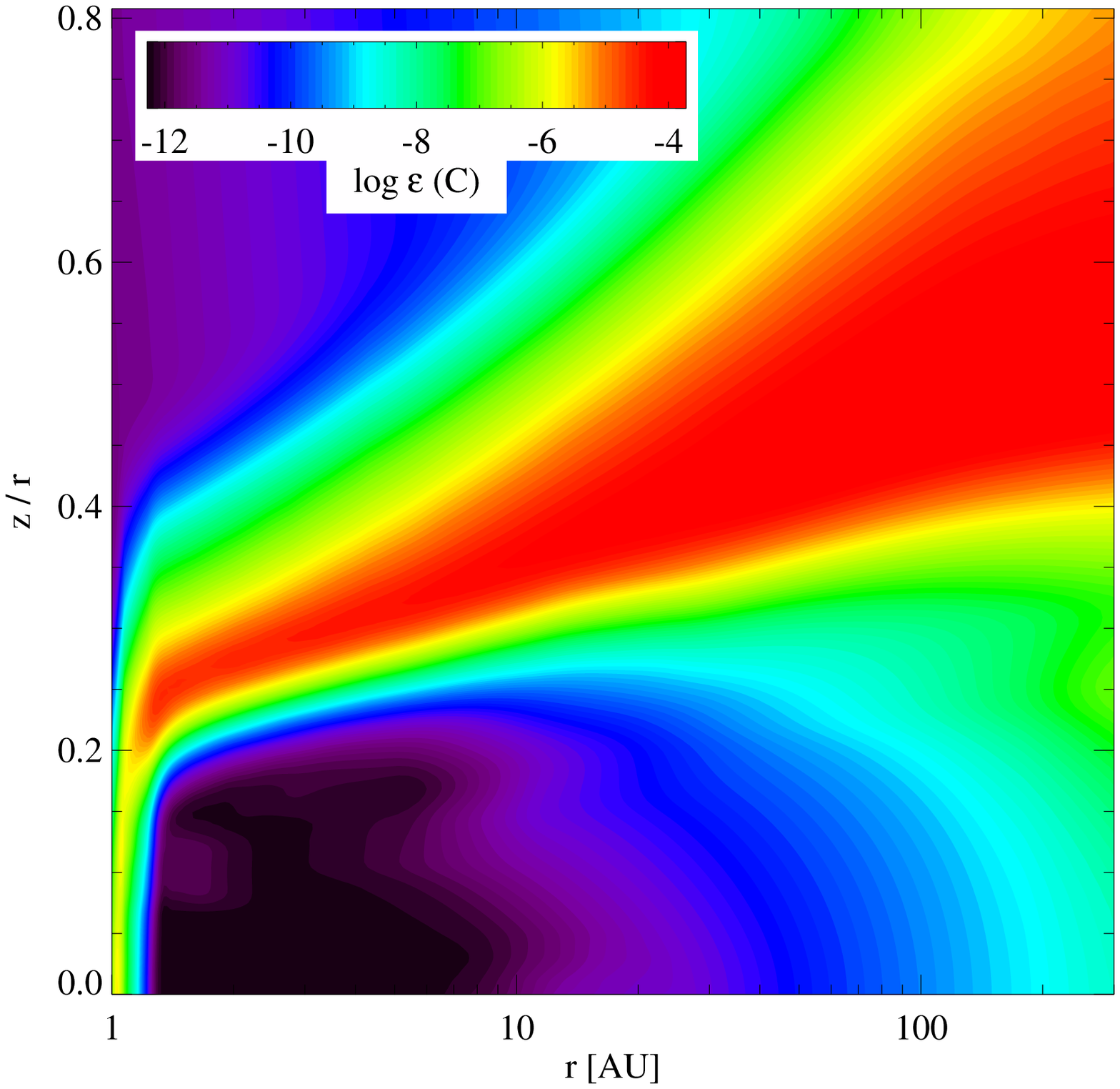}
  \includegraphics[scale=0.3]{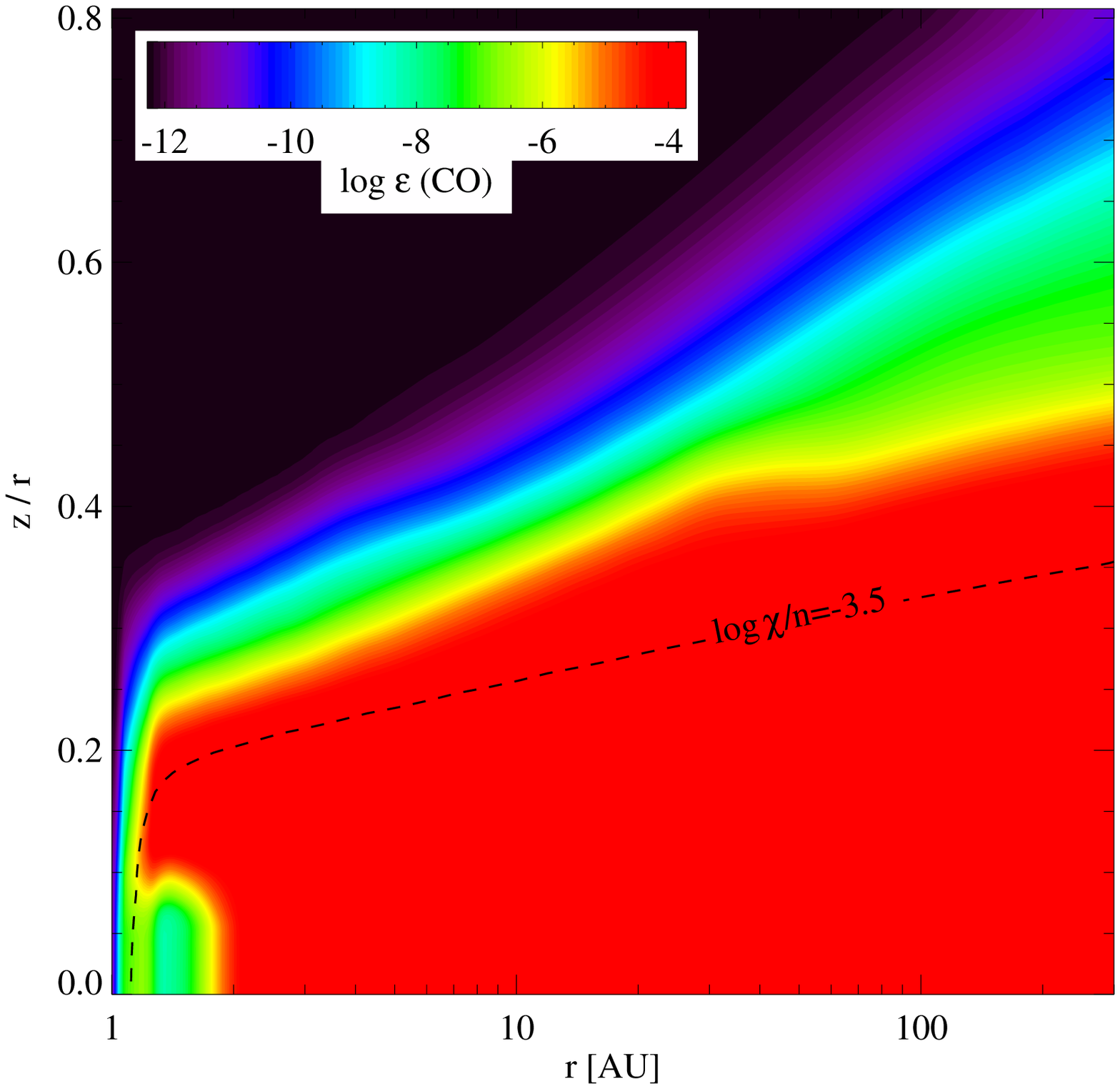}
  \includegraphics[scale=0.3]{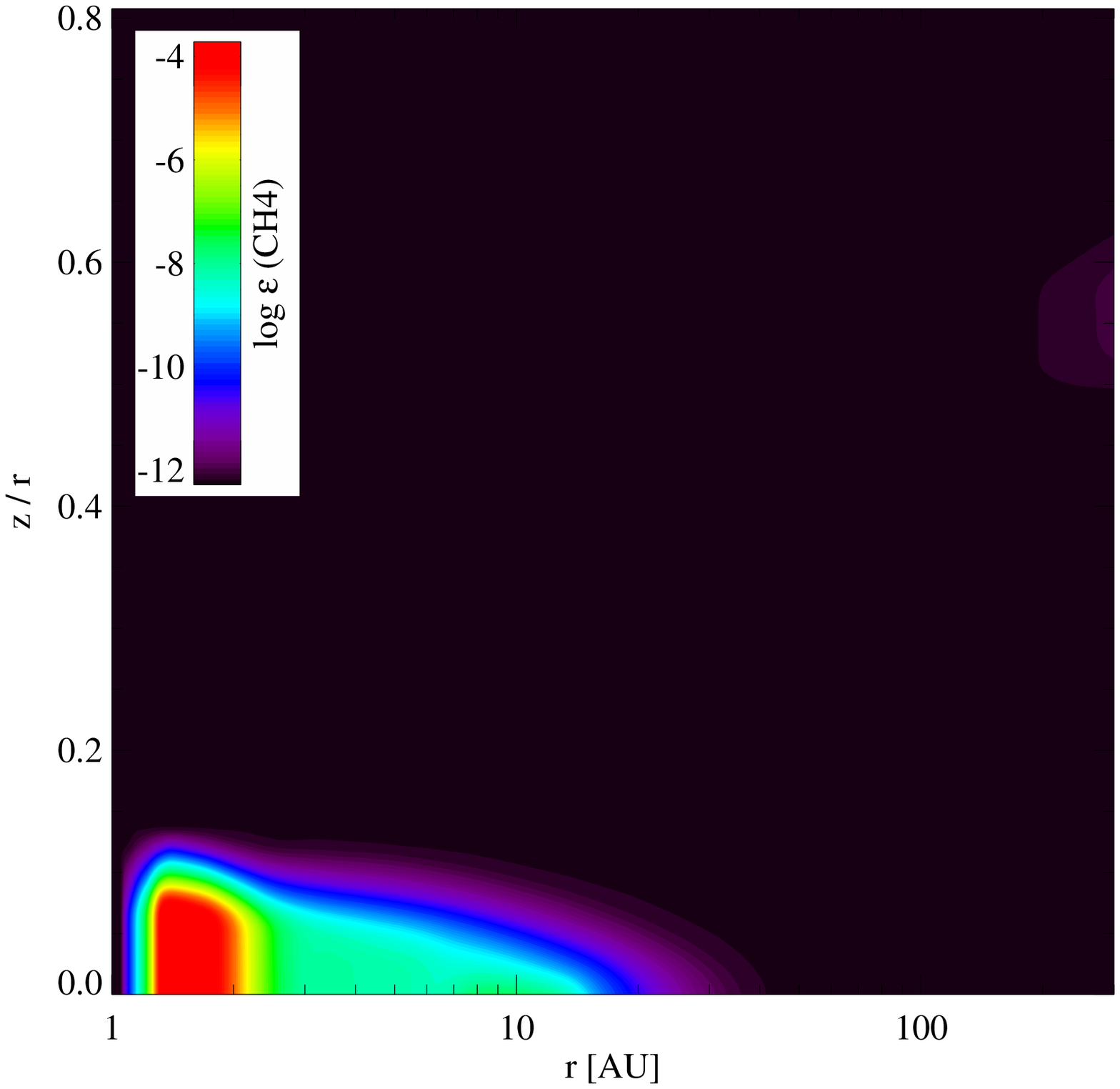}
  \includegraphics[scale=0.3]{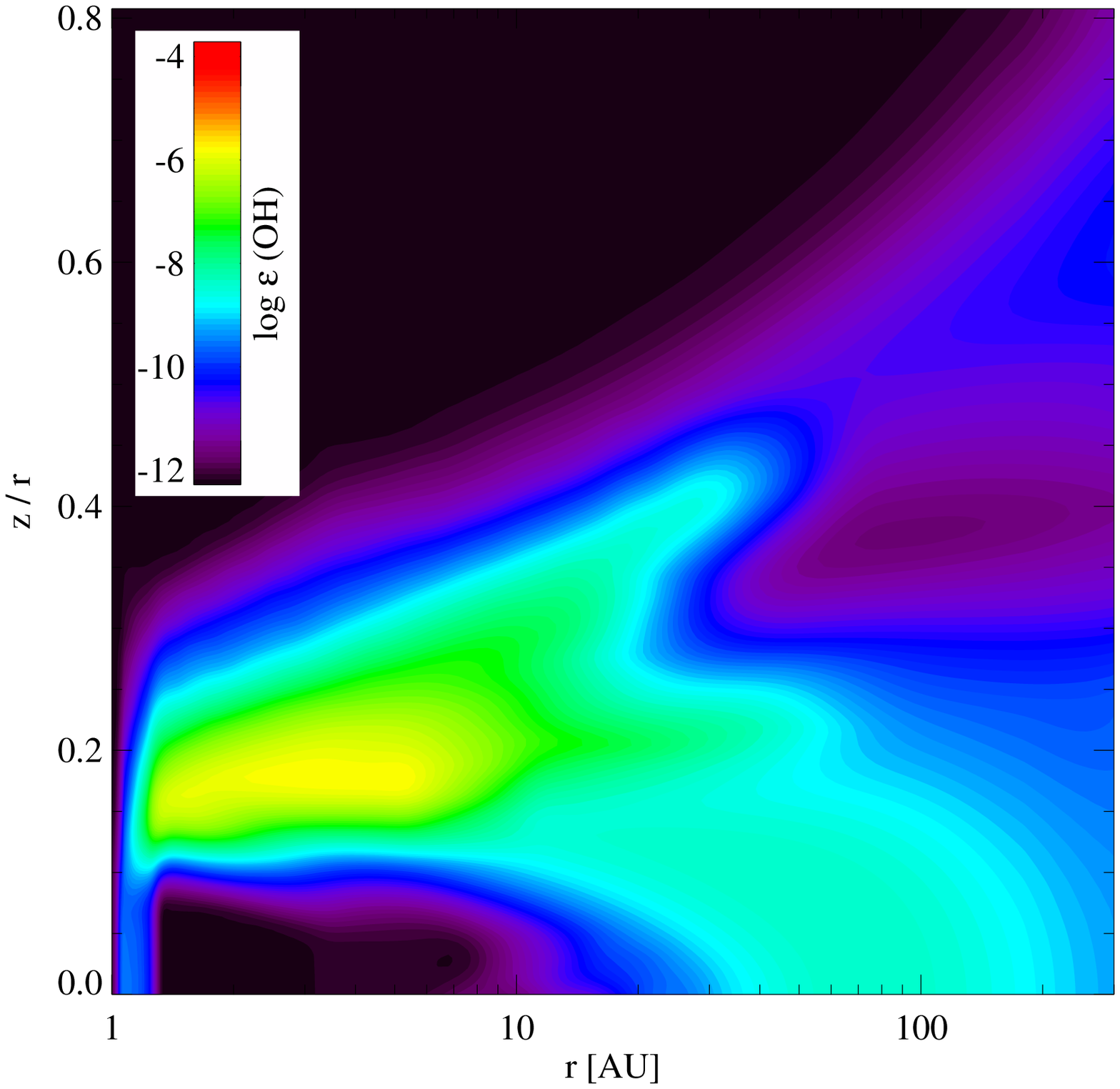}
  \includegraphics[scale=0.3]{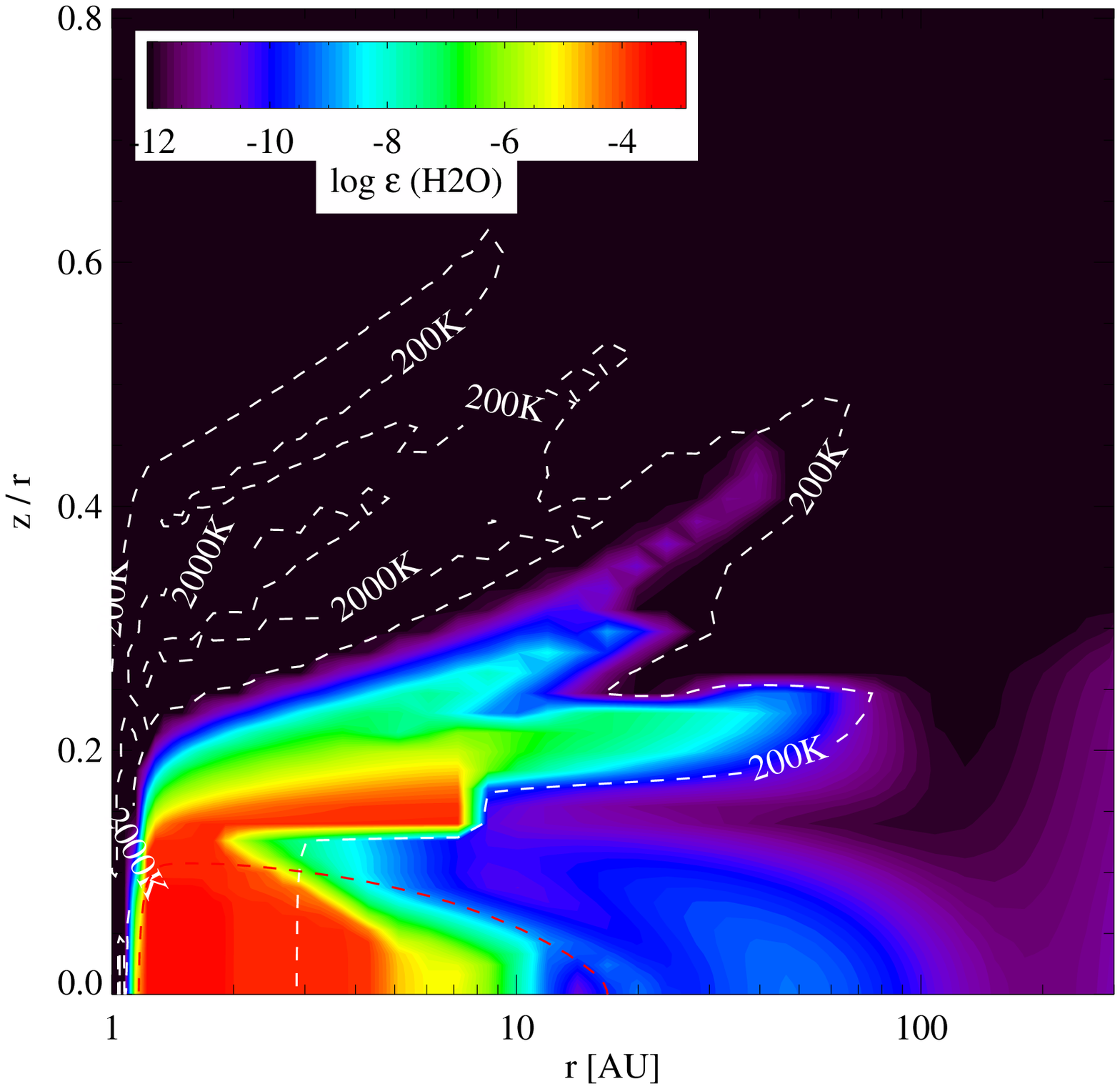}
\centering  
\caption{\label{CO_15panels} Model 1a $\chi$ structures and abundances
  relative to H-nuclei for atomic He, electron, H, H$_2$, H$_2$exc
  ($v>$0), C$^+$, C, CO, CH$_4$, OH, and H$_2$O. The contour plots in
  the ionized carbon abundance plot show $\log(\chi/n$), where $\chi$
  is the enhancement with respect to the interstellar UV and $n$ the
  total number of H-nuclei. In a few panels, dust temperature
  ($T_{\mathrm{d}}$), gas temperature (water abundance panel), or
  extinction $A_{\mathrm{V}}$ contours are overplotted by dashed
  lines.}
\end{figure*}
\begin{figure*}[ht] 
  \centering 
  \includegraphics[scale=0.5]{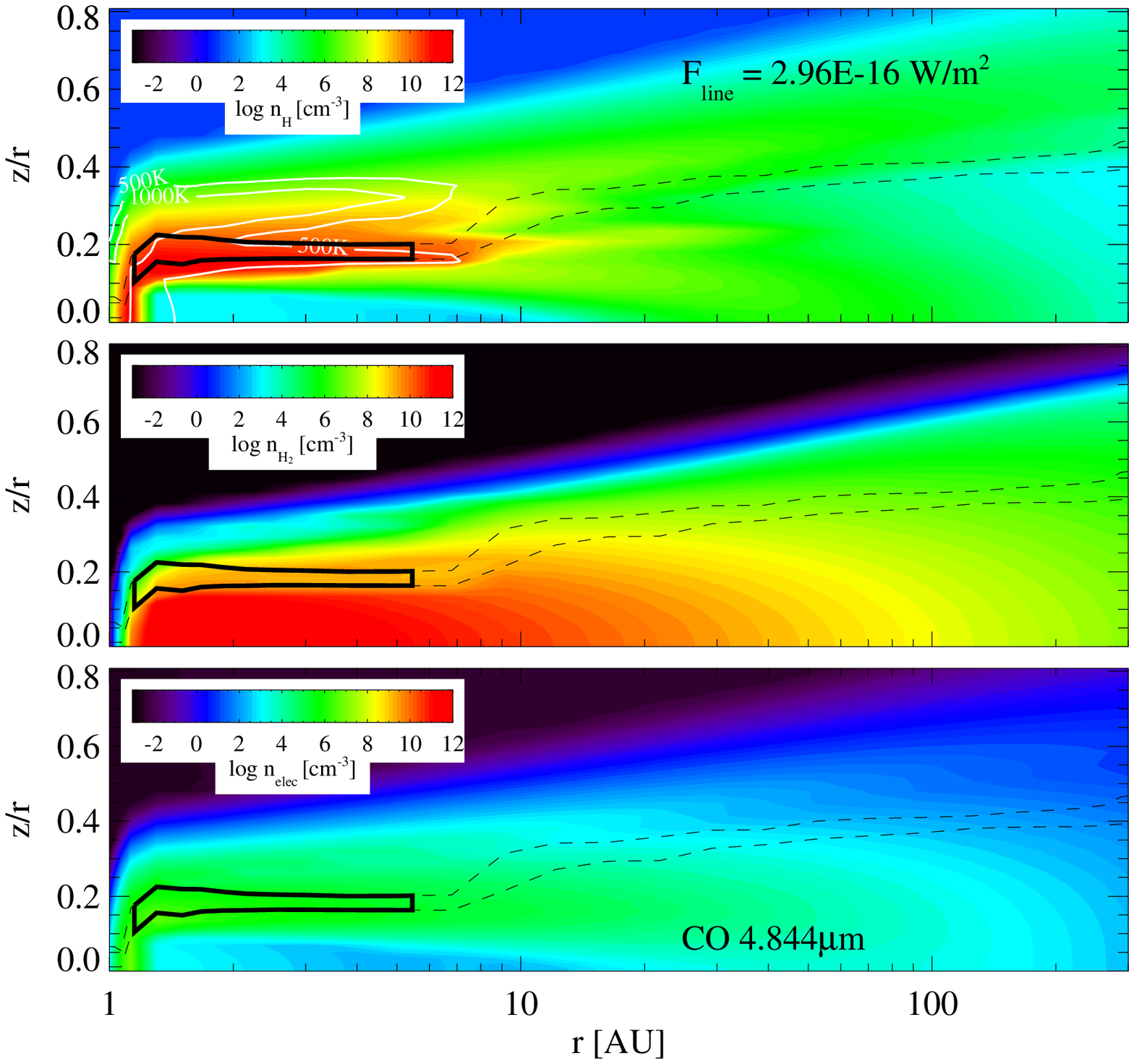}
  \includegraphics[scale=0.5]{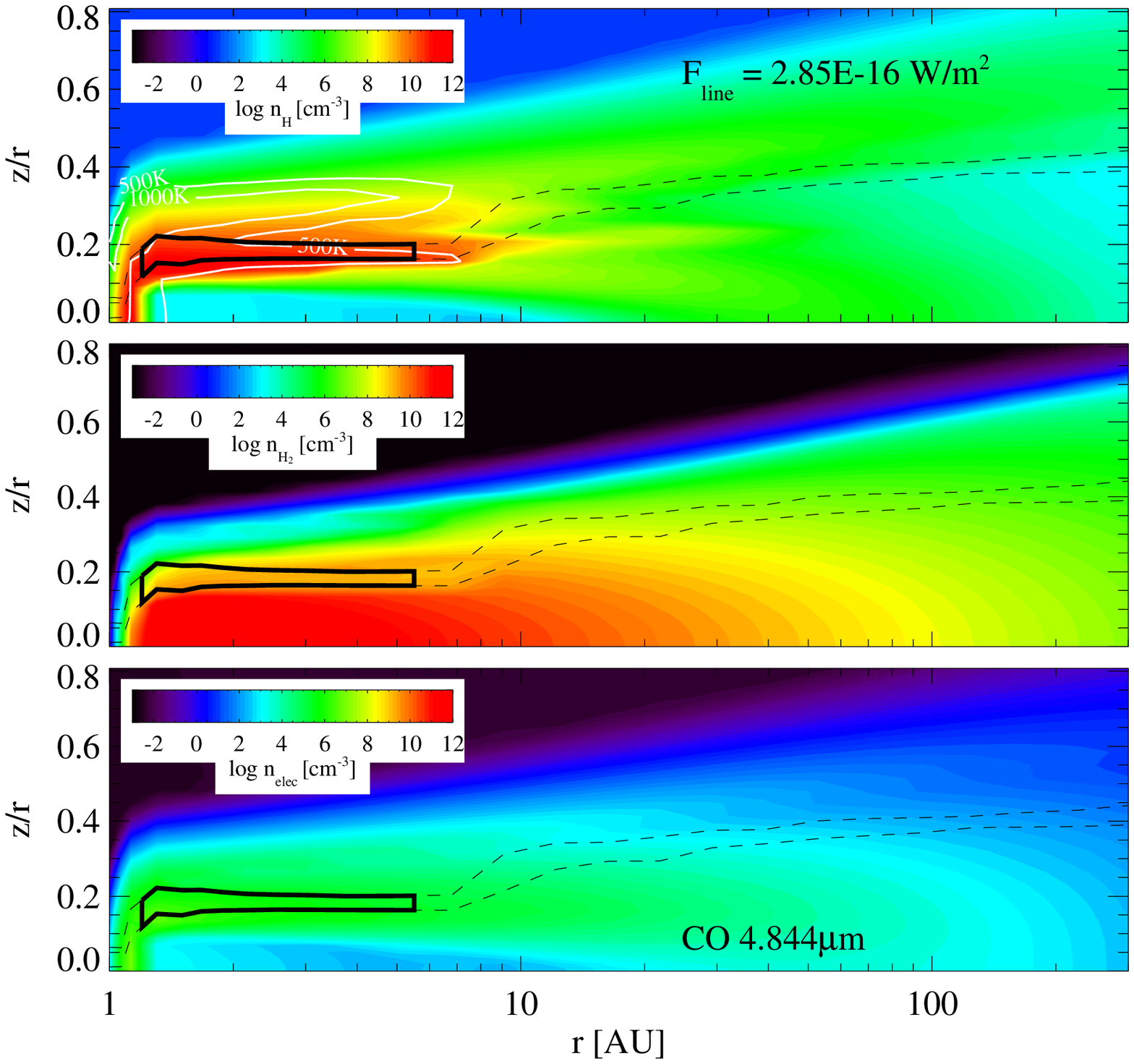}
  \includegraphics[scale=0.5]{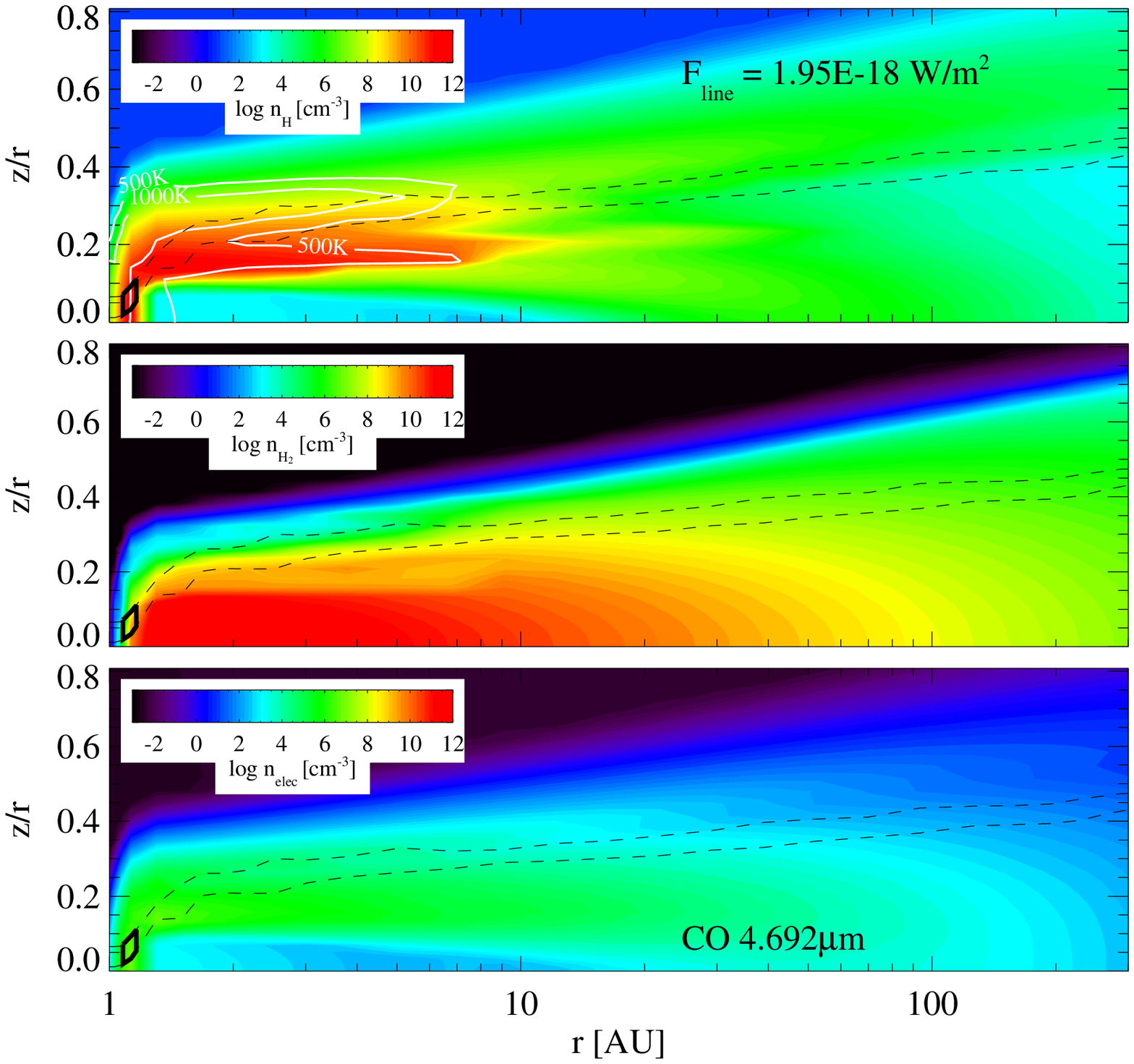}
  \includegraphics[scale=0.5]{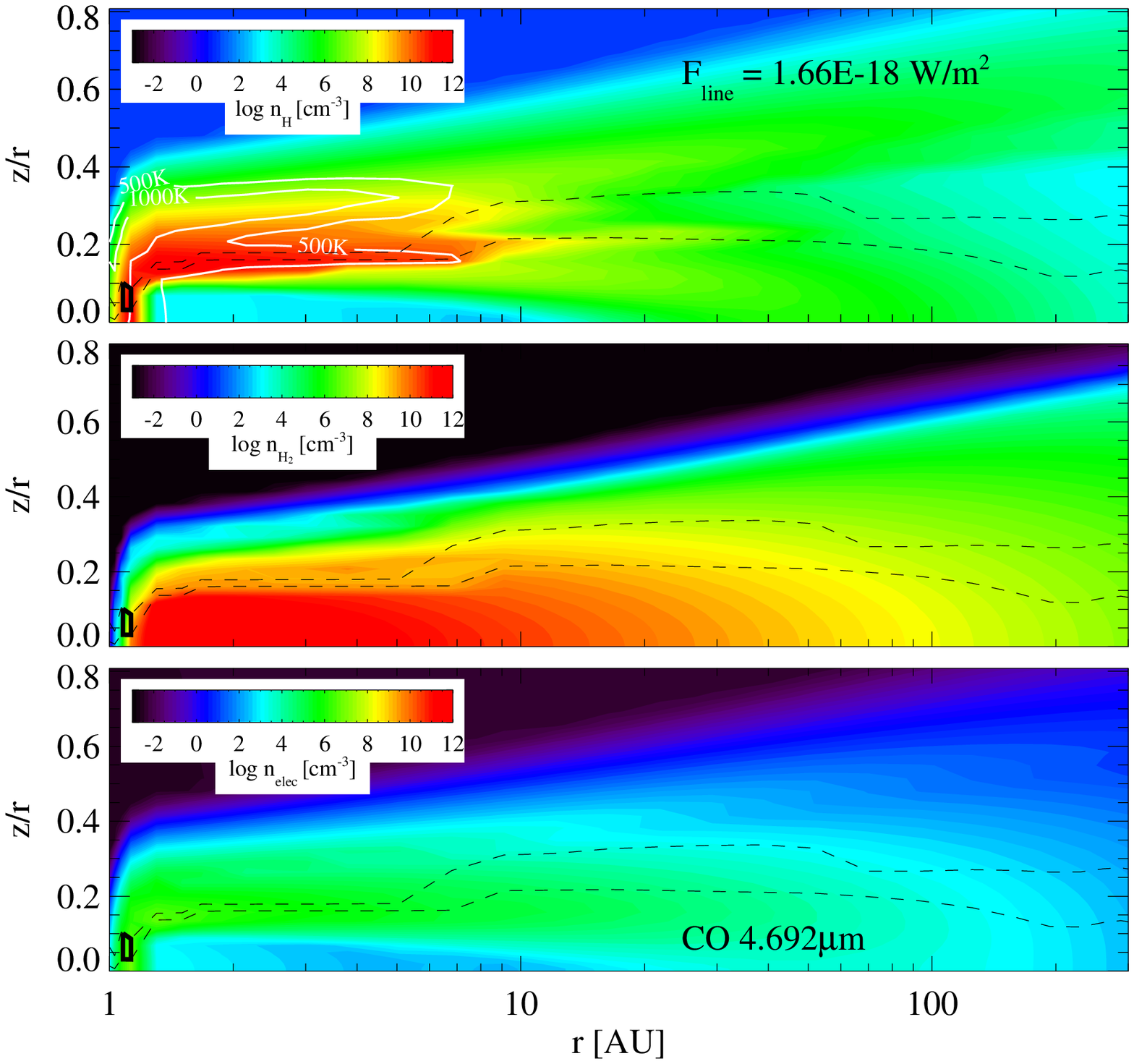}
\centering
\caption{\label{CO_emission_high_mass_disk_4panels} Location of the CO
  $v=1-0$ $J=19-18$ $P$(19) at 4.844 $\mu$m and CO $v=4-3$ $J=20-19$
  $P$(20) at 4.692 $\mu$m line emissions for the
  $M_{\mathrm{disc}}$=10$^{-2}$ M$_\odot$, $R_{\mathrm{in}}$=1~AU disc
  models with (left panels, model 1a) and without (right panels, model
  1b) UV-pumping.  The solid black contours in the CO density panels
  encompass the regions that emit 49\% of the fluxes. The black
  dashed-line contours contain 70\% of the fluxes in the vertical
  direction. The white contours are at gas temperatures of 500 and
  1000K. Each panel is divided into three sub-panels. The upper
  sub-panels show in color scale the atomic hydrogen density
  structure, the middle-panels show the molecular hydrogen density
  structure, and the lower-panels show the electron density
  structure.}
\end{figure*}

\begin{figure*}[ht] 
  \centering 
  \includegraphics[scale=0.5]{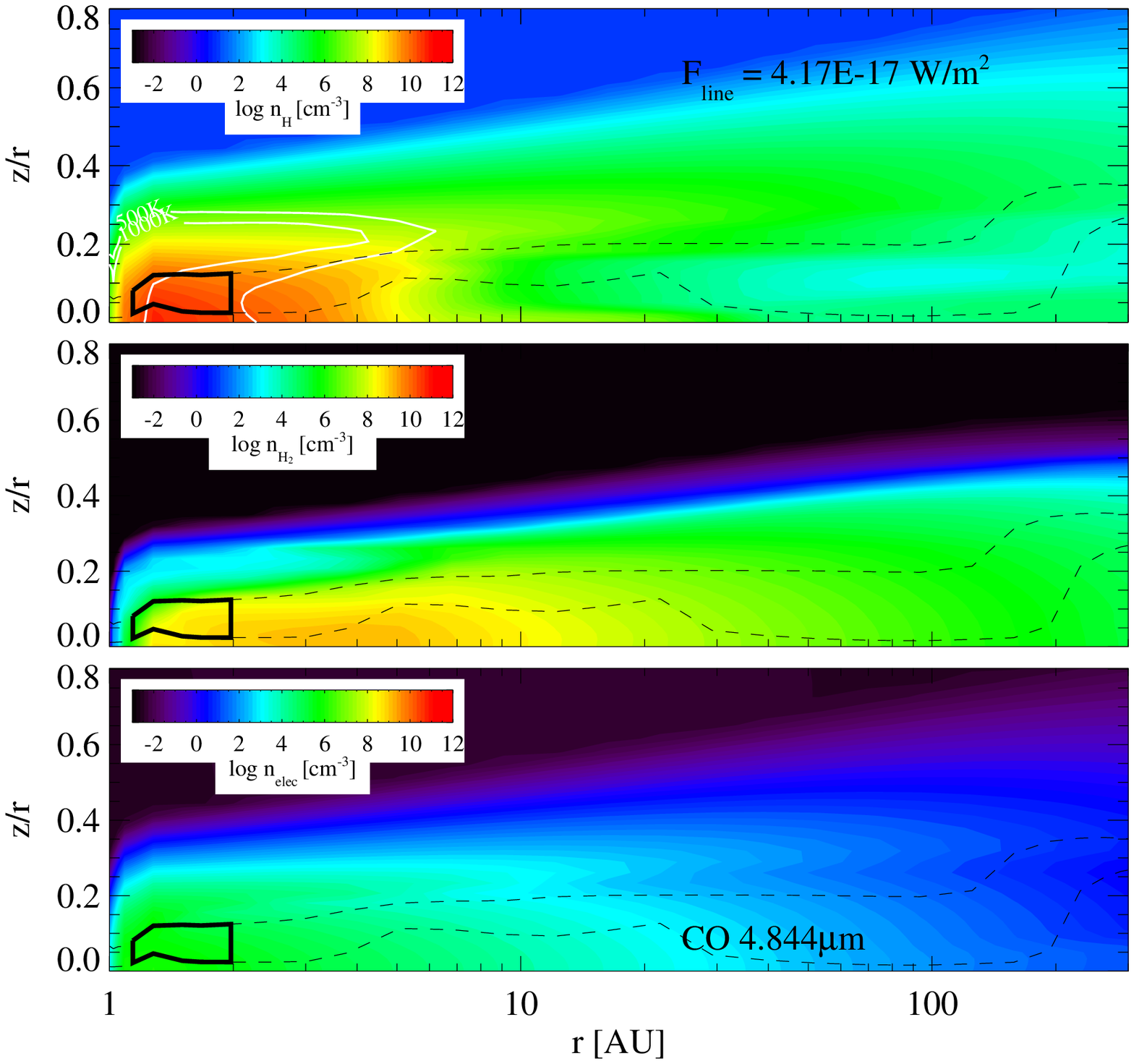}
  \includegraphics[scale=0.5]{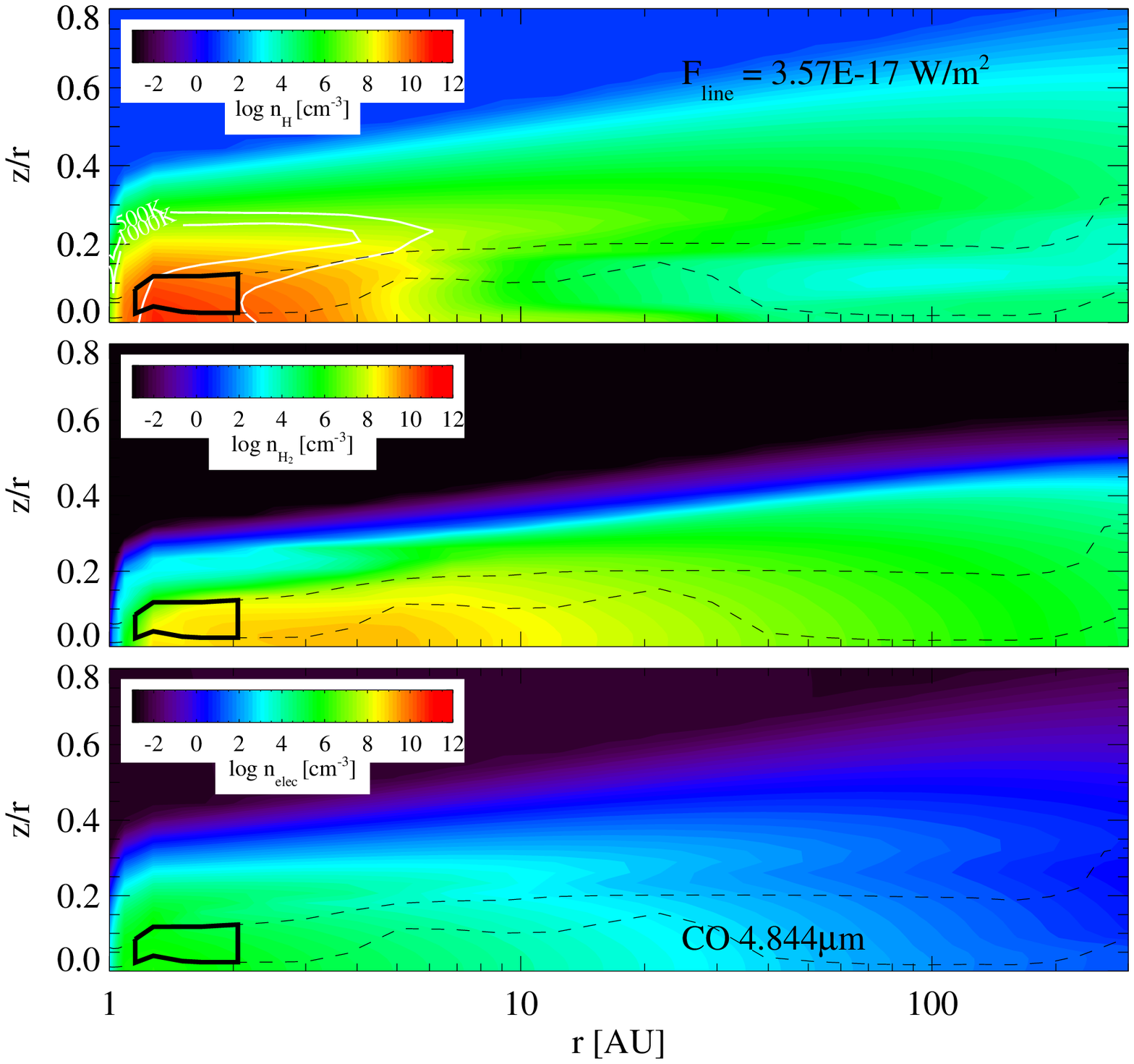}
  \includegraphics[scale=0.5]{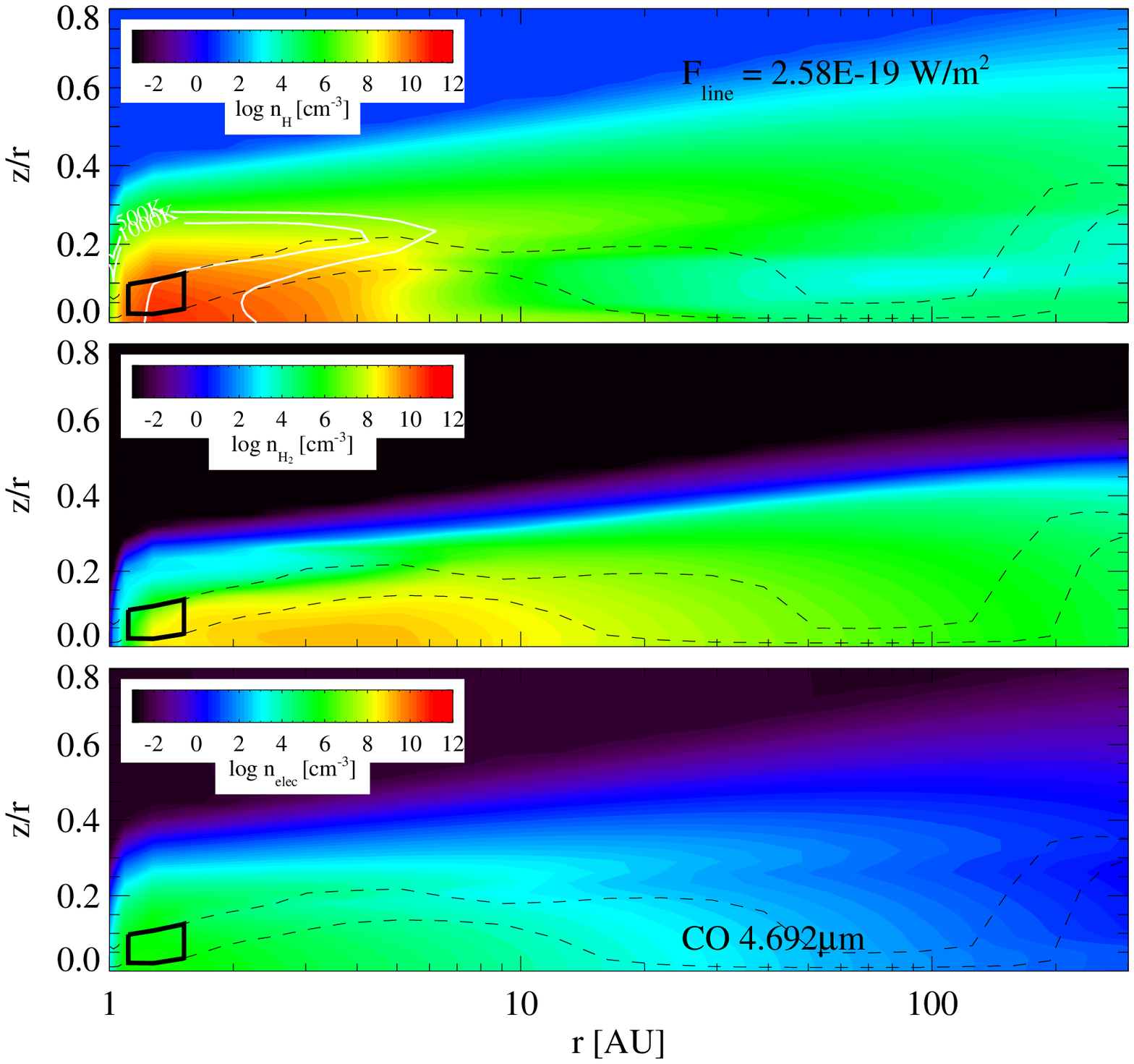}
  \includegraphics[scale=0.5]{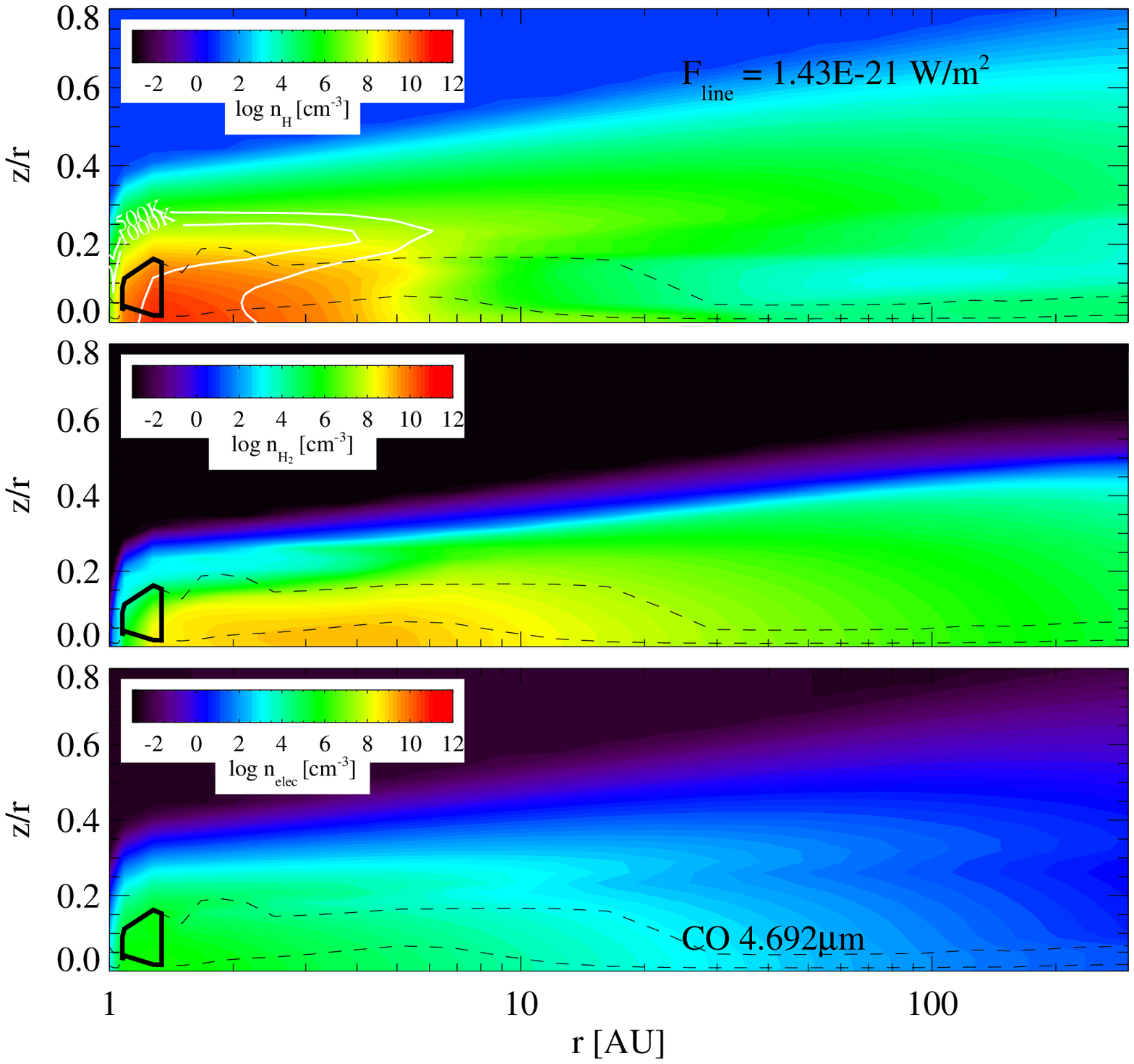}
\centering
\caption{\label{CO_emission_low_mass_disk_4panels} Similar to
  Fig.~\ref{CO_emission_high_mass_disk_4panels} but for the
  $M_{\mathrm{disc}}$=10$^{-4}$ M$_\odot$, $R_{\mathrm{in}}$=1~AU disc
  models (3a and 3b).}
\end{figure*}

\begin{figure*}[ht]  
  \centering  
  \includegraphics[scale=0.37,angle=90]{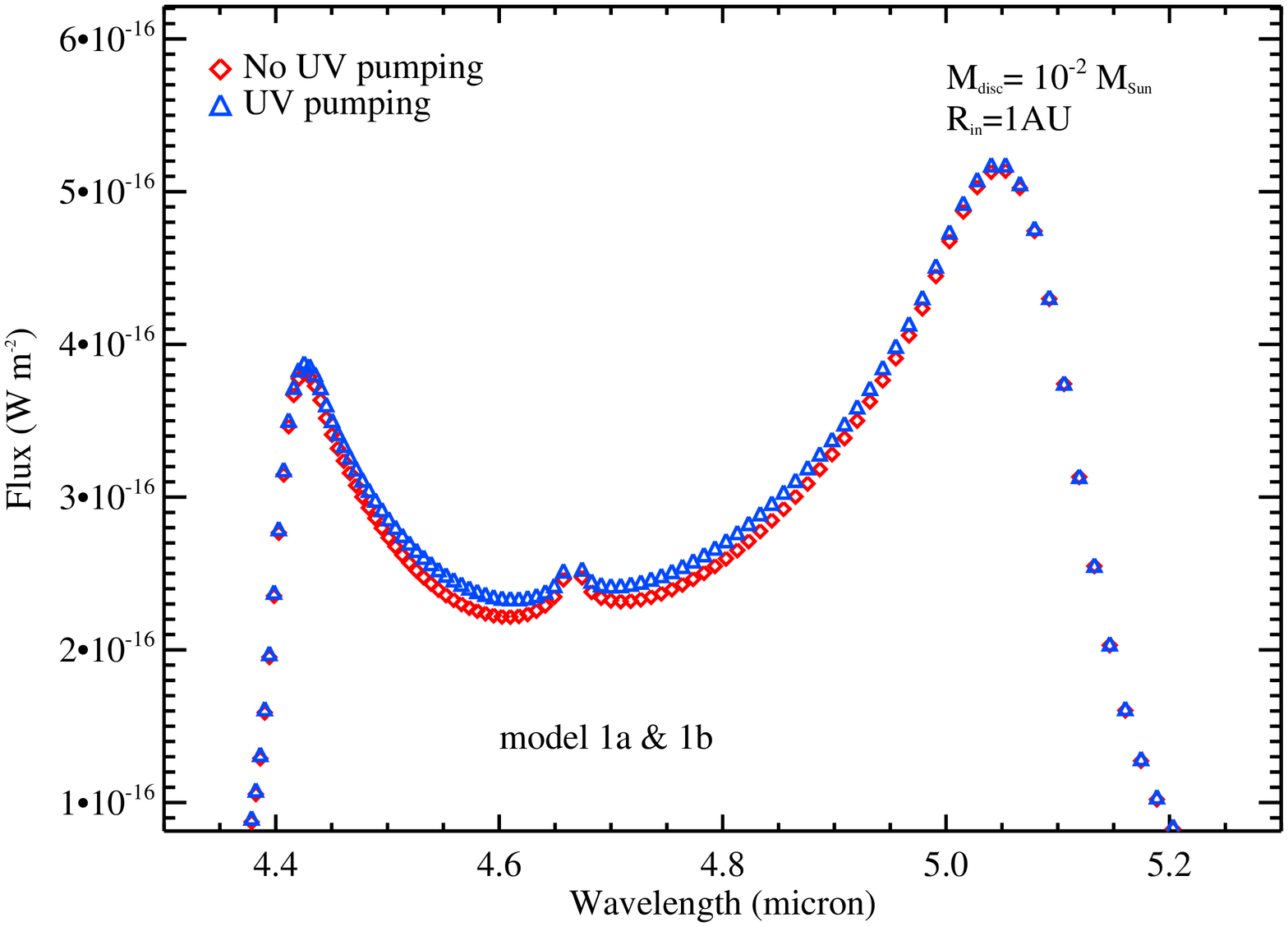}
  \includegraphics[scale=0.37,angle=90]{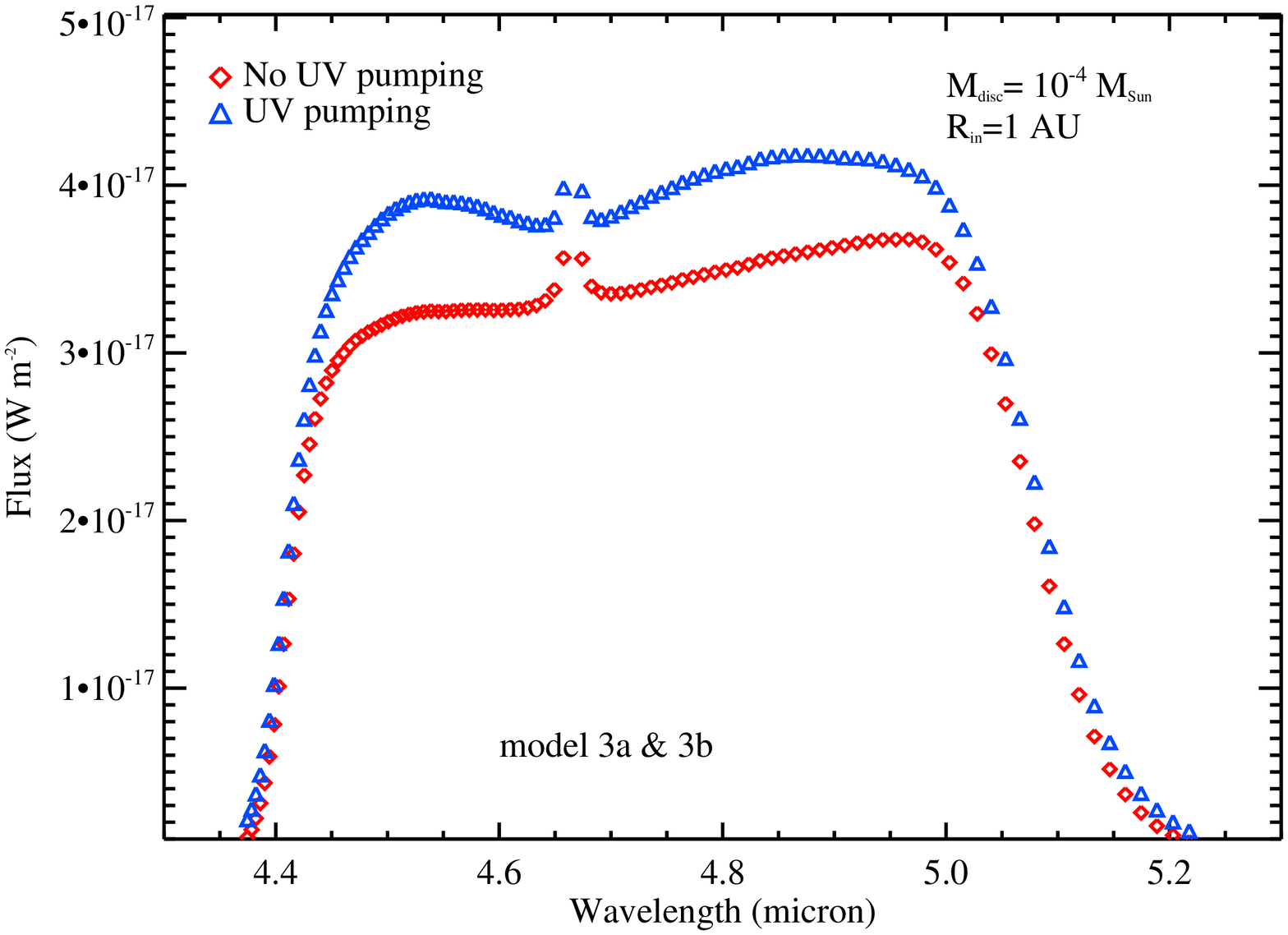}
  \includegraphics[scale=0.37,angle=90]{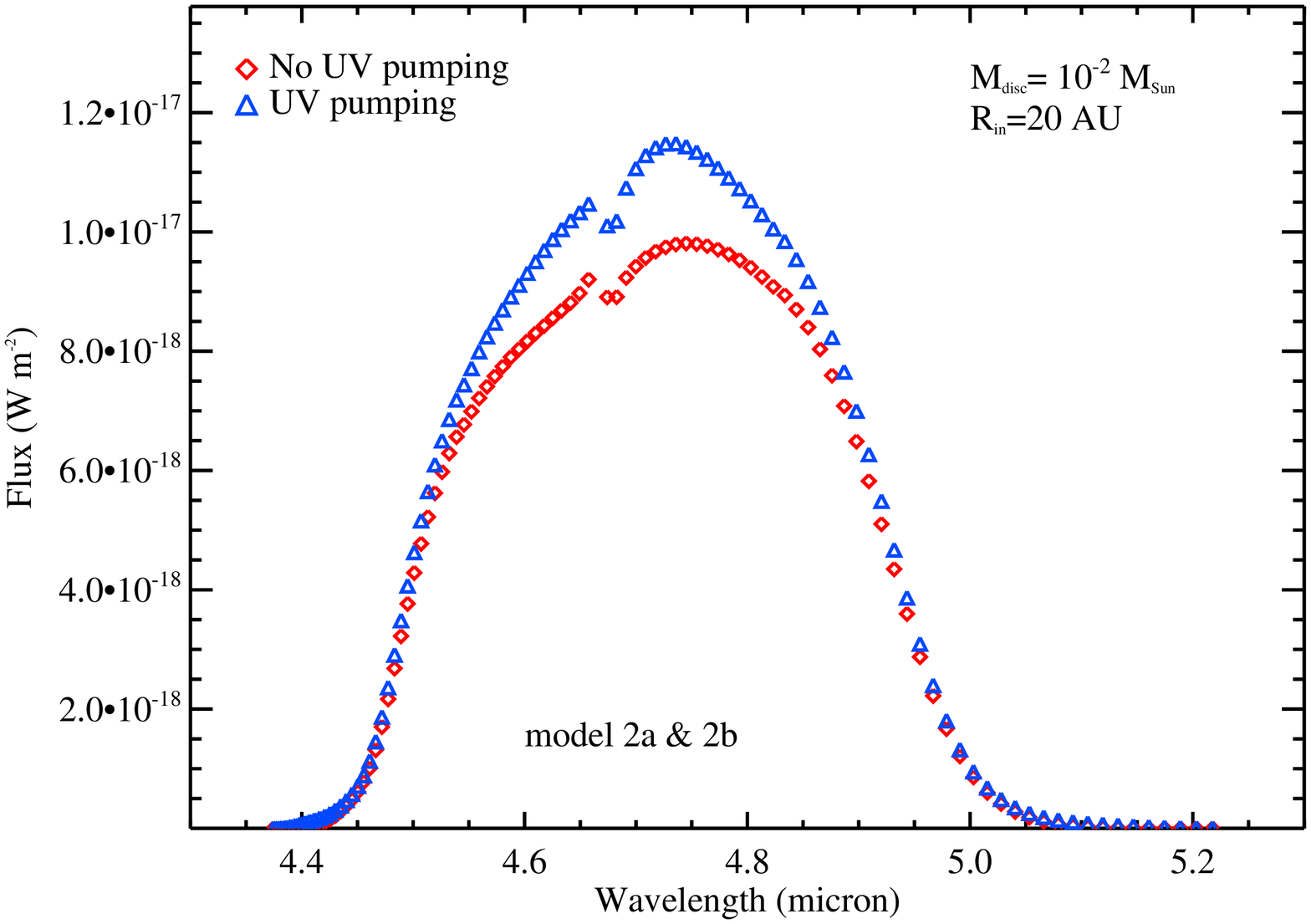}
  \includegraphics[scale=0.37,angle=90]{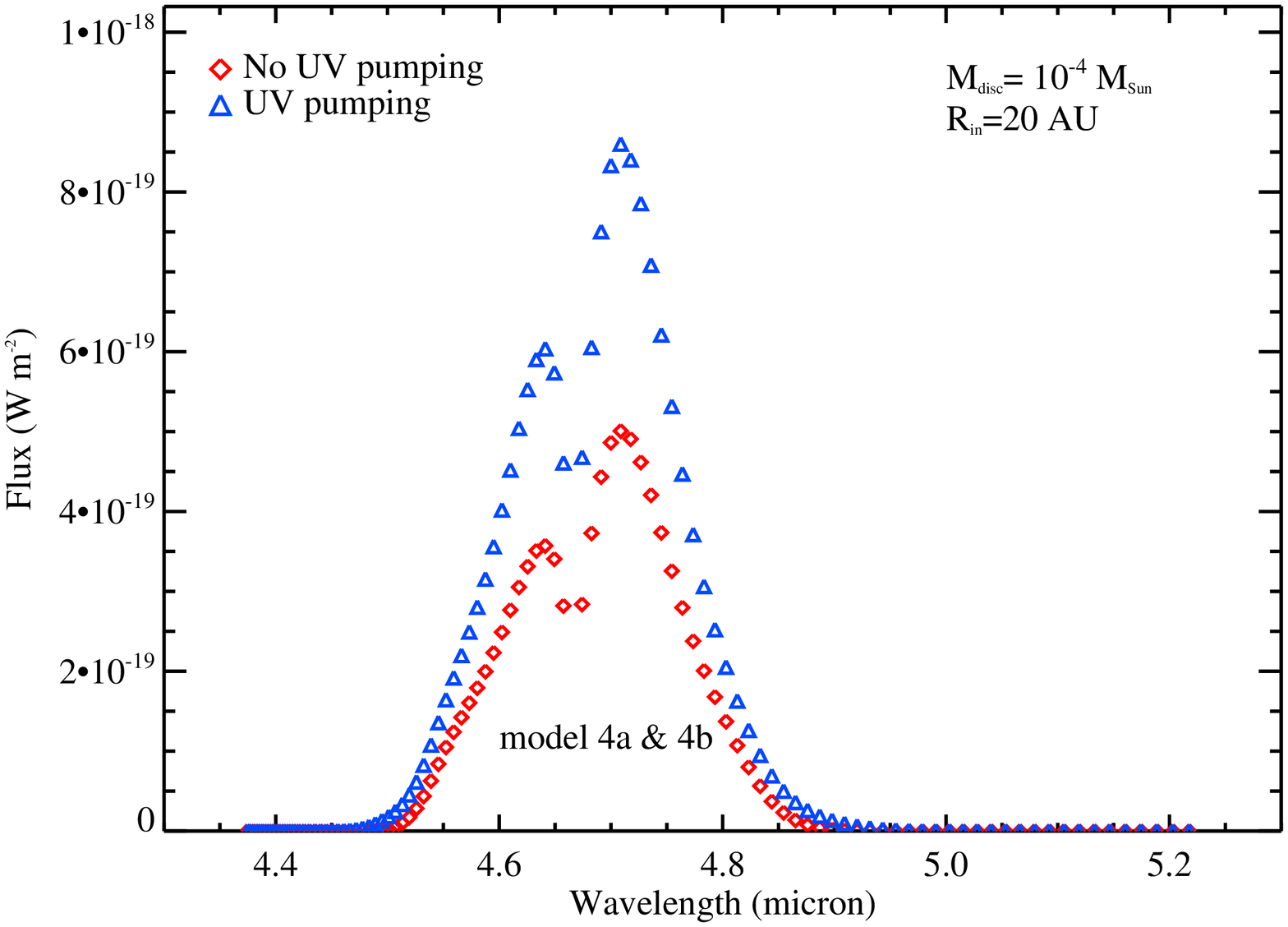}
\centering
\caption{\label{Linefluxes_4panels} CO $v=1-0$ continuum-subtracted
  line fluxes. The blue triangles show the fluxes for models with
  UV-pumping, whereas the red diamonds show the fluxes for models
  without UV-pumping.}
\end{figure*}
\begin{figure*}[ht]
  \centering  
 \includegraphics[angle=90,scale=0.35]{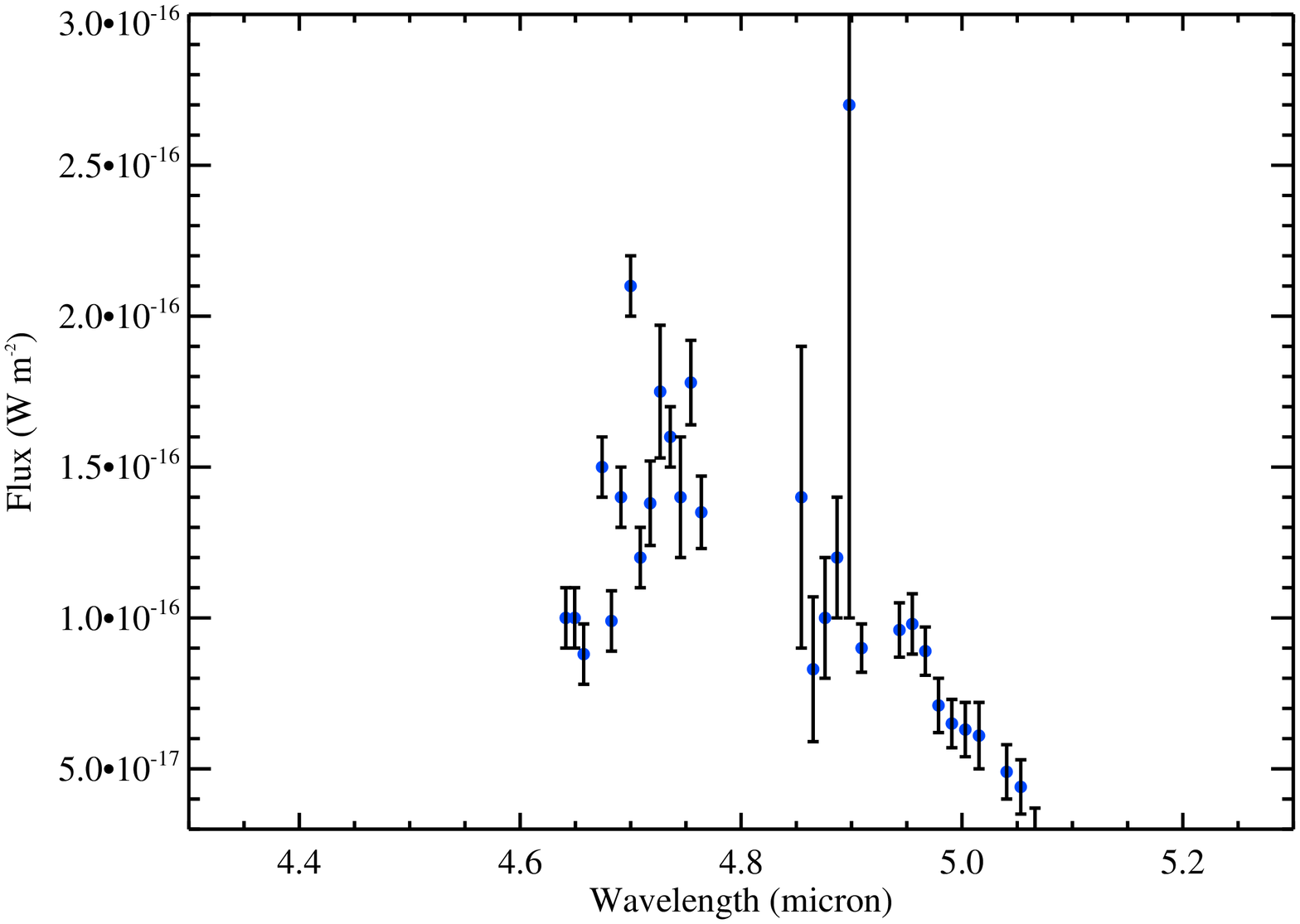}
 \includegraphics[angle=90,scale=0.375]{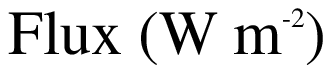}
 \caption{CO SLED from \object{AB~Aur} (left panel) and
   \object{HD~141569A} discs (with the fluxes scaled to a distance of
   140~pc, right panel). Both observations are taken from
   \citet{Brittain2003ApJ...588..535B}.}
\label{SLED_ABAur_HD141569A}
\end{figure*}  
\begin{figure*}[ht]  
  \centering    
  \includegraphics[scale=0.37,angle=90]{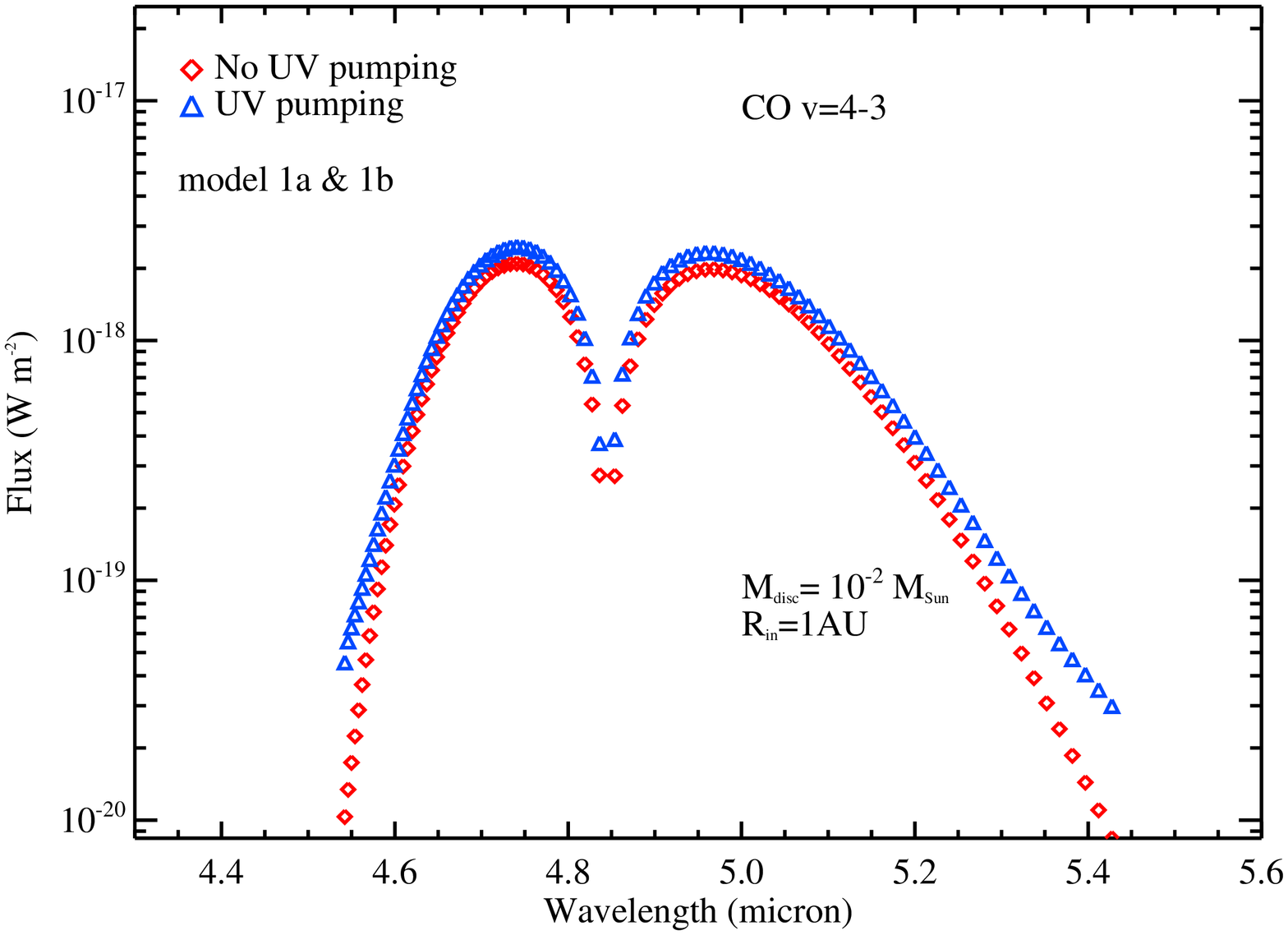}
  \includegraphics[scale=0.37,angle=90]{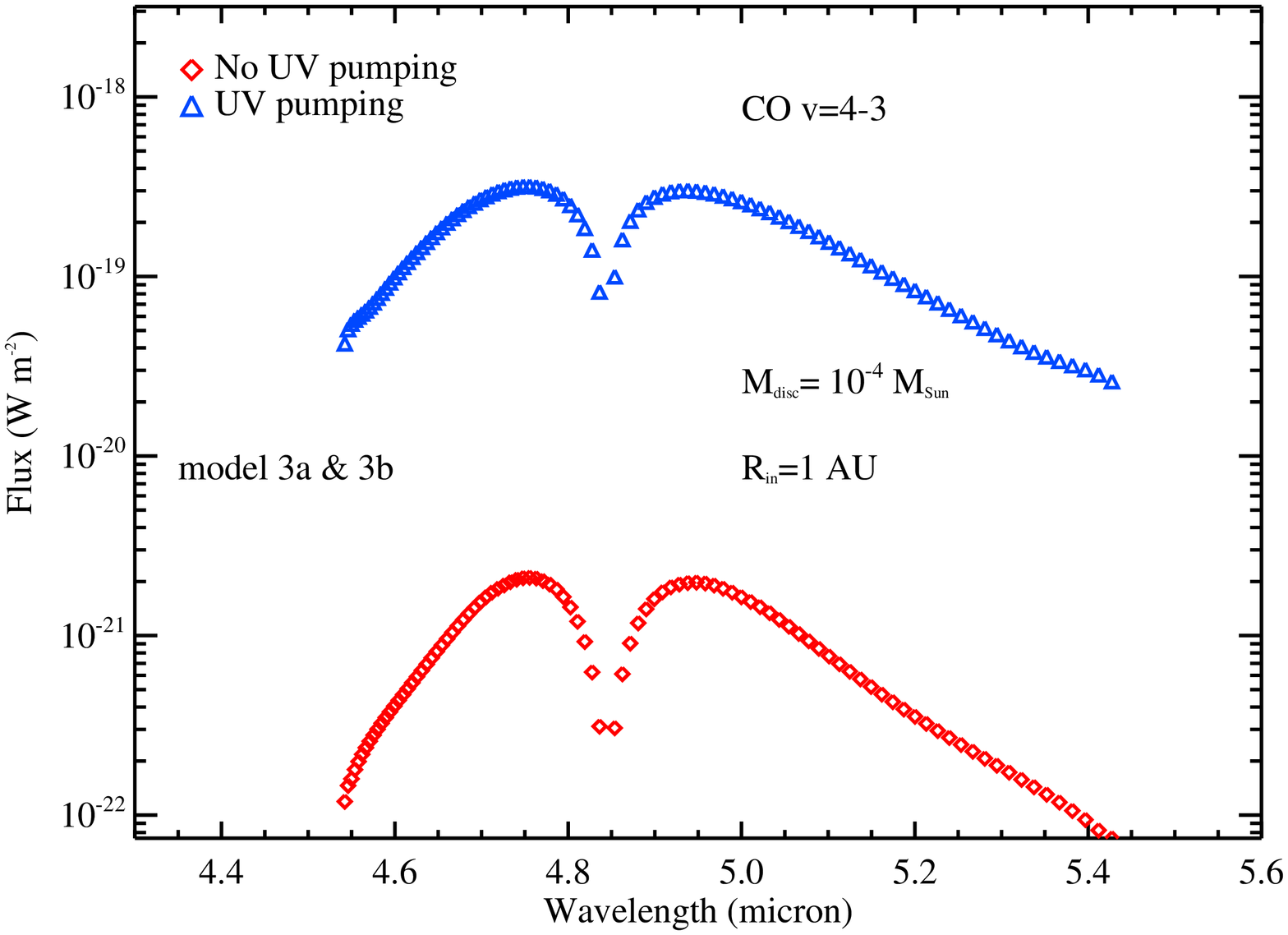}
  \includegraphics[scale=0.37,angle=90]{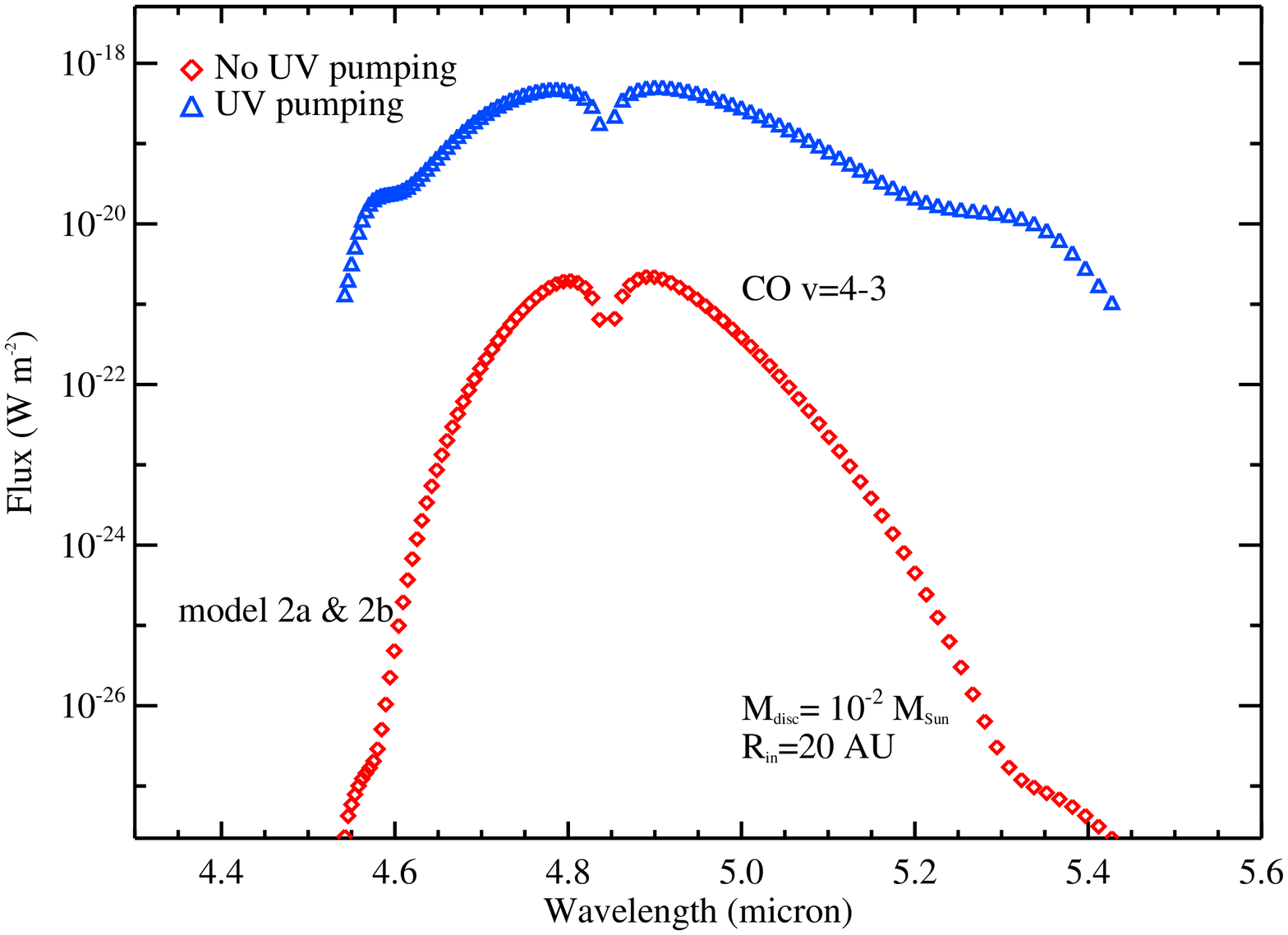}
  \includegraphics[scale=0.37,angle=90]{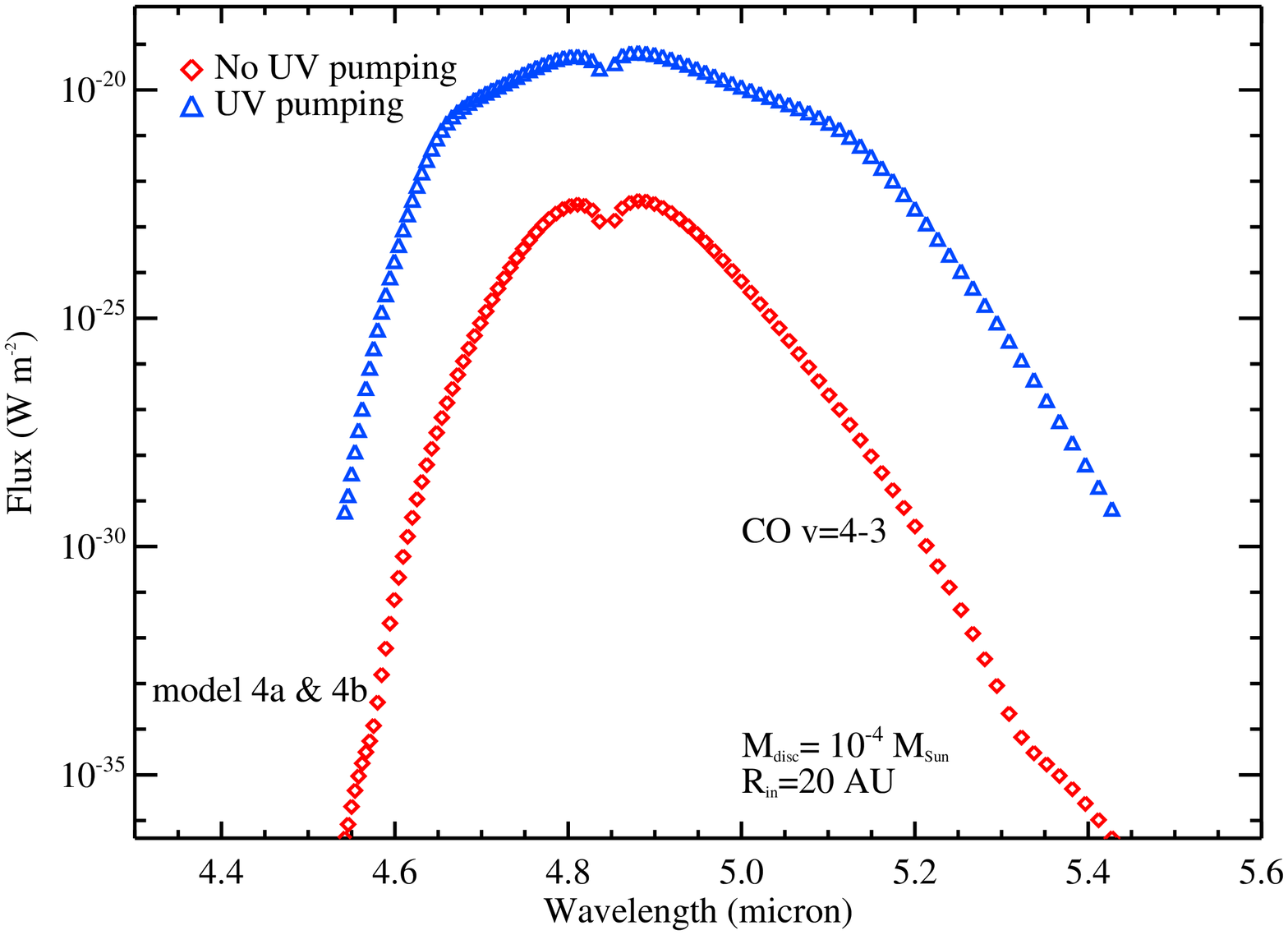}
\centering
\caption{\label{Linefluxes_hot_lines_4panels} CO $v=4-3$ continuum-subtracted line
  fluxes. The blue triangles show the fluxes for models with UV-pumping whereas the red diamonds show the fluxes for models without UV-pumping.}
\end{figure*}

\begin{figure*}[ht] 
  \centering 
  \includegraphics[scale=0.5]{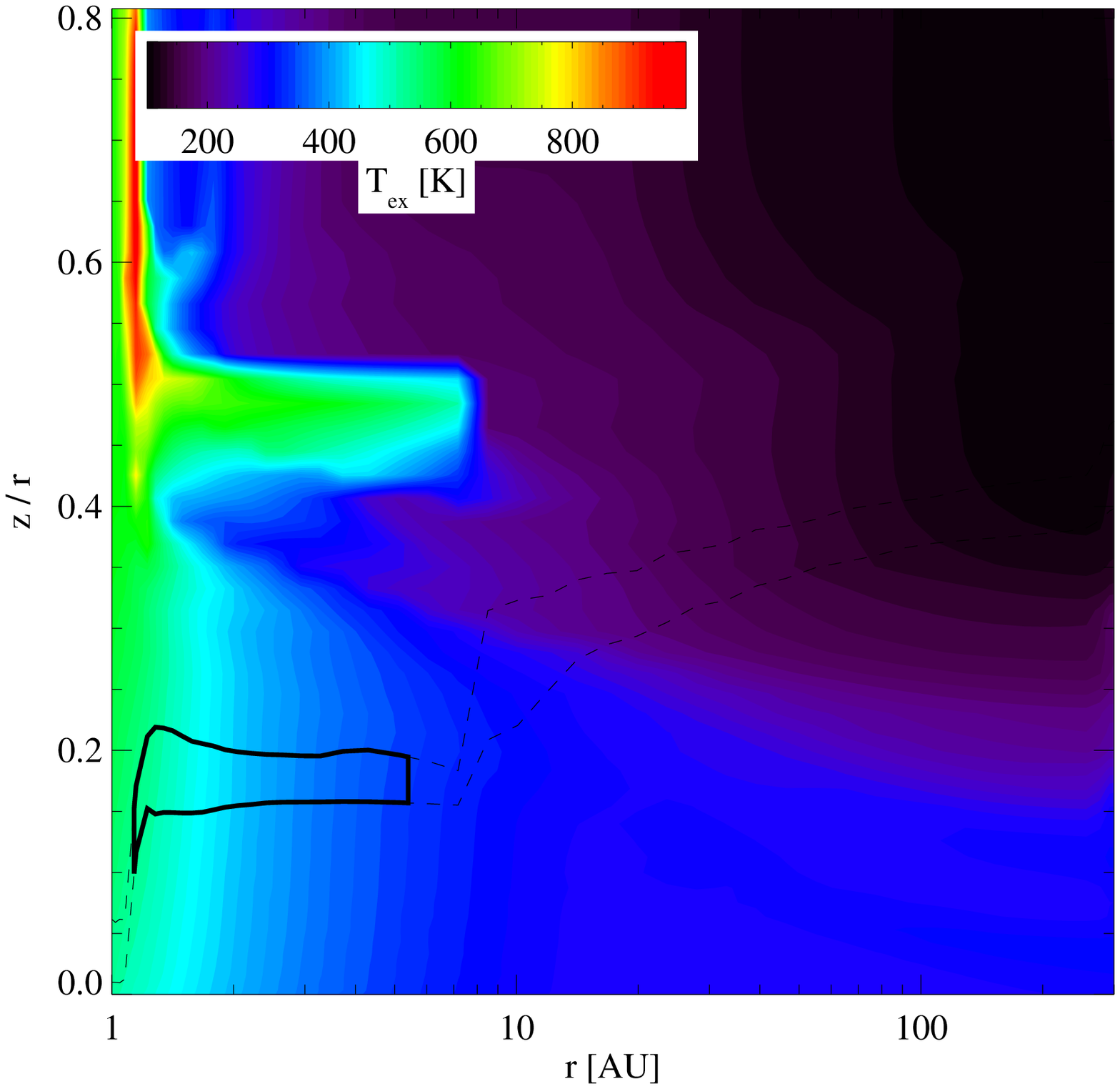}
  \includegraphics[scale=0.5]{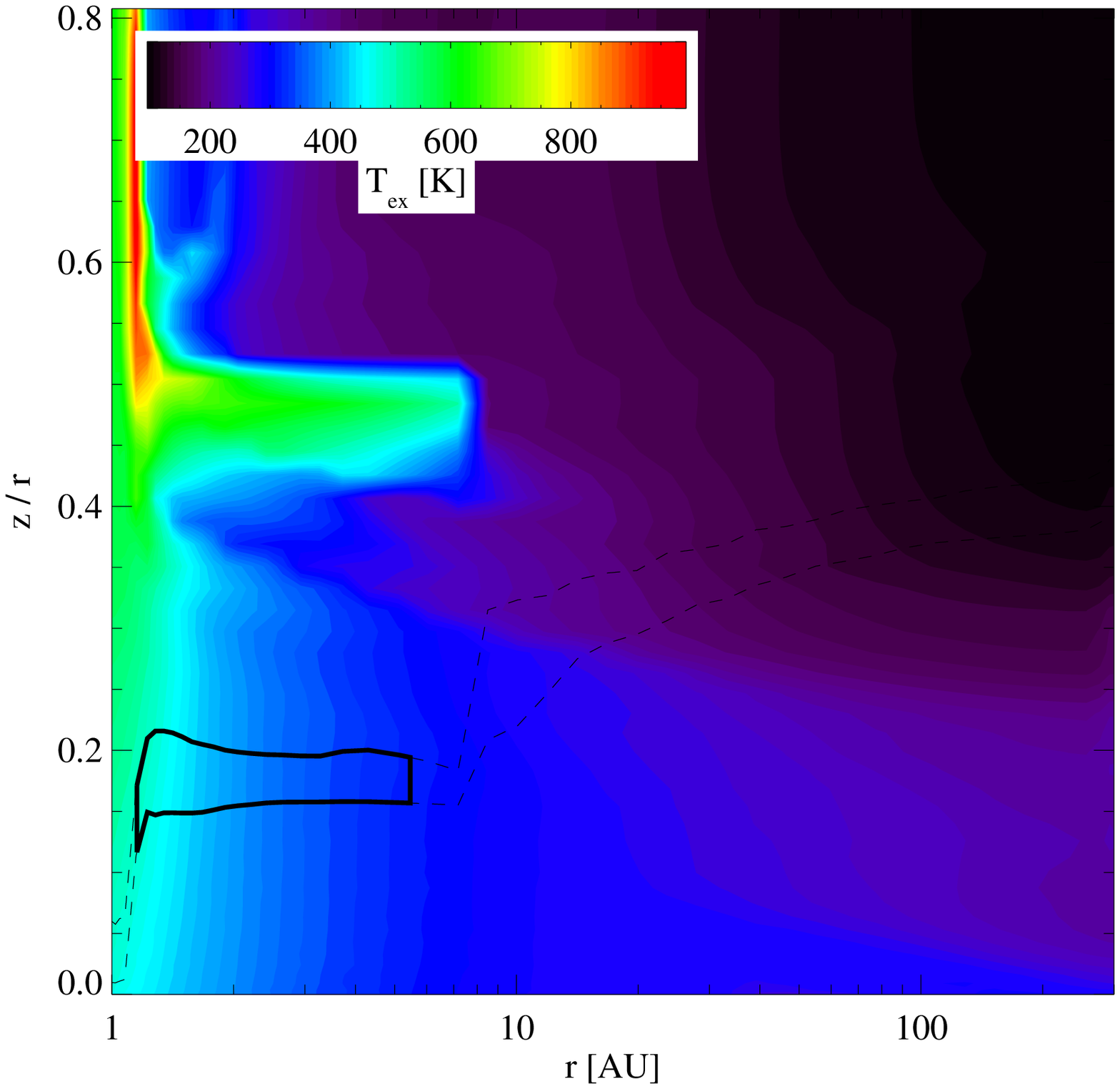}
  \includegraphics[scale=0.5]{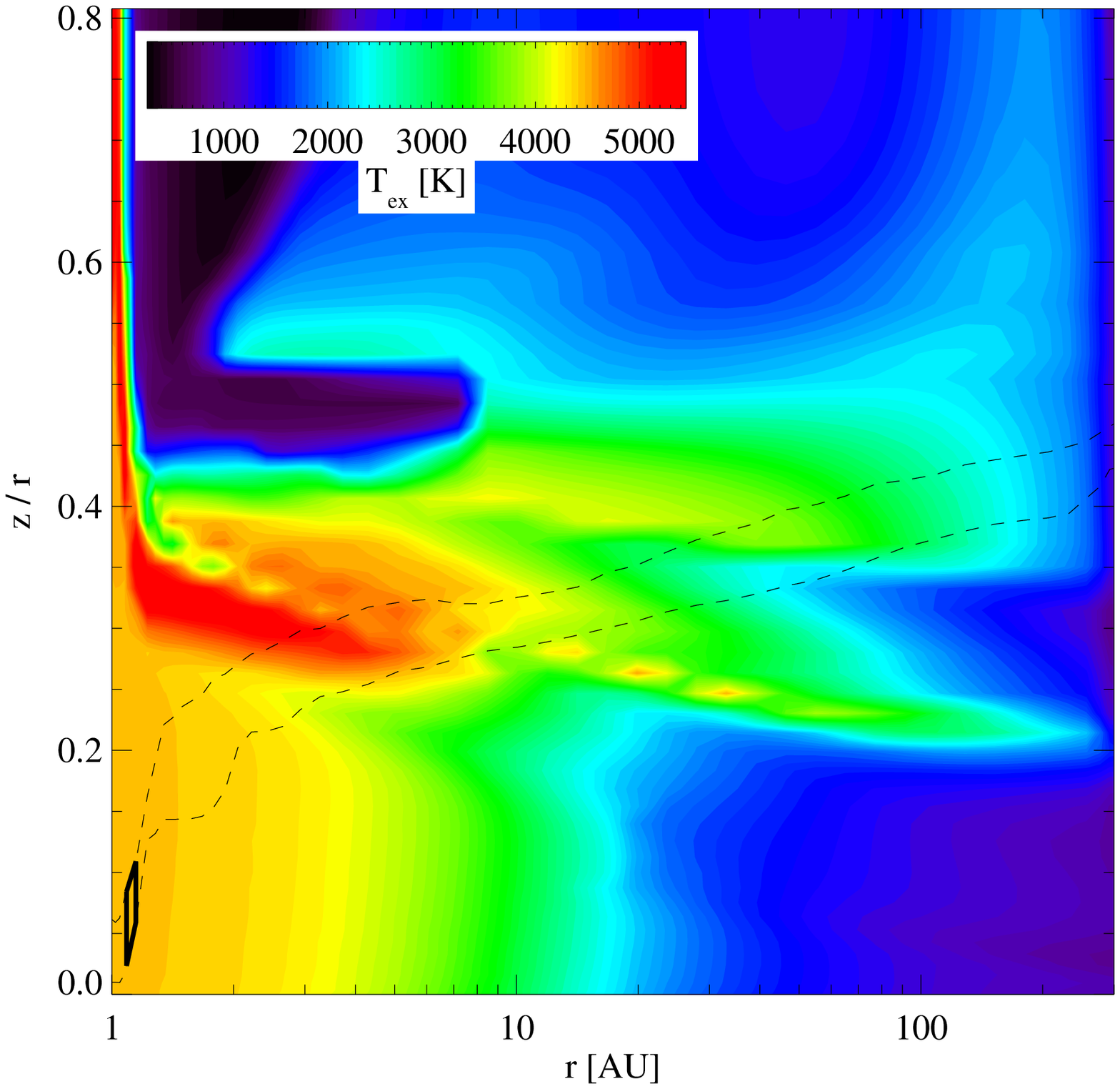}
  \includegraphics[scale=0.5]{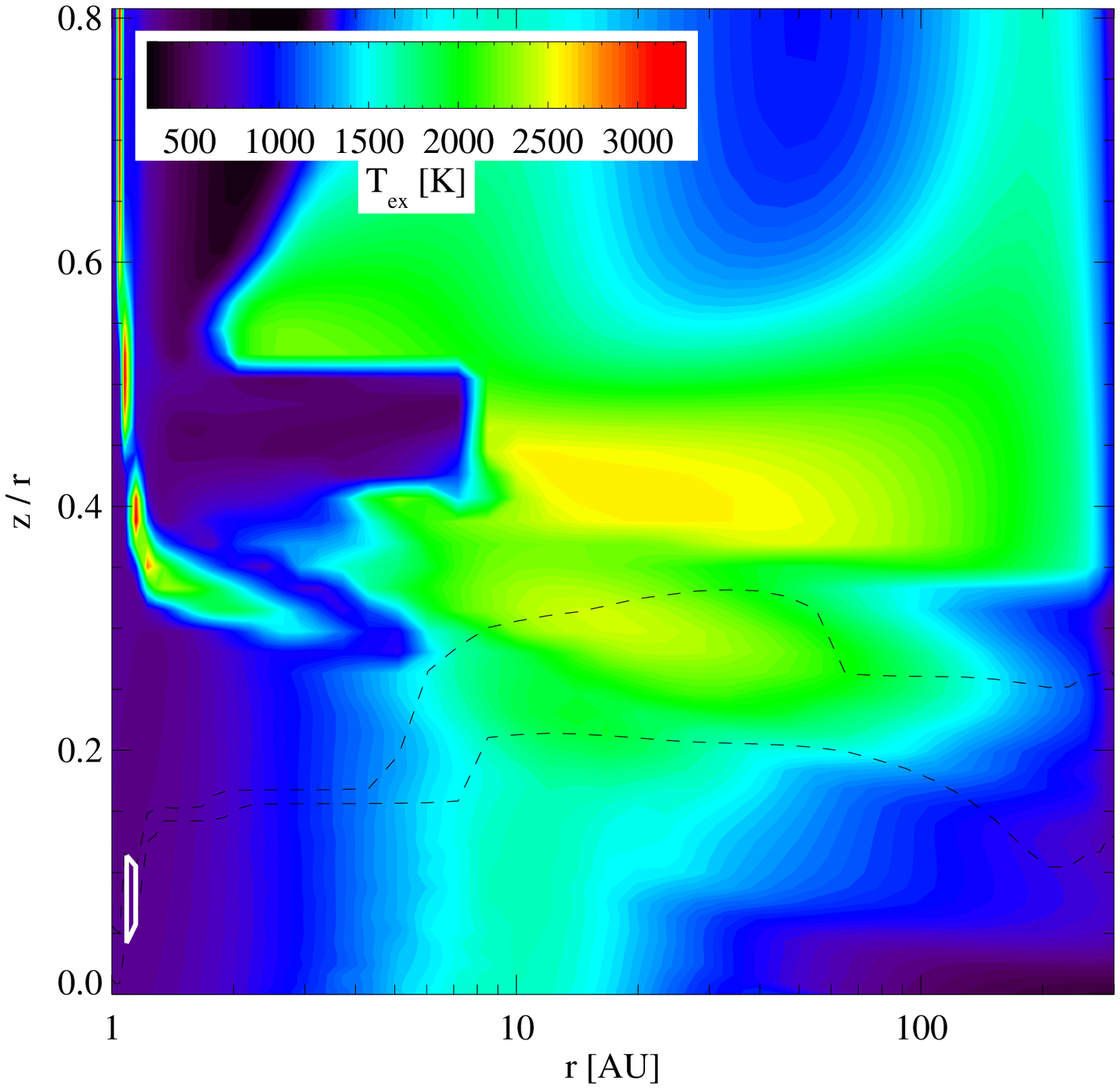}
\centering
\caption{\label{Tex_high_mass_disk_4panels}. Excitation temperatures
  $T_{\mathrm{ex}}$ between the upper and lower level of the CO
  $v=1-0$ $P$(19) transition at 4.844 $\mu$m $v=1, J=19$ (upper two
  panels) and that of the CO $v=4-3$ $P$(20) transition at 4.692
  $\mu$m $v=4, J=20$ (lower two panels) transitions for the
  $M_{\mathrm{disc}}$=10$^{-2}$ M$_\odot$, $R_{\mathrm{in}}$=1~AU disc
  models with (left panels, model 1a) and without (right panels, model
  1b) UV-pumping. The solid contours (in white or black) in the CO
  density panels encompass the regions that emit 49\% of the
  fluxes. The black dashed-line contours contain 70\% of the fluxes in
  the vertical direction. The white contours are at gas temperatures
  of 500 and 1000K.}
\end{figure*}
\begin{figure*}[ht] 
  \centering 
  \includegraphics[scale=0.5]{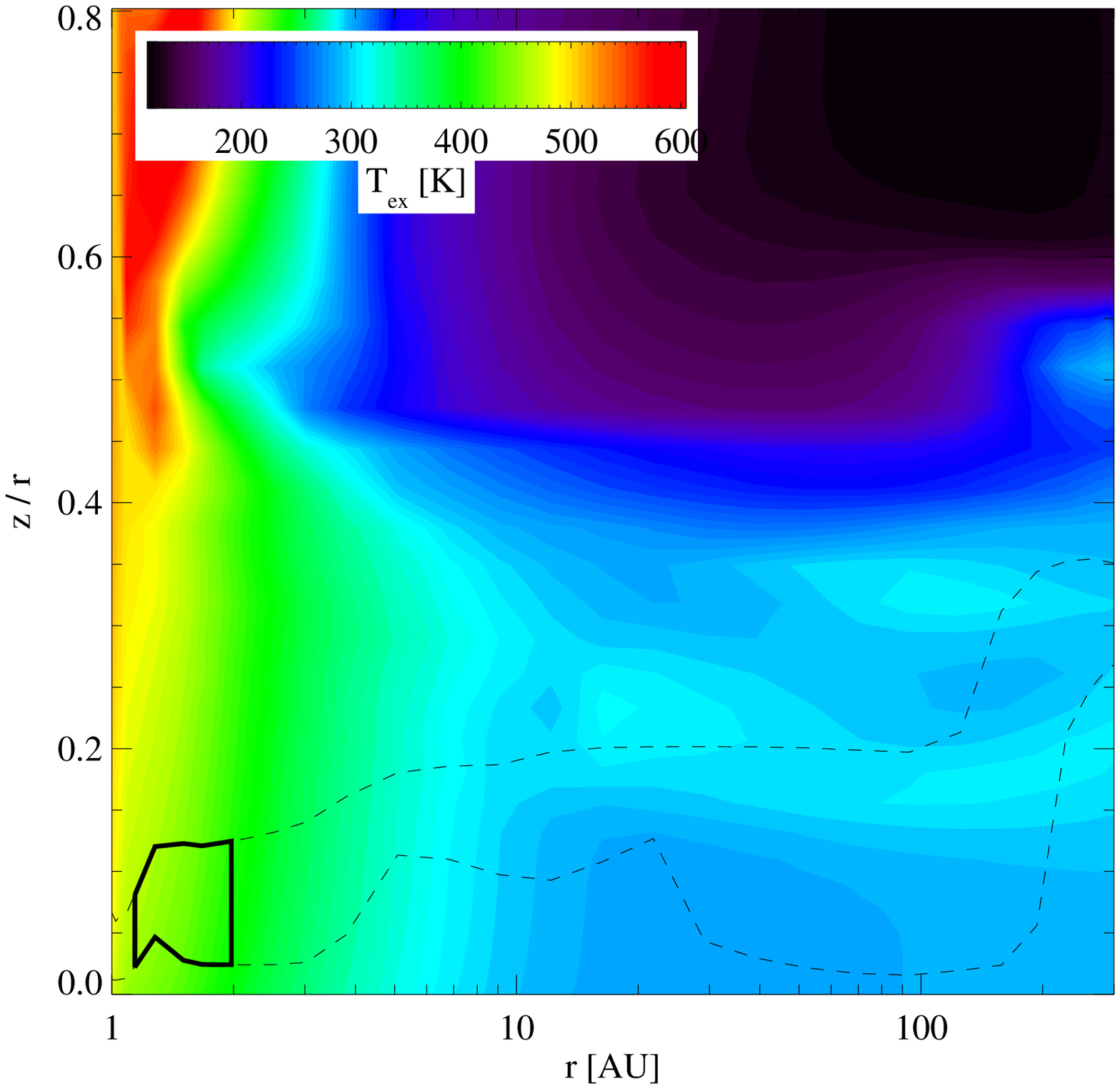}
  \includegraphics[scale=0.5]{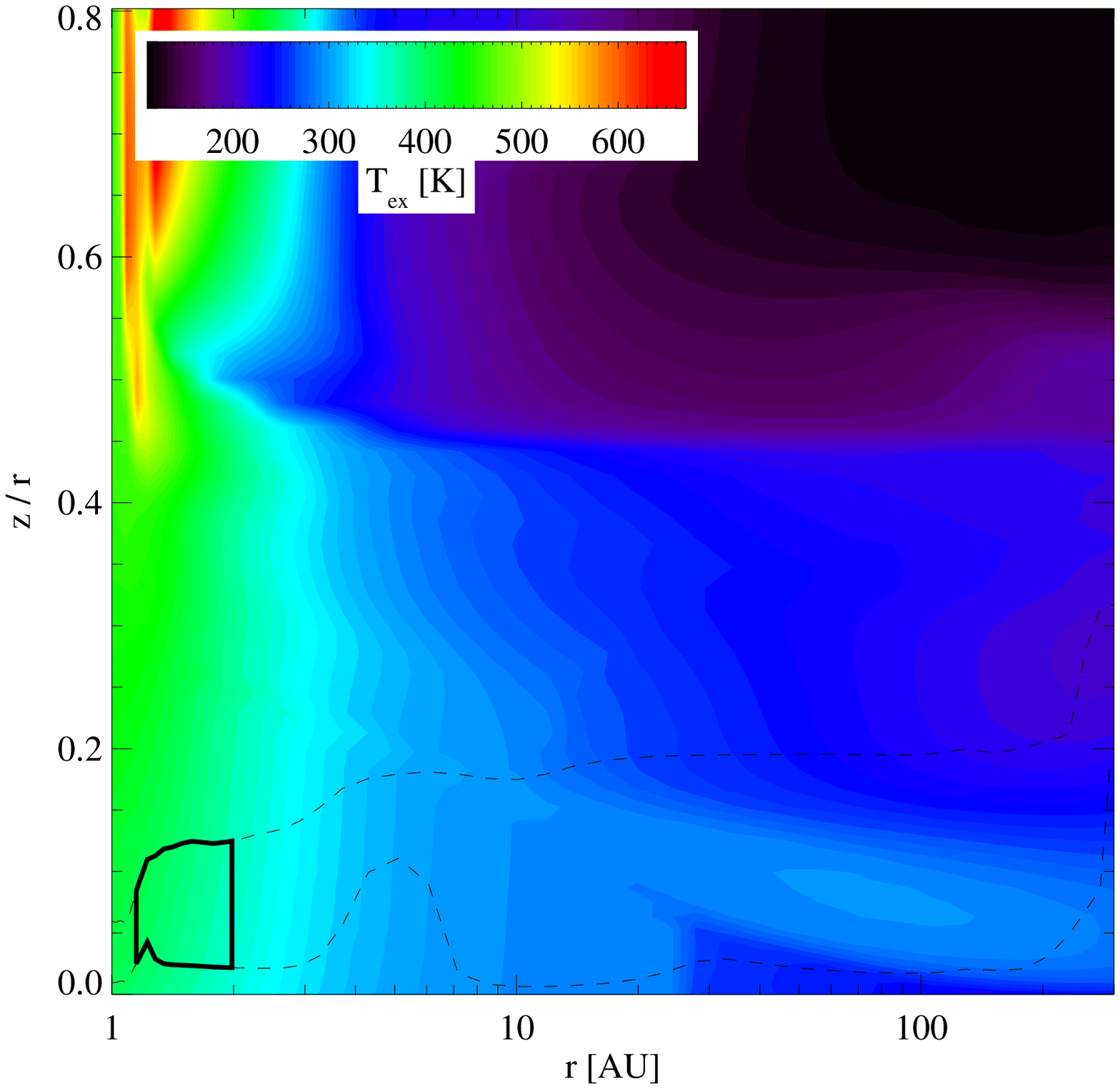}
  \includegraphics[scale=0.5]{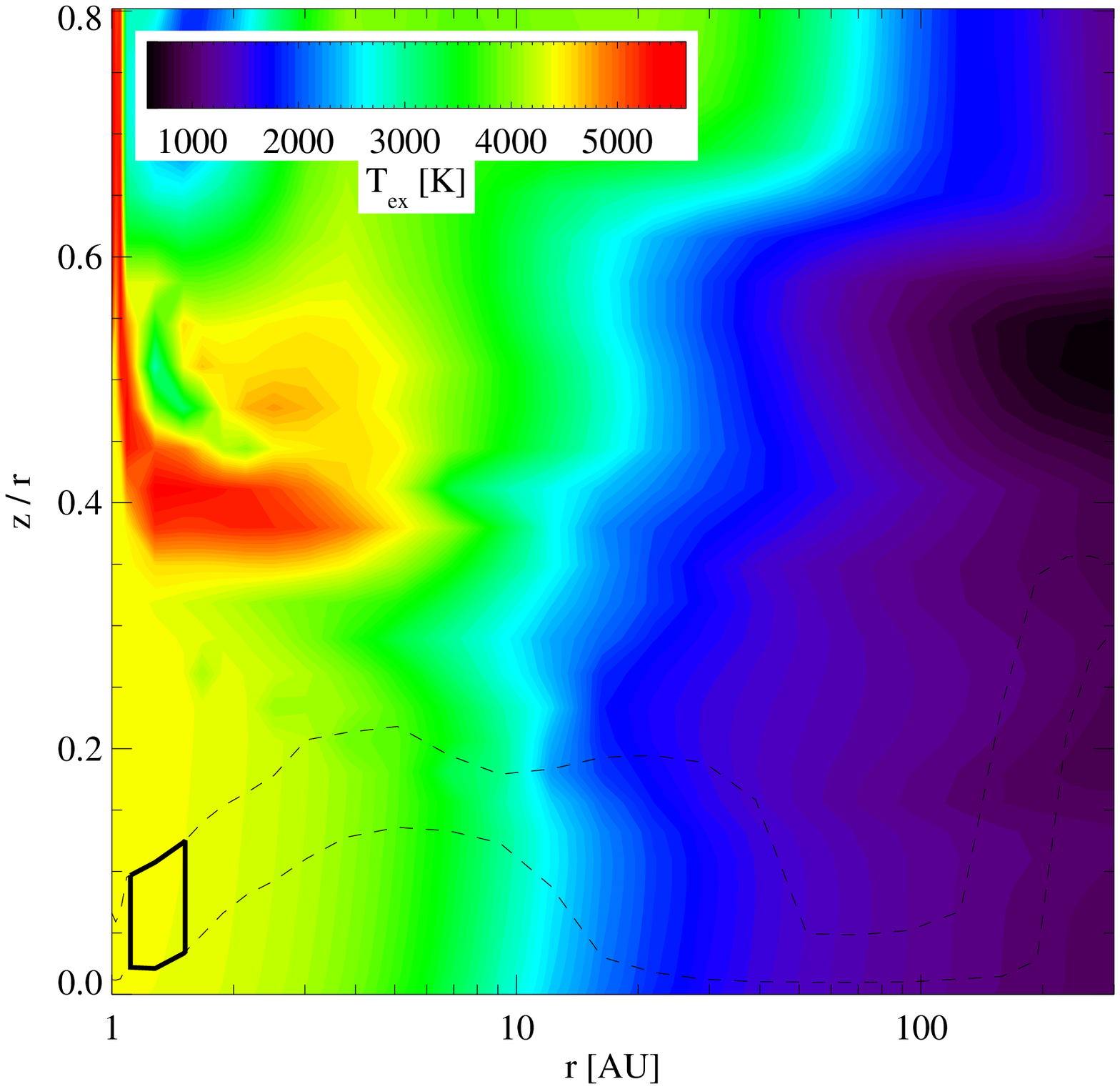}
  \includegraphics[scale=0.5]{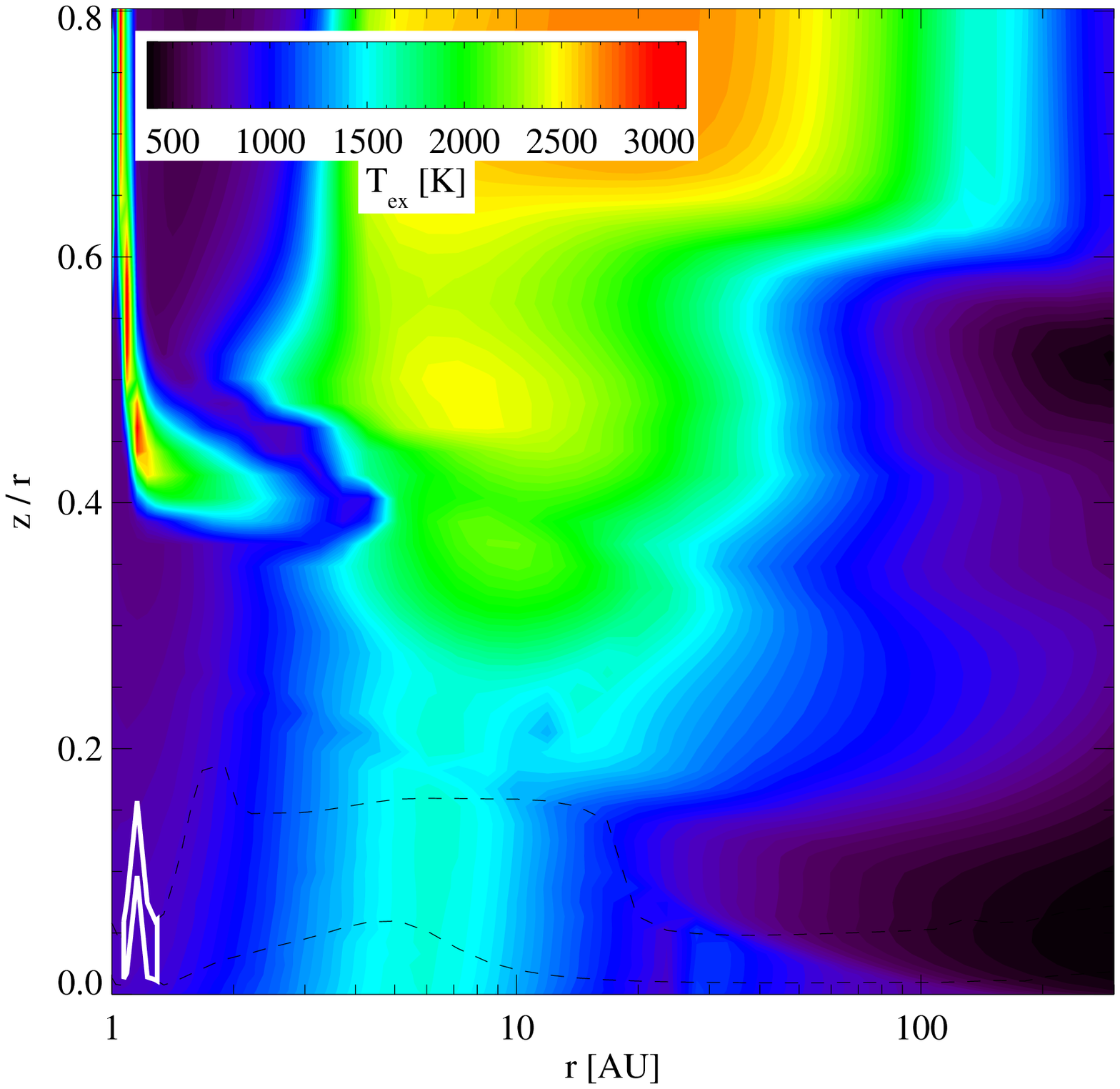}
\centering
\caption{\label{Tex_low_mass_disk_4panels}. Similar to
  Fig.~\ref{Tex_high_mass_disk_4panels} but for the
  $M_{\mathrm{disc}}$=10$^{-4}$ M$_\odot$, $R_{\mathrm{in}}$=1~AU
  disc models (3a and 3b).}
\end{figure*}
  

\begin{figure*}[ht]
  \centering
  \includegraphics[scale=0.37,angle=90]{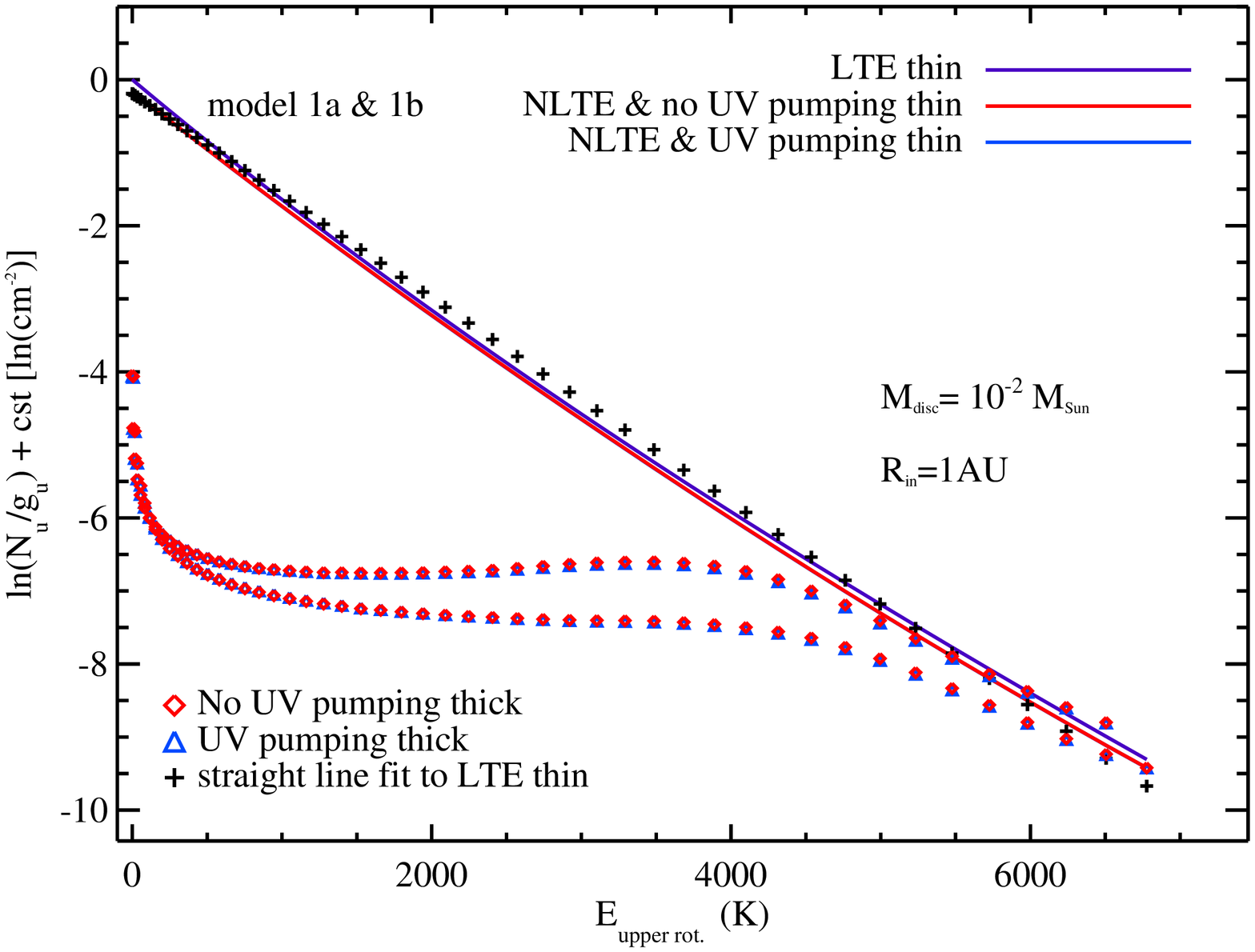}
  \includegraphics[scale=0.37,angle=90]{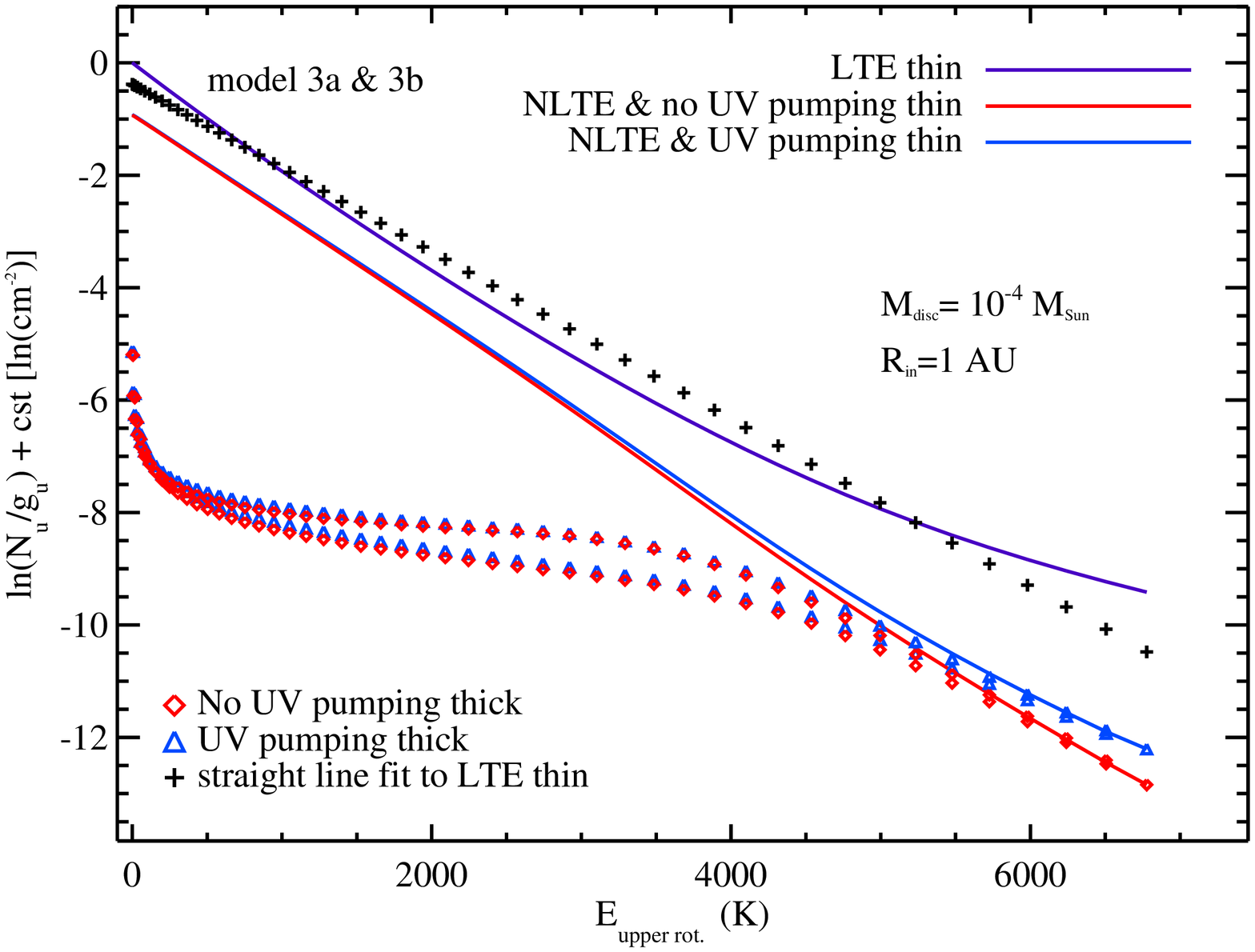}
  \includegraphics[scale=0.37,angle=90]{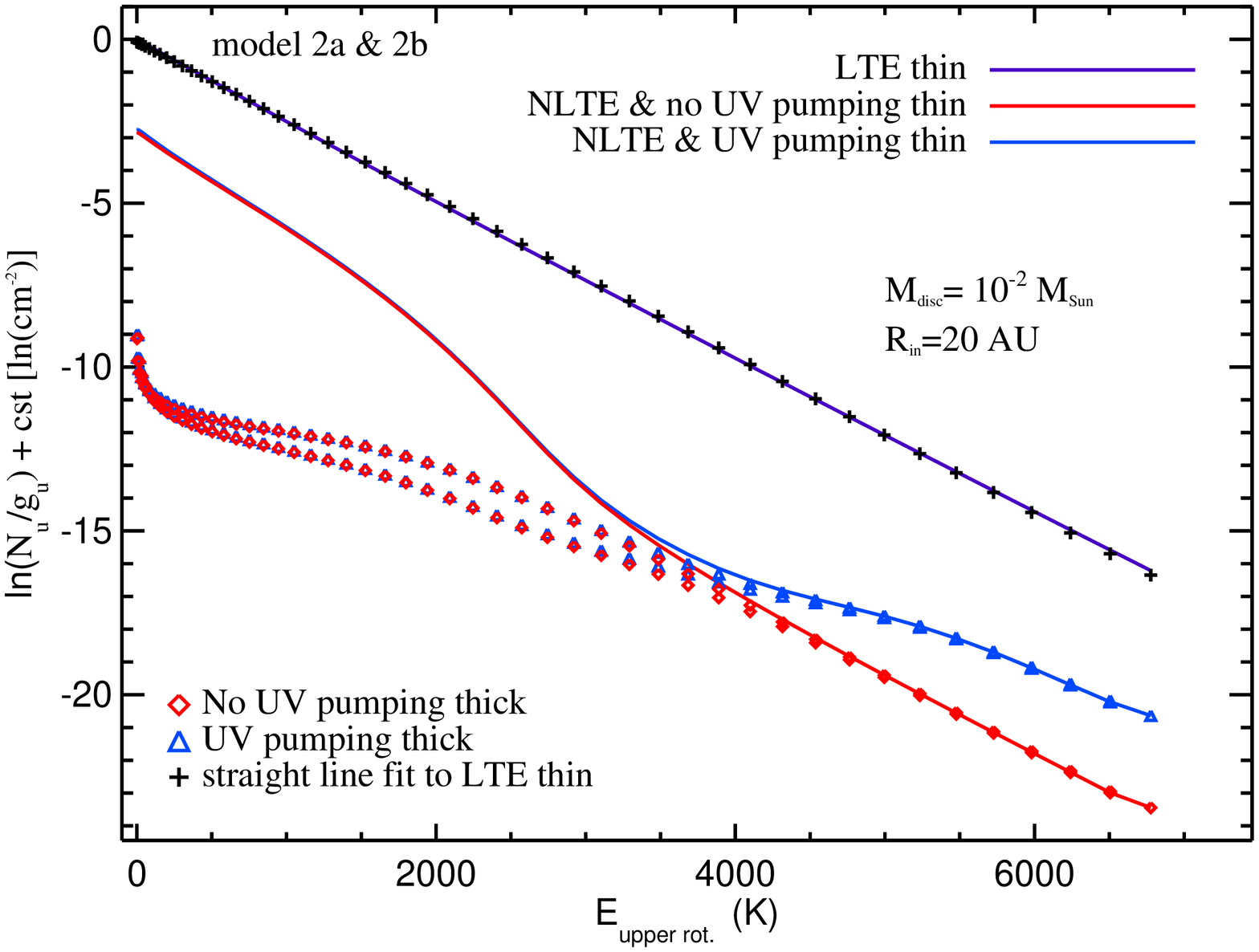}
  \includegraphics[scale=0.37,angle=90]{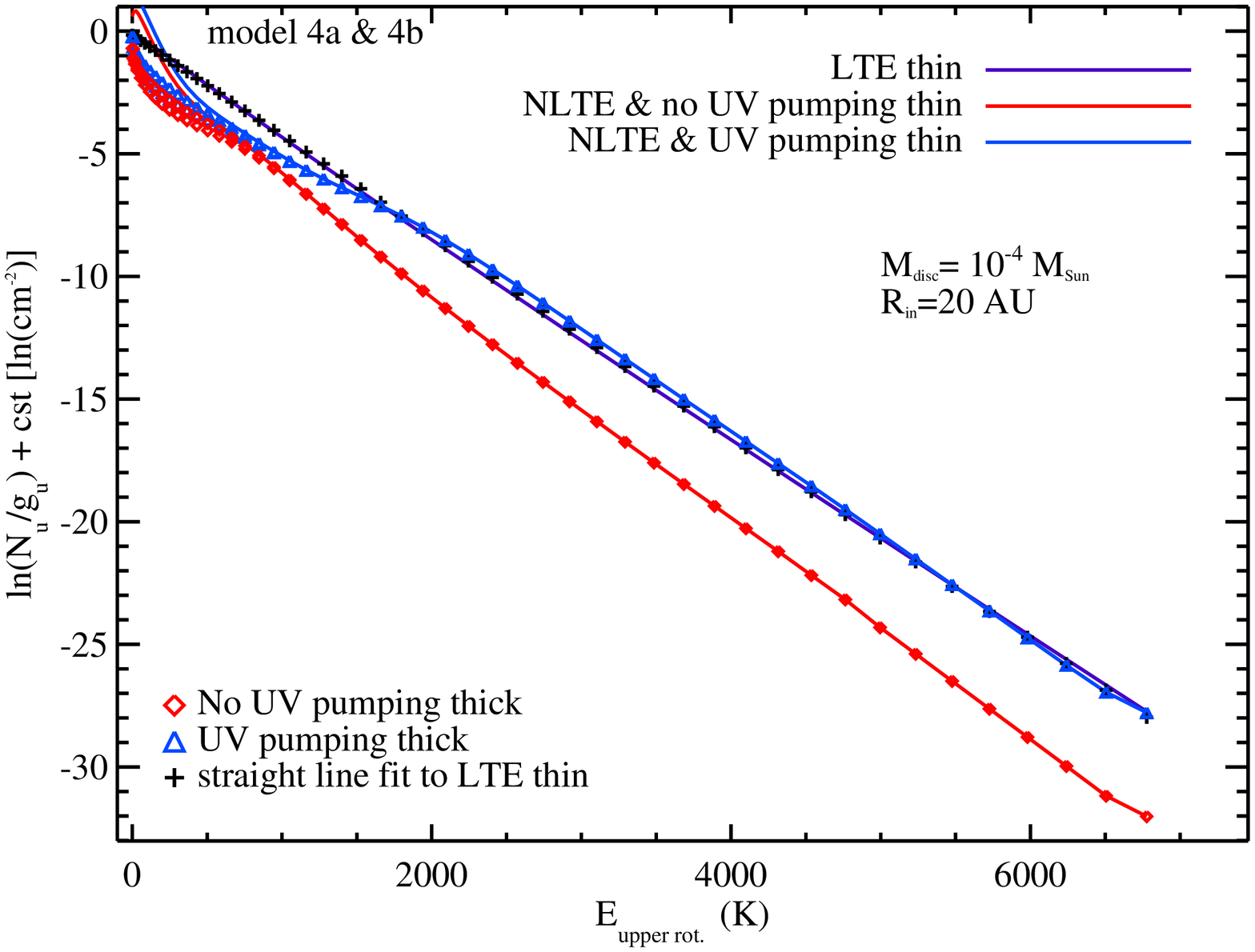}
\centering  
\caption{\label{rotDiag_4panels} Disc model CO $v=1$ population
  diagrams: LTE and NLTE with and without UV pumping in the optically
  thin case (solid lines) and optically thick case (symbols).  All the
  data have been normalized to the $J$=0 population in LTE.  The blue
  diamonds correspond to the column densities after line transfer in
  models with UV pumping whereas the red triangles correspond to
  column densities in line-transfer model results without UV
  pumping. The LTE populations are fitted by straight lines (black
  plus signs). The separations between $P-$ and $R-$ branches are
  clearly seen in the NLTE optically thick cases.}
\end{figure*}

Excited vibrational levels can also be populated by the absorption of
an IR photon emitted by the star and the dust grains, a phenomenon
called IR pumping.  We also checked the effect of IR pumping on the
vibrational ground rotational population by modelling two discs with
$R_\mathrm{in}$=1~AU, one with $M_{\mathrm{disc}}$=10$^{-2}$ M$_\odot$
(models 1c, 1d, and 1e) and with $M_{\mathrm{disc}}$=10$^{-4}$
M$_\odot$ (models 3c, 3d, and 3e). For each disc model, we computed
the rotational level population by taking into account: (1)
ro-vibrational levels up to $v$=5 (models 1c and 3c), (2)
ro-vibrational levels with $v$=0 and $v$=1 (models 1d and 3d), and (3)
the rotational levels in the ground vibrational only (models 1e and
3e).

In total, we ran four series of models with and without the electronic
levels and with different numbers of vibrational levels to assess the
importance of UV fluorescence and IR pumping (see
Table~\ref{ModelTable}). Complete spectra from four to five microns
are generated for discs seen face-on (zero inclination).
\begin{figure*}[ht] 
  \centering
  \includegraphics[scale=0.5,angle=90]{vibCO_diag_Mdisk1e-2_Rin1AU.ps}
  \includegraphics[scale=0.5,angle=90]{vibCO_diag_Mdisk1e-4_Rin1AU.ps}
  \includegraphics[scale=0.5,angle=90]{vibCO_diag_Mdisk1e-2_Rin20AU.ps}
  \includegraphics[scale=0.5,angle=90]{vibCO_diag_Mdisk1e-4_Rin20AU.ps}
\centering
\caption{\label{vibDiag_4panels} CO vibrational diagram with and
  without UV pumping for the four models. We used the $v$=1--6
  population after line radiative transfer.  The blue diamonds
  correspond to models with UV pumping whereas the red triangles
  correspond to models without UV pumping. The purple stars are models
  without UV pumping and 5 vibrational levels. Straight lines have
  been fitted to each dataset. The derived vibrational temperatures
  are indicated in each panel. The blue lines are for models with
  UV-pumping whereas the red dashed-lines are for models without
  UV-pumping. In the upper left panel, massive disc including
  UV-pumping show the same vibrational temperature
  ($T_{\mathrm{vib}}$=1413~K) with either 5 or 9 electronic
  ground-state vibrational levels included in the model. On the other
  hand, the vibrational temperatures of a low-mass disc depend on the
  number of assumed vibrational levels in the ground electronic state
  with higher vibrational temperatures as more levels are included
  (upper-right panel). The UV-pumping is the most efficient for the
  lowest disc mass model with the largest inner hole (model 4a in the
  lowest-right panel with $M_{\mathrm{disc}}=10^{-4}$~M$_\odot$,
  $R_{\mathrm{disc}}$=~20~AU).}
\end{figure*}      
\begin{figure*}[ht]   
  \centering
  \includegraphics[scale=0.5,angle=90]{vibCO_diag_Mdisk1e-3_Rin1AU.ps}
  \includegraphics[scale=0.5,angle=90]{vibCO_diag_Mdisk1e-3_Rin5AU.ps}
  \includegraphics[scale=0.5,angle=90]{vibCO_diag_Mdisk2.5e-3_Rin5AU.ps}
  \includegraphics[scale=0.5,angle=90]{vibCO_diag_Mdisk5e-3_Rin5AU.ps}
  \caption{\label{vibDiag_other_models} Similar to
    Fig.~\ref{vibDiag_4panels} but for models 5a, 5b to 8a, 8b. Among
    the models shown in this figure, the vibrational temperature
    differences between models with and without UV-pumping are the
    largest in the lowest mass disc with a inner hole (upper-right panel, model 6a with
    $M_{\mathrm{disc}}=10^{-3}$~M$_\odot$,
    $R_{\mathrm{disc}}$=~5~AU).}
  \centering
\end{figure*}
  
\subsection{Results \& Discussion}\label{results_discussion}

\subsubsection{CO chemistry and location of the CO ro-vibrational
  emission}\label{COemission_location}

We focus on the CO chemistry of model 1a. The disc density, dust
temperature, and gas temperature structures are shown in
Fig.~\ref{disc_structur_3panels}. The panels in Fig.~\ref{CO_15panels}
show the enhancement ($\chi$) with respect to the interstellar UV, the
abundance relative to the total number of hydrogen-nuclei for He,
electron, atomic hydrogen, molecular hydrogen, vibrationally excited
H$_2$ (H2exc), C$^+$, C, CO, CH$_4$, OH, and H$_2$O. The transition
from C$^+$ to C occurs at $\log$($\chi$/n) between -1 and -2, and the
transition between C and CO occurs at $\log$($\chi$/n)$\sim$ -2.5 to
-3.

In Fig.~\ref{CO_emission_high_mass_disk_4panels}, we show the location
of the CO $v=1-0$ $P$(19) and $v=4-3$ $P$(20) emissions overlaid with
the number density of the three main collision partners (H, H$_2$, and
electrons) for a 10$^{-2}$ M$_\odot$ disc model taking UV pumping or
not into account (models 1a and 1b).  Similar plots are displayed in
Fig.~\ref{CO_emission_low_mass_disk_4panels} for models 3a and 3b. The
CO lines are emitted up to a few AU after $R_{\mathrm{in}}$, where the
density is high enough such that the level population is the closest
to the LTE population for the fundamental transitions. The emitting
area becomes more and more confined to the inner rim as the upper
energy of the transition increases. In the 10$^{-2}$ M$_\odot$ disc
model the fluxes are a few percents higher when UV pumping is
switched on. The fundamental $P$(19) line is emitted in region where
the gas is between 500 and 1000K. In the vertical direction, the CO
ro-vibrational lines are mostly emitted in the disc region with
$z/r$$\sim$0.2 and $R$$<$~5 AU.

In the emitting region, the extinction in the visible $A_{\mathrm{V}}$
is below 1 and some H$_2$ molecules are excited by UV and/or hot
enough to be vibrationally excited (the excited H$_2$ abundance
reaches 10$^{-12}$-10$^{-9}$). As soon as it is formed, the warm or
vibrationally excited H$_2$ reacts quickly with atomic oxygen to form
OH. In turn, OH reacts with H$_2$ to form H$_2$O or with atomic carbon
to form CO.

The OH and water vapour abundance are high for $z/r$$\sim$0.2 and
$R$$<$~5 AU. The central role played by OH in hot gas chemistry is
discussed in model details in \citet{Thi2005A&A...438..557T}.  The 1a
and 1b disc models are massive enough such that in the inner disc
midplane ($z/r<$0.1) the most abundant carbon-bearing gas-phase
species are methane and C$_2$H$_2$ ($R$$<$~3~AU), whereas most of the
oxygen is locked in water vapour up to $R$$\sim$~5~AU. The remnant
oxygen is the form of atomic oxygen, CO, and other minor species. The
gas in the inner disc is dense enough such that UV fluorescence is
quenched by collisional de-excitations between the ro-vibrational
levels.

\subsubsection{CO Spectral Line Energy Distribution}\label{SLED}

The analysis of the CO ro-vibrational lines can be performed either by
comparing directly the observed and modelled line fluxes or by using
rotational diagrams. In Fig.~\ref{Linefluxes_4panels} and
\ref{Linefluxes_hot_lines_4panels} we plotted the continuum-subtracted
line fluxes for model 1 to 4 (spectral line energy distribution, SLED)
for the fundamental ($v$=1-0) and one ``hot'' transition
($v$=4-3). The shape of the SLEDs varies from model to model and
between models with and without UV pumping. The large variety of the
SLEDs illustrates the sensitivity of the CO ro-vibrational line fluxes
to the disc parameters ($M_{\mathrm{disc}}$, $R_{\mathrm{in}}$, etc
...). The line fluxes range from 10$^{-19}$ W m$^{-2}$ to a few
10$^{-16}$ W m$^{-2}$ for a low-mass disc (10$^{-4}$ M$_\odot$) with
an inner hole to a few 10$^{-16}$ W m$^{-2}$ to for a massive disc
(10$^{-2}$ M$_\odot$) at a distance of 140~pc. In comparison, we show
in Fig.~\ref{SLED_ABAur_HD141569A} the $v$=1-0 CO line fluxes
\citep{Brittain2003ApJ...588..535B} for the massive disc \object{AB
  Aur} and the low-mass disc \object{HD141569A}.The observed fluxes
from \object{HD141569A} should be scaled from its distance of 99~pc
\citep{vandenancker1997A&A...324L..33V} to 140~pc. Fluxes for the hot
lines from the \object{HD141569A} disc are $\sim$ 10$^{-18}$ W
m$^{-2}$ after scaling the fluxes to 140~pc. When UV pumping is
included, the hot line fluxes from our models are of the order of
10$^{-19}$-10$^{-17}$ W m$^{-2}$. The shapes of the SLEDs for models
with a small inner radius ($R_{\mathrm{in}}$=~1~AU, models 1 \& 3)
reflect the effects of highly optically thick lines. On the other
hand, the shapes of SLEDS of models with a large inner radius
($R_{\mathrm{in}}$=~20~AU, models 2 and 4) are typical of optically
thin or moderately optically thick lines. The observed line fluxes are
best matched by models with a small inner radius whereas the shapes of
the observed SLED are closer to models with a large inner radius. To
obtain fluxes that are close to the observed values, the emitting gas
should be close to the star, but at the same time lower surface
densities are needed to have moderate optical depths for the CO
lines. Interestingly, UV pumping will increase the line $v$=1-0 fluxes
only by a factor of few and the $v$=4-3 fluxes by orders of
magnitude. In our models, UV pumping does not dramatically change the
shape of the SLEDS for $v=$1-0 and $v$=4-3 transitions.

We ran models with inclination from 1 to 89 degrees with the line
ray-tracing module. The line fluxes vary by up to 30\% for
inclinations below 70 degrees. At high inclination, the cold CO in the
outer flaring disc starts to reabsorb the emission from the inner
disc.

  \subsubsection{CO ro-vibrational level
    excitation}\label{CO_excitation} 

  The CO vibrational levels in the ground electronic state can be
  populated by collisions with the main chemical species (H, H$_2$,
  electrons, He), by formation pumping (i.e. formation of CO in an
  vibrationally excited level), by pumping by dust emission, or by
  UV-fluorescence pumping.  The region of high warm CO abundance is H
  rich and H$_2$ poor \citep{Kamp2004ApJ...615..991K} (see
  Fig.~\ref{CO_emission_high_mass_disk_4panels} and
  \ref{CO_emission_low_mass_disk_4panels}). We also plotted the
  excitation temperatures. We chose two typical transitions, one from
  the $v=1$ level and one from the $v=4$ level, to illustrate the
  efficiency of the excitation mechanisms. The relative population
  between the $v=1, J=19$ and $v=1, J=18$ and between $v=4, J=20$ and
  $v=4, J=19$ levels are shown as excitation temperatures in
  Fig.~\ref{Tex_high_mass_disk_4panels} and
  \ref{Tex_low_mass_disk_4panels} for the high- and low-mass disc
  models, respectively. The excitation temperatures between the $v=1,
  J=19$ and $v=0$, $J=18$ levels in the line-emitting region are below
  the gas kinetic temperatures but higher than the radiation
  temperature at the line frequency 4.844 micron (subthermal
  population, see Fig.~\ref{Trad_fundamental},
  \ref{Tex_Trad_Tgas_massive_disc_ratio} and
  \ref{Tex_Trad_Tgas_low_mass_disc_ratio} in the appendix). However,
  the excitation temperatures between the $v=4, J=20$ and $v=3, J=19$
  levels are much higher than the gas kinetic and radiation (at 4.692
  micron) temperatures (suprathermal population). We can distinguish
  between the excitation mechanisms of the $v=1$ level and that of the
  $v>1$ levels.

  The $v=1$ level is mostly populated by collisions and IR pumping
  because the disc densities in those emitting regions are close to
  the critical densities (10$^{10}$-10$^{12}$cm$^{-3}$). The $v=1$
  level population is weakly affected by UV-fluorescence pumping (see
  the upper panels of Fig.~\ref{Tex_high_mass_disk_4panels} and
  \ref{Tex_low_mass_disk_4panels}). Likewise the line fluxes for the
  fundamental transitions change by at most a factor four when
  UV-fluorescence pumping is taken into account
  (Fig.~\ref{Linefluxes_4panels}).

  The collision rates with electrons are $\sim$~100 times higher than
  with H but the electron abundance is $\sim$~10$^{-6}$ lower than the
  H-atom abundance. The collision rates with He are of the order of
  10$^{-16}$ cm$^{3}$ s$^{-1}$ at 500~K. The abundance is 0.075 times
  less than H+H$_2$. Therefore He is not a major collision
  partner. The main collision partner in the CO ro-vibrational
  emitting region is atomic hydrogen, whose de-excitation collision
  rates with CO are two orders of magnitude (10$^{-12}$ cm$^{3}$
  s$^{-1}$ at 500~K) higher than rates with H$_2$.
    
  The $v>1$ levels have higher critical densities than the $v=1$
  levels and require very high gas temperatures
  $T_{\mathrm{gas}}>$1000~K and high densities to be populated
  efficiently by collisions. Pumping of the $v>1$ levels by UV
  fluorescence becomes the prominent excitation mechanism. The
  excitation temperatures of UV-pumped levels (lower left-hand panels
  in Fig.~\ref{Tex_high_mass_disk_4panels} and
  \ref{Tex_low_mass_disk_4panels}) tend towards the effective
  radiation temperature (brightness temperature) in the UV
  (Fig.~\ref{Trad_hot}, the radiation temperatures at 0.15 micron).
 
  The effect of UV pumping becomes more pronounced in the 10$^{-4}$
  M$_\odot$ disc models (3a and 3b,
  Fig.\ref{CO_emission_low_mass_disk_4panels}), where the gas
  densities and hence the UV-pumped high-$v$ level quenching
  efficiencies are lower and the UV photons are less absorbed by the
  dust grains (Fig.~\ref{Trad_hot}). In the
  $M_{\mathrm{disc}}$=10$^{-2}$ M$_\odot$ disc model, the UV pumping
  is confined to a very thin layer at the inner disc rim, whereas in
  the low-mass disc models ($M_{\mathrm{disc}}$=10$^{-4}$ M$_\odot$),
  the UV pumping occurs till $\sim$~0.5~AU inside the disc because the
  UV photons can penetrate deeper in the disc.

    The UV-fluorescence pumping can affect dramatically the line
    fluxes (Fig.~\ref{Linefluxes_hot_lines_4panels}).  In particular
    the hot line $v=4-3$ $P$(20) is two orders of magnitude stronger
    with UV-pumping. The effect of UV pumping is better captured by
    computing the vibrational temperature (see
    Sect.~\ref{vibrational_temperature}). The relative rotational
    level populations within a vibrational level are less affected by
    UV pumping than the relative vibrational level populations (see
    the shape of the SLED in Fig.~\ref{Linefluxes_4panels}).

    An alternative excitation mechanism is the formation pumping of
    CO. The CO formation reaction via C + OH has an exothermicity of
    -6.5 eV and a rate of $k_{\mathrm{chem}}\sim$10$^{-10}$ cm$^{3}$
    s$^{-1}$ \citep{Zanchet2007JChPh.126r4308Z}. The excess energy is
    used to pump CO to high vibrational levels with a maximum
    population at $v$=10 \citep{Bulut2011JChPh.135j4307B}. The
    formation of CO constitutes a pumping mechanism for CO.

    We can compare the efficiency of the chemical pumping with the
    efficiency of the population of the $v$=1 level by collisions with
    atomic hydrogen by noting that OH reaches a maximum abundance of
    $\sim$10$^{-6}$ (Fig.~\ref{CO_15panels}). The chemical pumping
    rate assuming atomic carbon abundance [C]$\sim$∼10$^{-4}$ is
    $k_{\mathrm{chem}}n_{\mathrm{C}}n_{\mathrm{OH}}\sim$ 10$^{-20}$
    $n^2$ cm$^{−3}$ s$^{−1}$ , where $n$ is the gas density. Collision
    rates with atomic hydrogen are of the order of 10$^{-16}n^2$
    cm$^{−3}$ s$^{−1}$, a few orders of magnitude more efficient than
    the chemical pumping.
    
\subsubsection{Rotational diagram for the fundamental transitions}\label{rotdiagram}

The analysis of the line fluxes using a rotational diagram can provide
more insight than the study of the shape of the SLEDs. We plotted in
Fig.~\ref{rotDiag_4panels} the fundamental transition rotational
diagram, i.e. the log of the level column density versus
$B_{\mathrm{CO}}J'(J'+1)$, where $B_{\mathrm{CO}}$ is the rotational
constant in Kelvins, for the 10$^{-2}$ and 10$^{-4}$ M$_\odot$
$R_{\mathrm{in}}$=1~AU and 20~AU models with (blue symbols) and
without UV pumping (red symbols). The shape of a rotational diagram is
affected by NLTE (subthermal population and IR/UV pumping) and optical
depth effects. To obtain the optically thin emission in LTE (purple
solid lines), we populated the CO ro-vibrational levels assuming
$T_{\mathrm{ex}}=T_{\mathrm{gas}}$. The line fluxes were computed in
the optically thin approximation, i.e. optically depth effects, which
would have prevented the flux from increasing indefinitely, were not
taken into account.

The optically thin CO LTE populations probe the actual kinetic
temperatures in the discs. The populations were calculated by summing
for each level all the CO molecules at that level in the models,
assuming that the population is in LTE.  For each model, the
rotational population of the $v=1$ level can be relatively easily
matched by a single temperature $T_{\mathrm{rot}}^{\mathrm{LTE}}$
because the CO fundamental lines probe a small region in discs where
the gas is warm and dense. Moreover, the densities are high enough for
the low- and medium-$J$ rotational levels within the $v=1$ level to be
close in rotational LTE.  The line profiles will be wider for high-$J$
than for low-$J$ transitions in discs not seen edge-on. The LTE
temperatures (fits to the purple lines shown with black crosses)
$T_{\mathrm{LTE}}$ are given in Table~\ref{ModelTable}. The gas is
warmer for lower mass, hence lower density discs because the
molecules, which are efficient gas coolants, are more rapidly formed
in dense gases.

Since the stellar radiation decreases with distance squared, it is
normal that the gas is cooler for discs with an inner radius starting
at 20~AU rather than at 1~AU.  Since CO ro-vibrational lines are
efficient coolant for the gas, the gas probed by the CO gas is warmer
(i.e. with higher $T_{\mathrm{rot}}^{\mathrm{LTE}}$) when the number
of ro-vibrational levels decreases (Table~\ref{ModelTable}).

\begin{figure*}[ht]  
  \centering
\includegraphics[scale=0.37,angle=90]{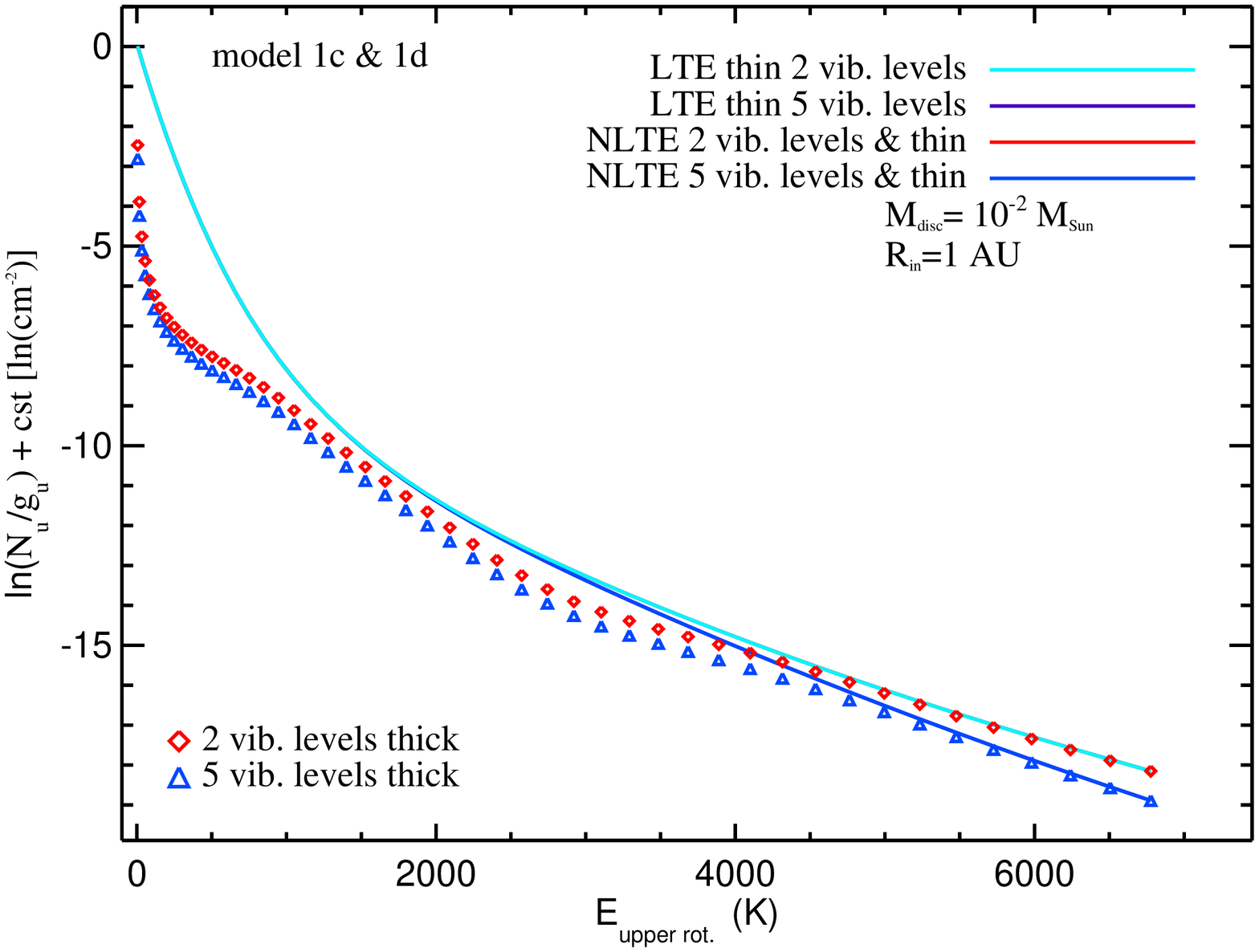}
\includegraphics[scale=0.37,angle=90]{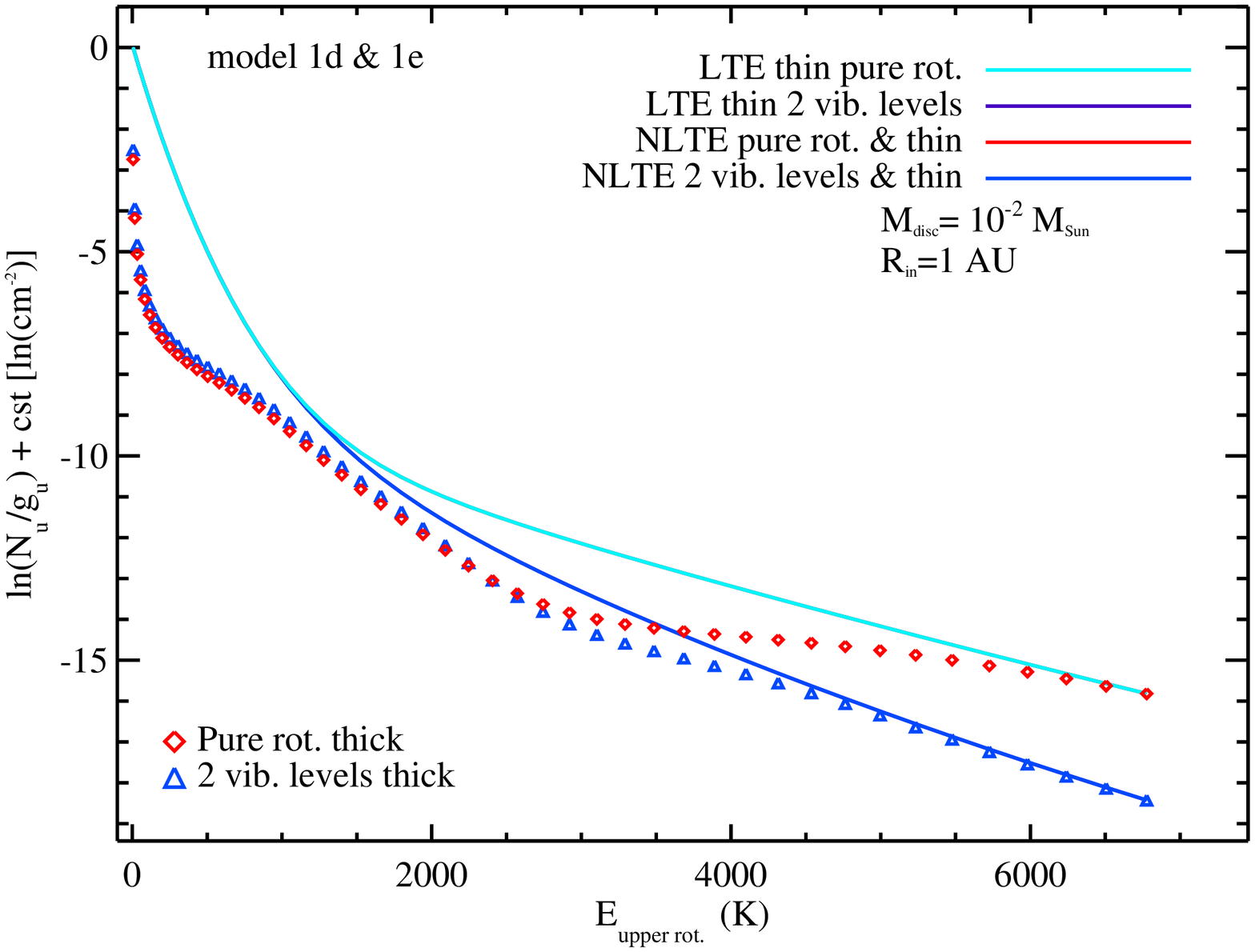}  
\includegraphics[scale=0.37,angle=90]{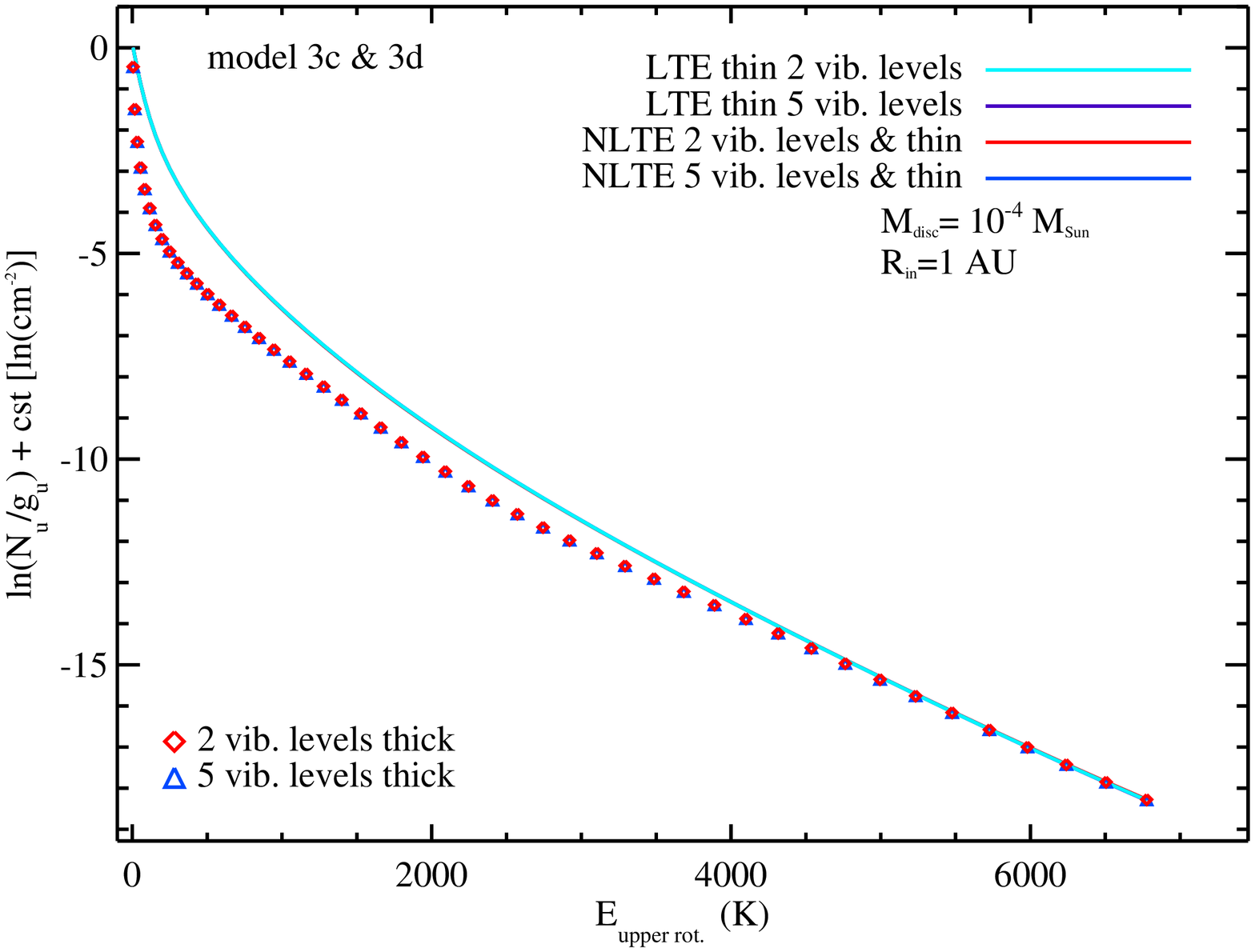}
\includegraphics[scale=0.37,angle=90]{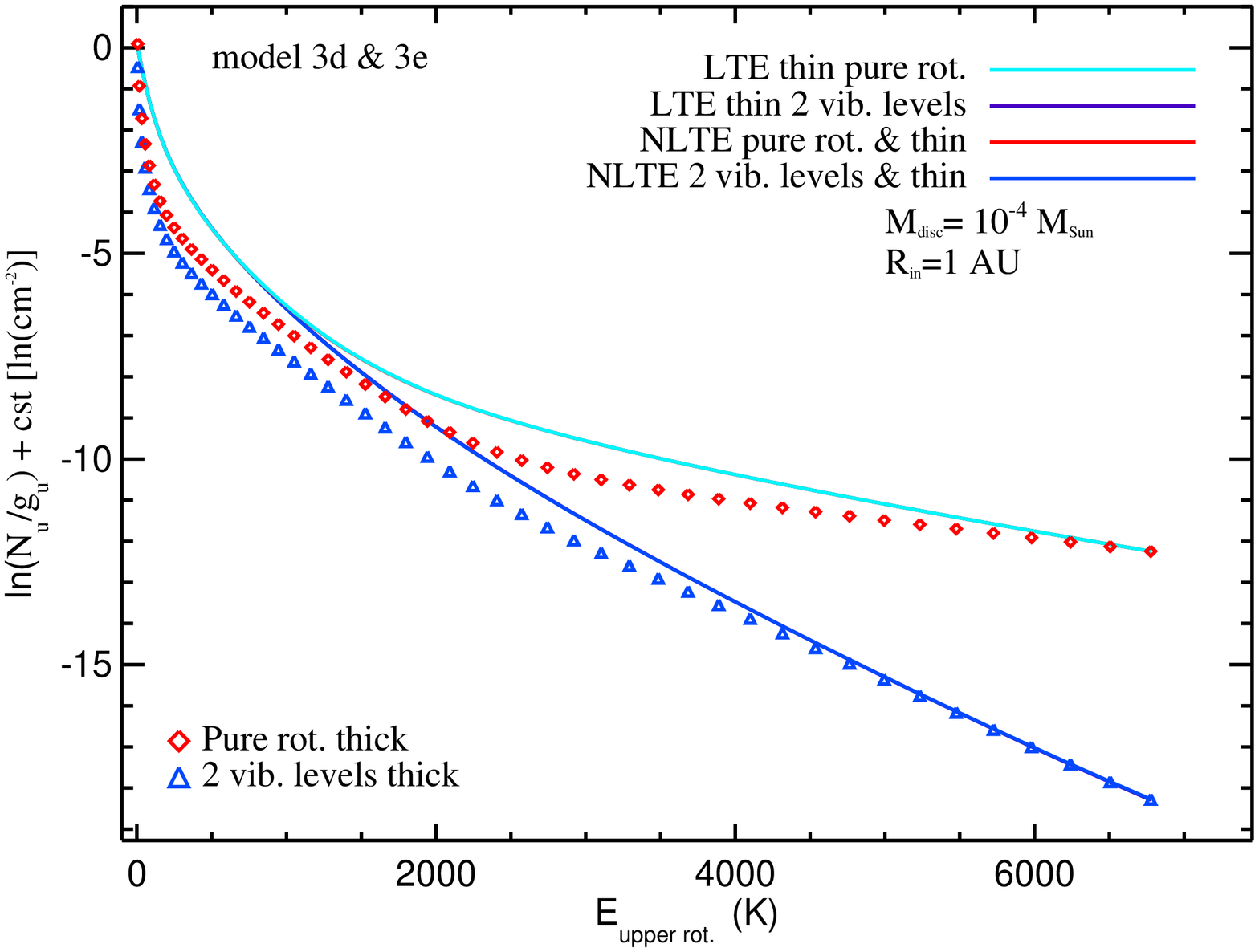}  
\centering  
\caption{\label{purerotDiag_4panels} Disc CO $v=0$ rotational
  population at LTE and at NLTE for the $R_{\mathrm{in}}$=1~AU models
  with only the $v=0$, $v=1,2$, and $v=1,2,3,4,5$ levels taken into
  account. All populations have been normalized.}
\end{figure*}
   
The red and blue lines in Fig.~\ref{rotDiag_4panels} correspond to the
CO ro-vibrational NLTE level population computed locally in the disc
using the 1+1D escape probability method. Those lines show that the
level populations are rotationally subthermal. The effects are
stronger for less massive (10$^{-4}$ M$_\odot$), hence less dense
discs.

The ro-vibrational NLTE population is orders of magnitude smaller than
the ro-vibrational LTE population for all models, pointing to a
subthermal population of the excited ro-vibrational levels: the
rotational level populations within a vibrational level can be in LTE,
while the vibrational level populations are subthermal. Assuming a
transition probability of 10 s$^{-1}$ and a rate coefficient of
10$^{-16}$ cm$^{3}$ s$^{-1}$ (H$_2$) to 10$^{-12}$ cm$^{3}$ s$^{-1}$
(H), the critical density is 10$^{11}$--10$^{15}$ cm$^{-3}$. Such high
densities are reached only close to the star in the inner disc
midplane of massive discs.

Radiative transfer of the optically thick lines decreases further the
populations derived from the analysis of rotational diagrams. One
consequence is that rotational temperatures derived from fitting
points from low-$J$ optically thick lines in rotational diagrams would
appear lower than the actual gas kinetic temperatures. In appendix
\ref{appendix1}, we provide a simple analytical method to analyse
observations of optically thick $^{12}$CO v=1-0 emission
lines. Optical thickness results in flux differences between lines of
transitions that have the same upper level but lower levels that
differ by $\Delta J$ and different transition probabilities: the $P$
($J"=J'+1$) and the $R$ ($J"=J'-1$) branches.  As expected, the column
density differences between LTE and NLTE population are strong in
models 1a and 1b and are weak in model 4a and 4b.

The $R$-branch transitions generally have higher transition
probabilities (i.e. with higher Einstein-$A$ coefficients) than
$P$-branch transitions. Hence, if the lines are optically thick, the
column densities derived from $R$-branch lines are the smallest
($N_{J'} \propto F_\nu/A_{ul}$ and $F_\nu$ are almost the same). Lines
are optically thick up to very high $J$, giving rise to a steep slope
for the low $J$ and to a shallow slope for the moderate $J$ lines
(symbols). Observed rotational diagrams of Herbig~Ae discs show
similar shapes
\citep{Blake2004ApJ...606L..73B,Brittain2007ApJ...659..685B,Brown2012ApJ...744..116B}. Only
lines from $J'>$~40 ($>$ 25 for the 10$^{-4}$ M$_\odot$,
$R_{\mathrm{in}}$=20~AU disc model) are optically thin: their
populations join the optically thin NLTE populations and the
differences between the $P$ and $R$ branch column densities fade away.
The optical thickness of the lines and rotationally subthermal
population of the levels can cause the lines to be actually optically
thick but effectively thin, especially in discs where the continuum
emission at 4-5 micron is optically thin.

UV pumping affects the level population for all levels with $v>1$
whatever the inner disc radius. The main effect of UV pumping is to
populate the high-$J$ and $v>1$ (see discussion below and
Fig.~\ref{vibDiag_4panels}) levels at the expense of the lower $v=0,J$
levels. We also ran the models with UV pumping but without collision
excitations between the ground and electronic excited states. However,
we noticed no difference over a few percents.

The shape of the rotational diagrams (symbols) shows that it would be
difficult to fit straight lines to estimate the excitation
temperatures. Moreover, they do not reflect the actual CO population
due to strong optical depth effects.

\subsubsection{Vibrational temperature and UV/IR fluorescence pumping}
\label{vibrational_temperature}

The vibrational temperature measures the relative population of the
vibrational levels. We plotted in Fig.~\ref{vibDiag_4panels} and
\ref{vibDiag_other_models} the log of the vibrational level population
($\ln(N_v)$) as function of the vibration energy $E_v=3122v$, where $v$
is the vibrational quantum number
\citep{Brittain2007ApJ...659..685B}. The vibrational level population
is the sum of all the rotational level populations at that vibrational
level.

We derived the vibrational temperatures by fitting straight lines
through the data in models with and without UV pumping (nine and five
vibrational levels) apart from the 10$^{-2}$ M$_\odot$ $R_{\mathrm
  {in}}$=1~AU model where only $v$=4, 5, and 6 levels are taken into
account because a straight line could not be fitted when those levels
were included. The values are reported in the last column of
Table~\ref{ModelTable}.

Even without UV pumping, the vibrational temperatures are higher than
the LTE rotational temperatures.  However, the rotational
temperatures derived using points in the shallow slope of the diagrams
will yield temperatures higher than the LTE values and closer to the
non-pumped vibrational temperatures. 
  
Two clear trends can be seen from the vibrational diagrams and the
temperatures in Table~\ref{ModelTable}. First, the vibrational
temperature is higher for the lower mass disc. Second, the vibrational
temperature increases with the inner hole size if UV pumping is
switched on, whereas it decreases if UV pumping is switched off.

When the UV pumping is not present, the high vibrational levels can be
populated by collisions and/or by IR pumping. The collision rates for
vibrational transitions are lower than for rotational transitions. If
the vibrational levels are populated by collision only, then the
vibrational temperatures should be lower then the rotational
temperatures. Therefore, collisions play a minor role in populating
the high-$v$ ro-vibrational levels, leaving IR pumping as the
dominating mechanism. In the absence of UV pumping, the vibrational
temperatures are still higher than the maximum dust temperatures in
the discs (max($T_{\mathrm{dust}}$) in Table~\ref{ModelTable}). The IR
continuum flux between 1 and 5 $\mu$m is dominated by the dust
emission for the models with $R_{\mathrm{in}}$=1~AU (models 1 and 3)
and by the stellar emission for models with an inner hole
($R_{\mathrm{in}}$=20~AU; models 2 and 4). The vibrational
temperatures are the same whether five or nine vibrational levels are
taken into account in the models, suggesting that the effect of IR
pumping does not depend on the vibrational level.

As the gas cools with increasing inner radius, the high-$v$ levels are
less populated.  Absorption of UV photons directly by CO molecules
will favour the population of high-$v$ levels unless high gas
densities quench the overpopulation. The gas density at the inner edge
decreases with the edge's distance to the star. Higher
$R_{\mathrm{in}}$ values and lower mass discs will have less
collisional de-excitations of UV-pumped high-$v$ levels.

\subsubsection{Effect of IR pumping on the rotational population of the $v=$1 level}
 
The left-hand panels in Fig.~\ref{purerotDiag_4panels} show the
differences between the case with two and five vibrational levels. For
the massive disc with $M_{\mathrm{disc}}$=10$^{-2}$ M$_\odot$, the
population decreases as more vibrational levels are included. The CO
$v>$1 levels can be pumped by absorptions in the hot and overtone
bands of stellar or dust IR photons, resulting in lower populations of
the $v$=1, $J$ levels (upper right-hand panel). The pure rotational
levels in LTE cannot be explained by a single excitation temperature
contrary to the rotational levels in the first vibrational excited
state.  The pure rotational transitions probe much larger disc areas
with wide-ranging physical conditions than do the fundamental
ro-vibrational transitions. This stems from the low critical densities
needed to populate the pure-rotational levels.

The right-hand panels in Fig.~\ref{purerotDiag_4panels} compare the
population in models where only rotational levels with $v=$0 are
included and where the levels $v=$0 and $v=$1 are included. In the
optically thin cases (solid lines), the differences between the cyan
and blue lines reflect the rotationally subthermal populations for the
high-$J$ populations. The low-$J$ levels are at LTE. However, the high
optical depths for the low- and mid-$J$ lines result in lower derived
population and the appearance of a change in the slope at
$J$=8--12. At high $J$ ($E_{\mathrm{rot}}>$2000 K), the
overpopulations when only the ground vibrational ($v$=0, $J$) are
modelling artifacts because overlapping energies occur between the
$v$=0, high-$J$ level energies and that of the $v$=1, low-$J$
levels. For a gas at temperature $T$, both the high $J$ and low $J$ of
the $v=$1 levels can and should be populated. In warm and hot gases,
models that do include the pure rotational levels only may
overestimate their populations.

\section{Conclusion}\label{conclusions}

We have implemented a complete CO ro-vibrational molecular model in
the ground $X^1\Sigma^+$ and $A^1\Pi$ electronic state in the
radiative chemo-physical code {\sc ProDiMo}. We gathered existing
collisional rate coefficients and used scaling rules to extrapolate
the missing ones. Collision rates of CO with atomic hydrogen are two
orders of magnitude higher than rates with molecular hydrogen and
three orders of magnitude higher than with He.

We ran models that include continuum, NLTE-line radiative transfer
assuming a Voigt line profile, chemistry, and gas thermal balance to
understand the effect of UV on the population of high-vibrational and
high-rotational levels in protoplanetary discs. CO ro-vibrational
lines are emitted within the first few AU and probe disc regions where
CO is quickly formed via C + OH, with OH being formed via the fast
reaction of vibrationally excited H$_2$ with atomic oxygen.

The amount of atomic hydrogen is high in the CO line-emitting
region. The $v=1$ level population is dominated by collisional
excitation with atomic hydrogen with contributions from IR-pumping and
UV-fluorescence pumping. Excitation of high-$J$ and high-$v$ levels by
UV and IR photons can be efficient, especially in low mass discs where
collisional quenching is less rapid than in high-mass discs. The
molecular gas is rotationally ``cool'' and subthermal but
vibrationally ``hot'' and suprathermal. However, only the detection of
hot lines ($v'\ge$~2, $\Delta v$=1) in discs can provide a test of the
efficiency of UV-fluorescence.

The pure rotational transitions probe the entire disc while the
ro-vibrational transitions probe the inner warm disc region. Therefore
the rotational temperature depends on the vibrational number of the
initial level. It is difficult to derive accurate temperatures and
column densities from the optically thick $^{12}$CO fundamental
emissions. Spatially and spectrally resolved CO, $^{13}$CO, C$^{18}$O,
and C$^{17}$O ro-vibrational observations are needed to disentangle
the optical depth effects and constrain the location of the line
emissions. The low- and medium-$J$ rotational levels on the ground and
first excited vibrational levels are populated in LTE in discs due to
the low critical densities of the pure rotational transitions.

Future studies will focus on studying the effects of disc parameters
(mass, size, flaring index, ...) on the CO ro-vibrational emissions
and will include larger numbers of vibration levels and electronic
excited states.
  

\begin{acknowledgements}
  We thank ANR (contract ANR-07-BLAN-0221) and PNPS of CNRS/INSU,
  France for support. IK, WFT and PW acknowledge funding from the EU
  FP7-2011 under Grant Agreement nr. 284405. GvdP thanks the
  Millennium Science Initiative (ICM) of the Chilean ministry of
  Economy. Computations presented in this paper were performed at the
  Service Commun de Calcul Intensif de l'Observatoire de Grenoble
  (SCCI) on the super-computer funded by Agence Nationale pour la
  Recherche under contracts ANR-07-BLAN-0221, ANR-2010-JCJC-0504-01
  and ANR-2010-JCJC-0501-01.We acknowledge discussions with Ch. Pinte,
  F. M\'{e}nard, and A. Carmona. We thank the referee and the editor
  for the useful comments that helped improve the paper.
\end{acknowledgements}

\bibliographystyle{aa} 
\bibliography{co_rovib.bib}

\begin{appendix}
\section{Appendix}

\subsection{Rotational diagram analysis for levels in LTE and optically thick lines}
\label{appendix1}

We adopted the rotational diagram analysis that takes optical depth
effects into account (also called population diagram analysis in
\citealt{Goldsmith1999ApJ...517..209G}). A similar approach was taken
by \citet{Goto2011ApJ...728....5G}. The line intensity emitted by a
slab at a single temperature reads

\begin{equation}
\log{I_\nu}=\log{\left(\tau_\nu B_\nu\right)} + \log{\left(\frac{1-e^{-\tau_\nu}}{\tau_\nu}\right)}.
\end{equation}

$B_\nu$ is the Planck function at the excitation
$T_{ex}$=$T_{rot}$=$T_{vib}$.  The second term in the right-hand side
is an optical depth correction factor to the optically thin rotational
diagram. The term $\beta=(1-\exp{(-\tau_\nu)})/\tau_\nu$ is akin to
the escape probability. The maximum optical depth occurs at $E_{rot}
\sim kT_{rot}$, where $E_{rot}$ is the rotational upper level energy
of an emission line. In the Born-Oppenheimer approximation, $E_u =
E_{rot} + E_{vib}$. The rotational energy for CO can be approximated
by $E_{rot}=B_{rot}J_u(J_u+1)$, with the rotational constant
$B_{rot}=$ 2.76 K. The maximum optical depth occurs when
$J_u\simeq\sqrt{T_{rot}/B_{rot}}$. The line optical depth is
\begin{equation}
  \tau_\nu=\left(\frac{c^2}{8\pi}\right)\left(\frac{A}{\nu^3}\right)\left(\frac{c}{\Delta v}\right)\left(e^{(h\nu/kT_{ex}}-1\right)x(T_{rot},T_{vib})N(\mathrm{CO})/\cos(i),
\end{equation}
where $\nu$ is the line frequency, $A$ the Einstein spontaneous
emission probability of the transition, $c$ the speed of light,
$\Delta v$ the turbulent width, $N$(CO) the CO column density, $i$ the
disc inclination, and
\begin{equation}
x(T_{rot},T_{vib})= \frac{e^{-E_{rot}/kT_{rot}}}{Q_{rot}}\times\frac{e^{-E_{vib}/kT_{vib}}}{Q_{vib}}
\end{equation}
is the CO fractional population in the initial level ($v'$,$J'$). For
CO, $Q_{rot}\simeq kT_{rot}/B_{\mathrm CO}+1/3$ ($B_{\mathrm
  CO}=2.76$~K) and $Q_{vib}= 1/(1-\exp(-3122/T_{vib})$ are the
rotational and vibrational partition function respectively \citep{Brittain2007ApJ...659..685B}.

\begin{figure}
  \centering
 \includegraphics[scale=0.35,angle=90]{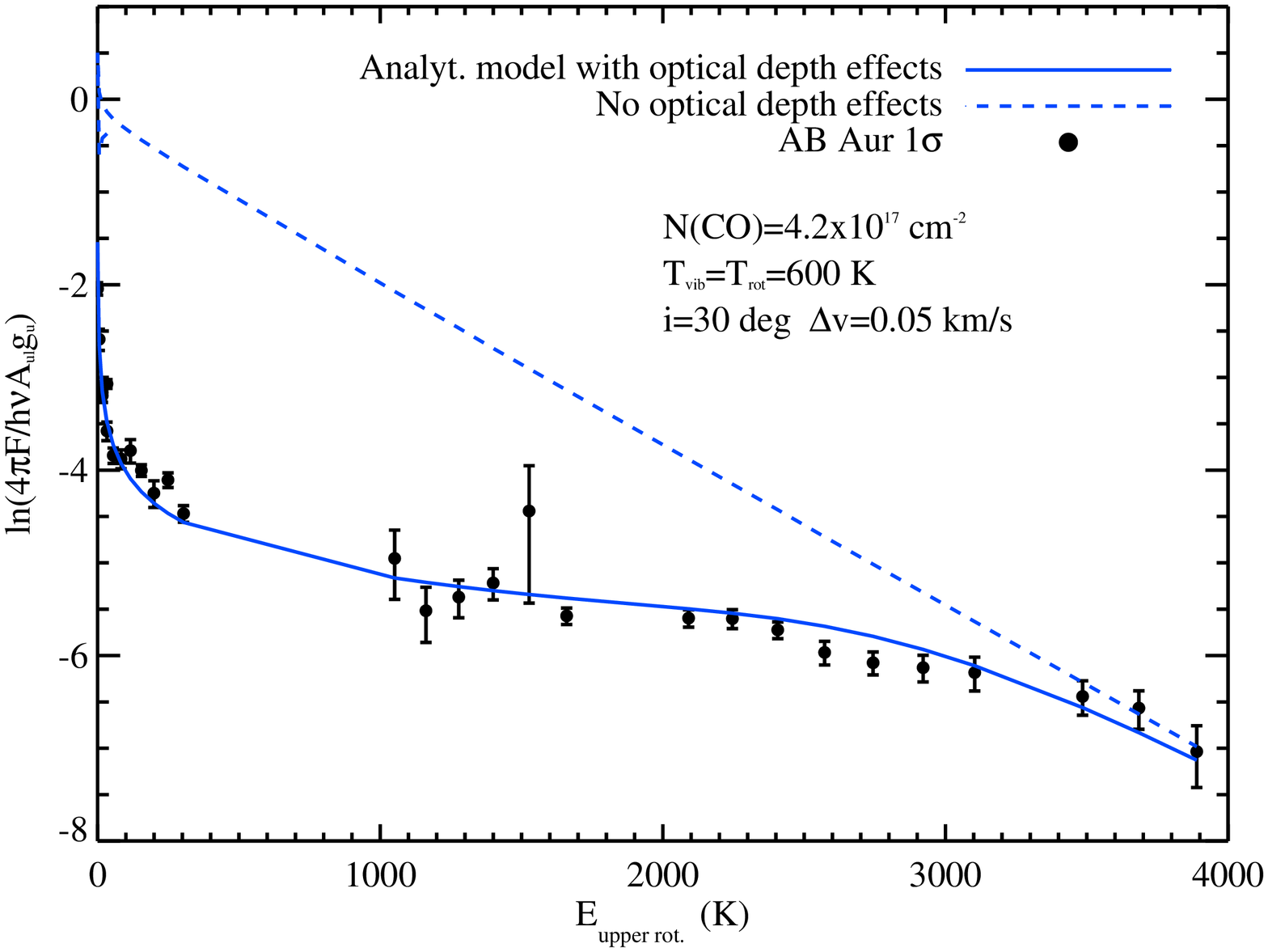}
 \includegraphics[scale=0.35,angle=90]{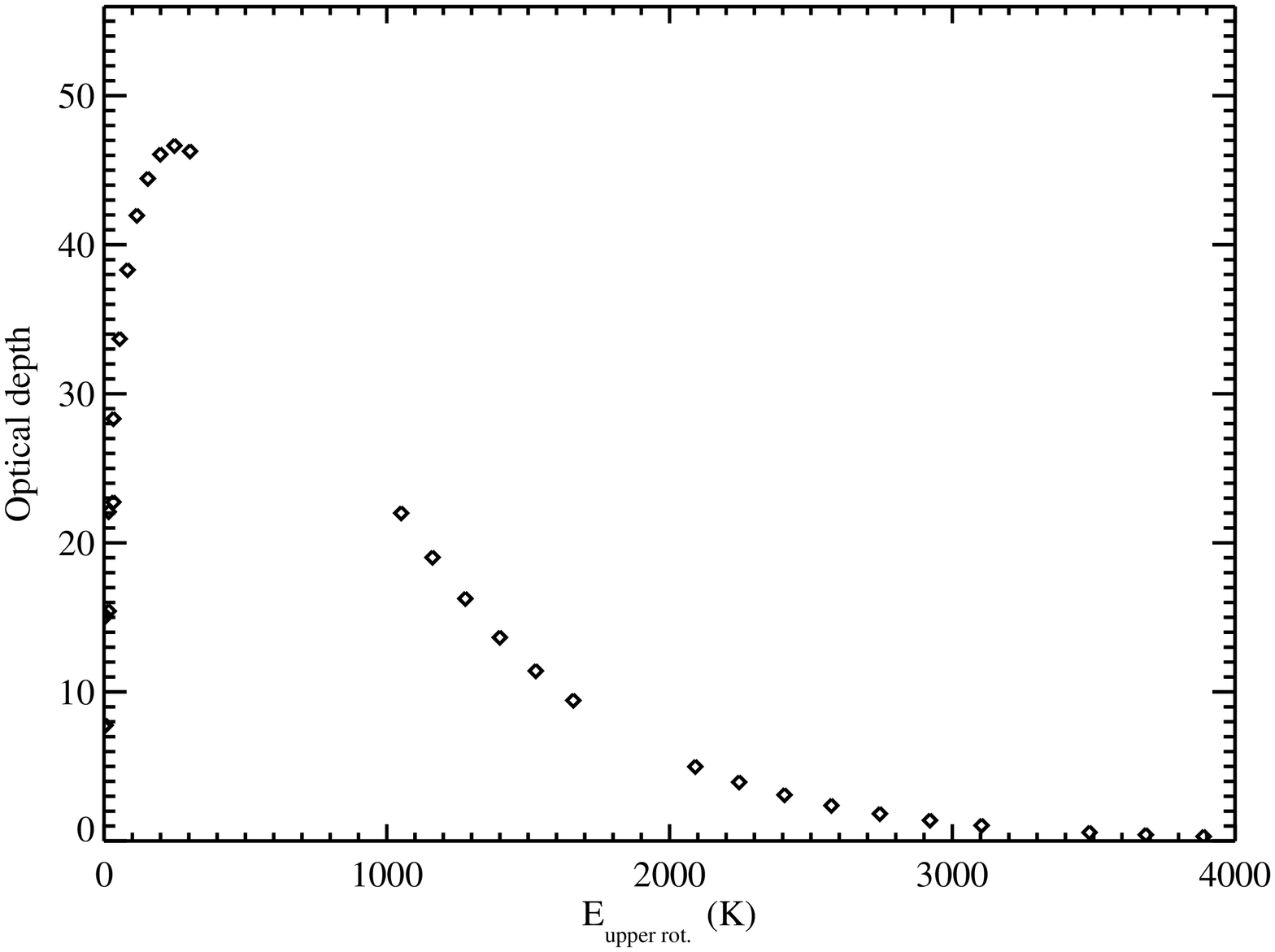}
 \caption{\label{ABAur_rotdiagram_analytical_fit} The upper panel
   shows the comparison between \object{AB Aur} and an analytical
   model rotational diagram for the $^{12}$CO $v$=1-0 transitions
   observed by \citet{Brittain2003ApJ...588..535B}. The solid blue
   line shows the fit by an analytical model that takes optical depths
   into account. The results from the same model but without the
   effect of optical depth are shown in dashed blue line. The lower
   panel shows the derived line optical depth.}
\end{figure}

The rotational diagram of \object{AB Aur}
(Fig.~\ref{ABAur_rotdiagram_analytical_fit}) shows three parts: a
steep slope corresponding to increasing optical depth until $E_{rot}
\sim kT_{ex}$, then a shallow slope where the line optical depth
decreases because the higher the rotational level, the less they are
populated. At very high $J$, the slope steepens again. This behaviour
also appears in our theoretical rotational diagrams, but the second
turning point occurs at higher $J$ than in the observations. The
second slope change corresponds to lines with $\tau<$1. Assuming that
the population is in LTE, we fitted the \object{AB Aur} rotational
diagram by a model that takes the optical depth effects into account
\citep{Goldsmith1999ApJ...517..209G}. The model parameters are the CO
column density $N$(CO)=4.2 $\times$ 10$^{17}$ cm$^{-2}$, the mean
excitation temperature $T_{ex}\simeq$~600~K (we assume that the
vibrational temperature is equal to the rotational temperature), the
turbulent width $\Delta v$=0.05, and the inclination $i$=30 degree. In
the upper panel of Fig.~\ref{ABAur_rotdiagram_analytical_fit}, the
analytical solid curve compares well with the observations. The
dashed-line curve shows the same model where optical depth effects are
not taken into account. The lower panel shows the derived optical
depths, which reach $\sim$~47.

\subsection{Excitation, radiation, and gas kinetic temperatures}

\begin{figure*}
  \centering 
  \includegraphics[scale=0.5,angle=0]{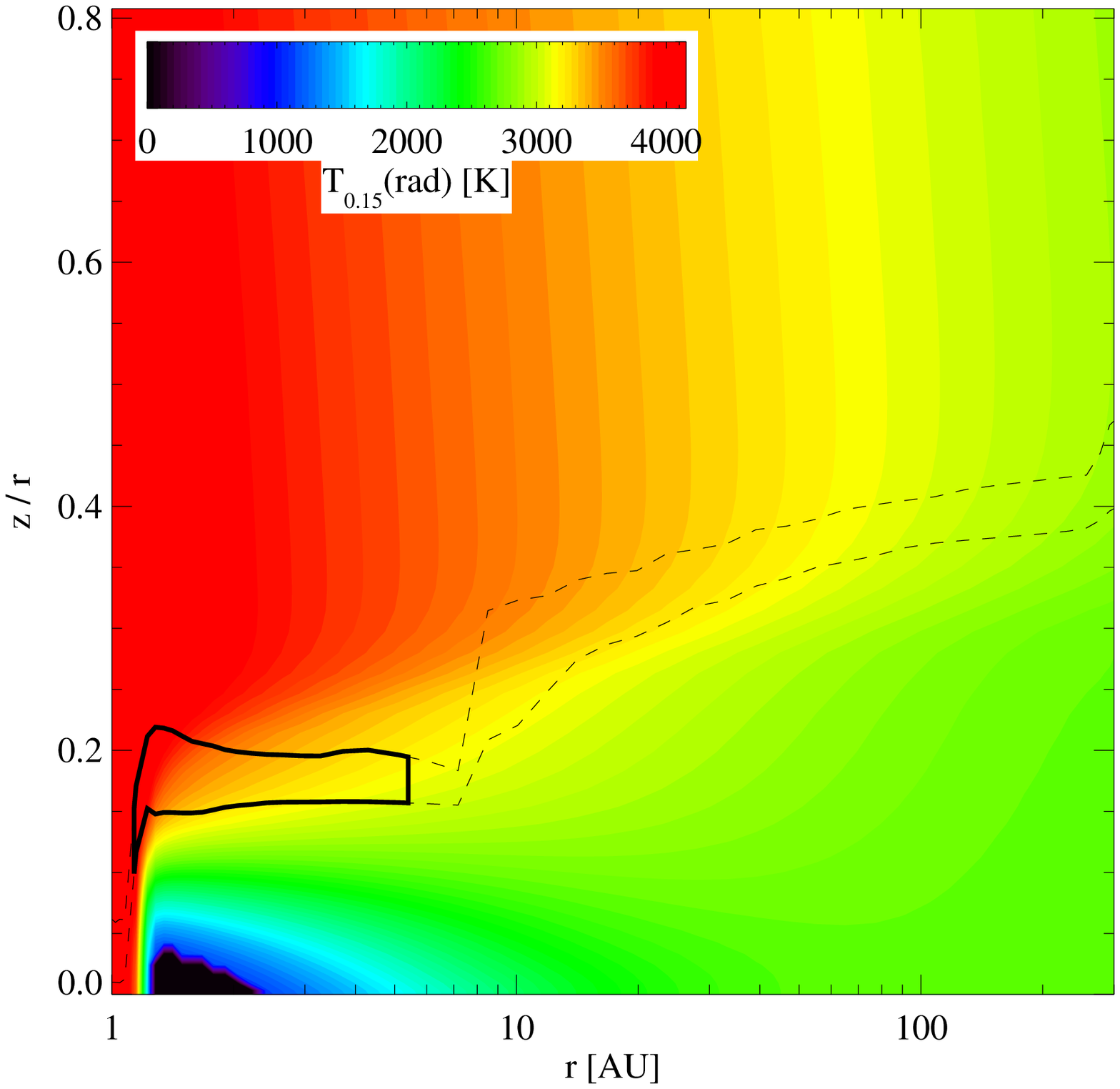}
  \includegraphics[scale=0.5,angle=0]{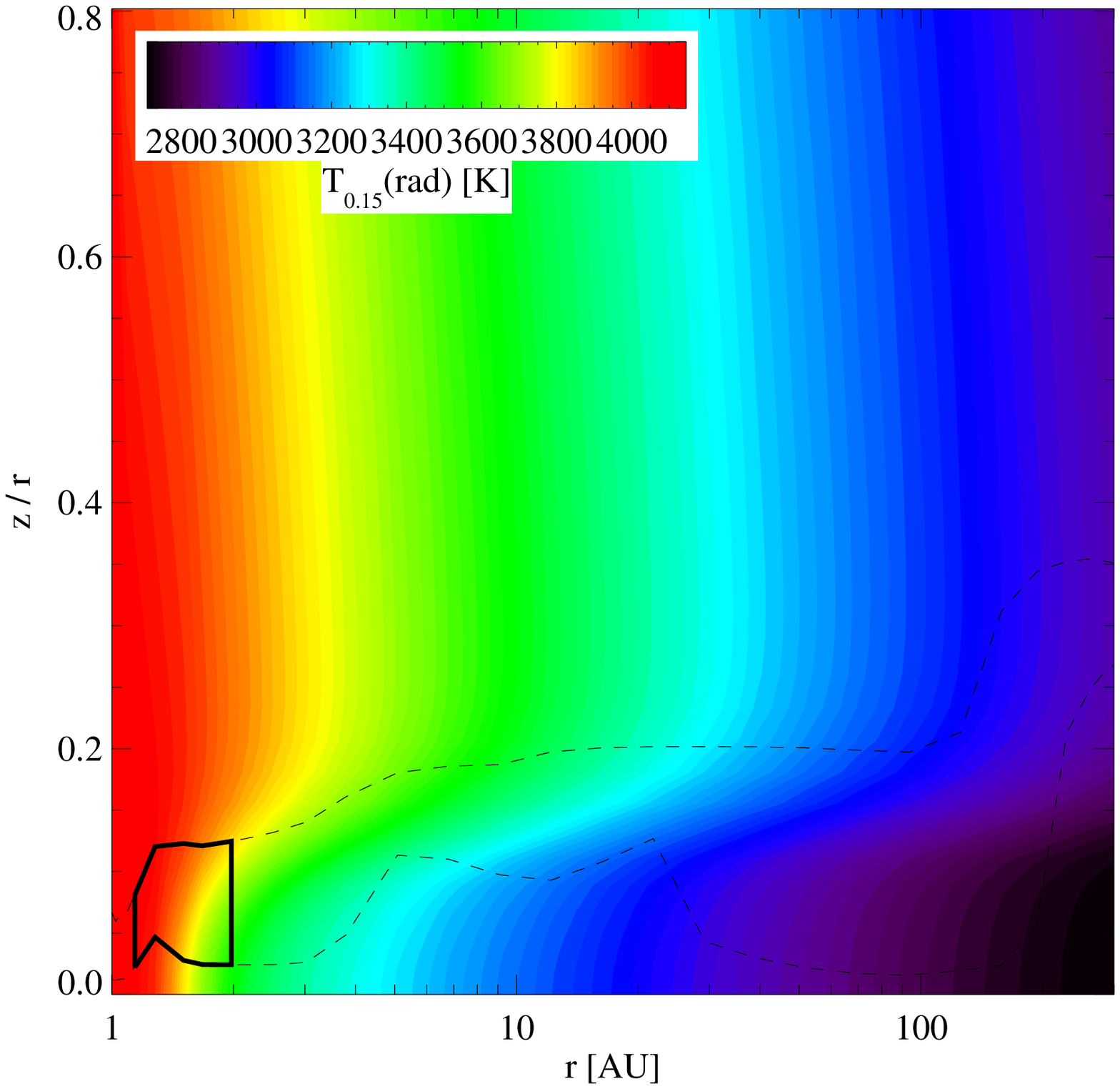}
  \includegraphics[scale=0.5,angle=0]{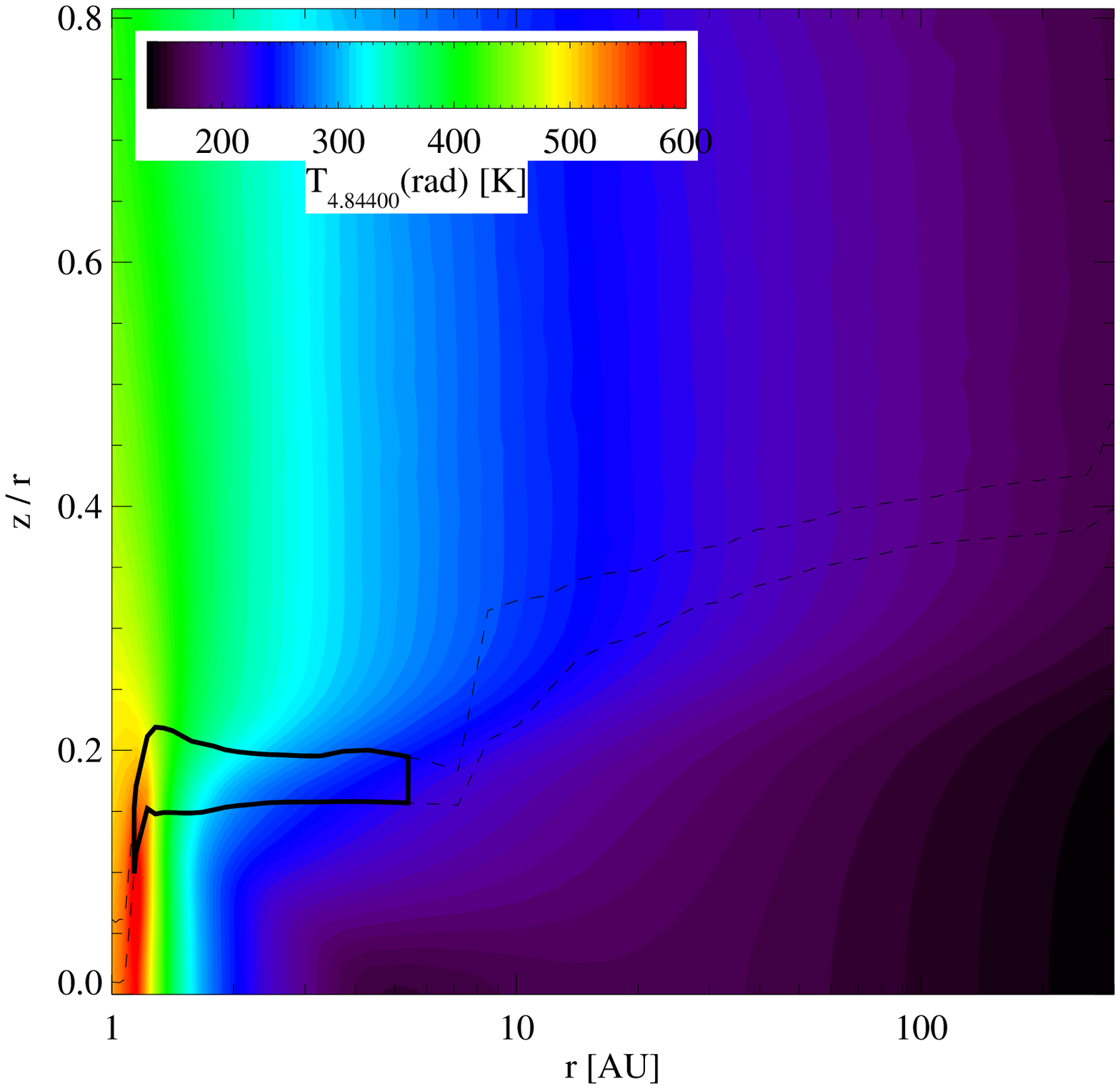}
  \includegraphics[scale=0.5,angle=0]{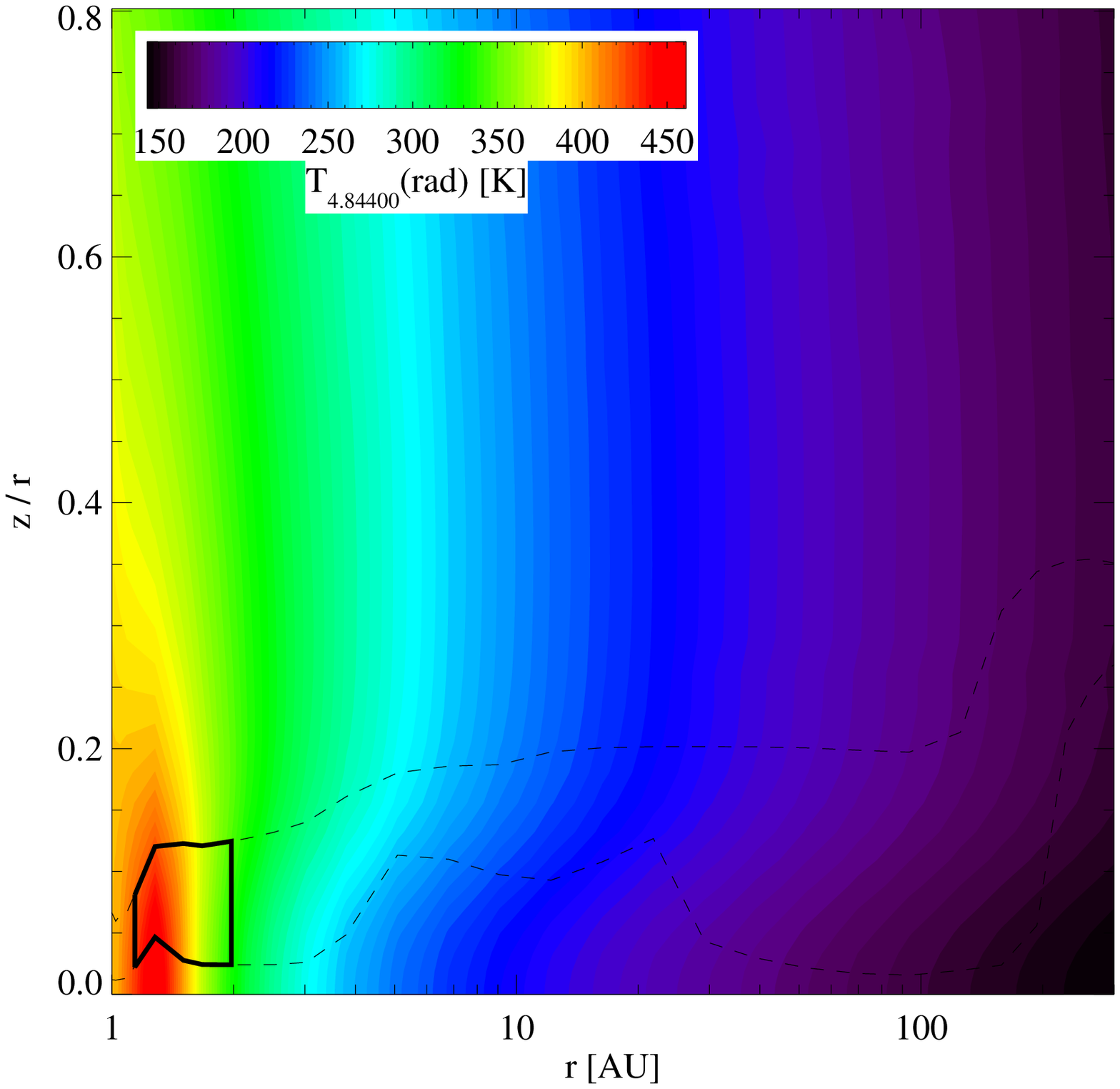}
  \caption{\label{Trad_fundamental} Radiation temperature
    $T_{\mathrm{rad}}$ at 0.15 micron and 4.844 micron for the
    $M_{\mathrm{disc}}=10^{-2}$ M$_\odot$, $R_{\mathrm{in}}$=1~AU disc
    models (left panel) and for the $M_{\mathrm{disc}}=10^{-4}$
    M$_\odot$, $R_{\mathrm{in}}$=1~AU disc models (right panel). The
    black contour shows the regions that emit 49\% of the fluxes at
    4.844 micron.  The black dashed-line contours contain 70\% of the
    fluxes in the vertical direction.}
\end{figure*}

\begin{figure*}
  \centering
  \includegraphics[scale=0.5,angle=0]{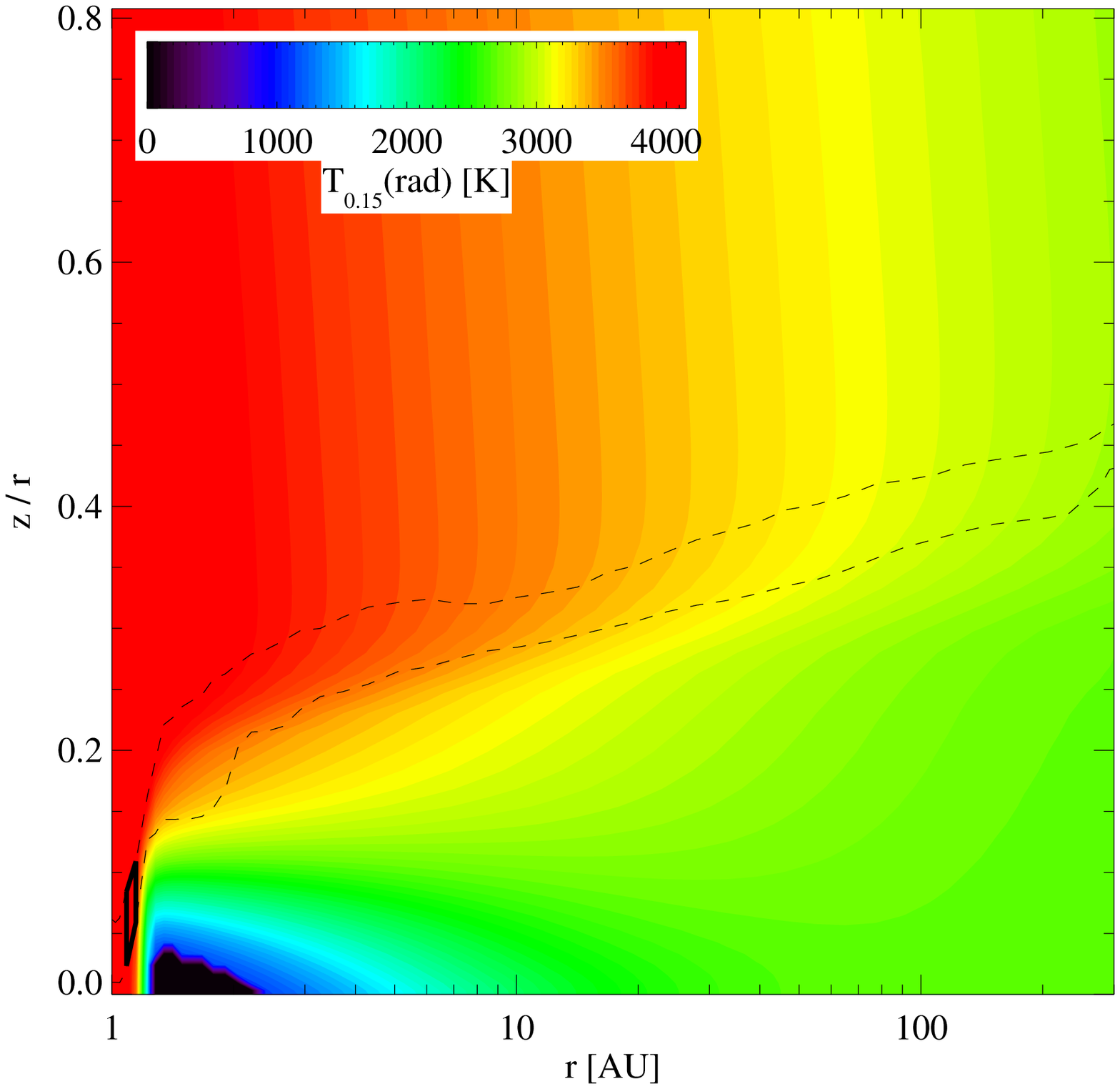}
  \includegraphics[scale=0.5,angle=0]{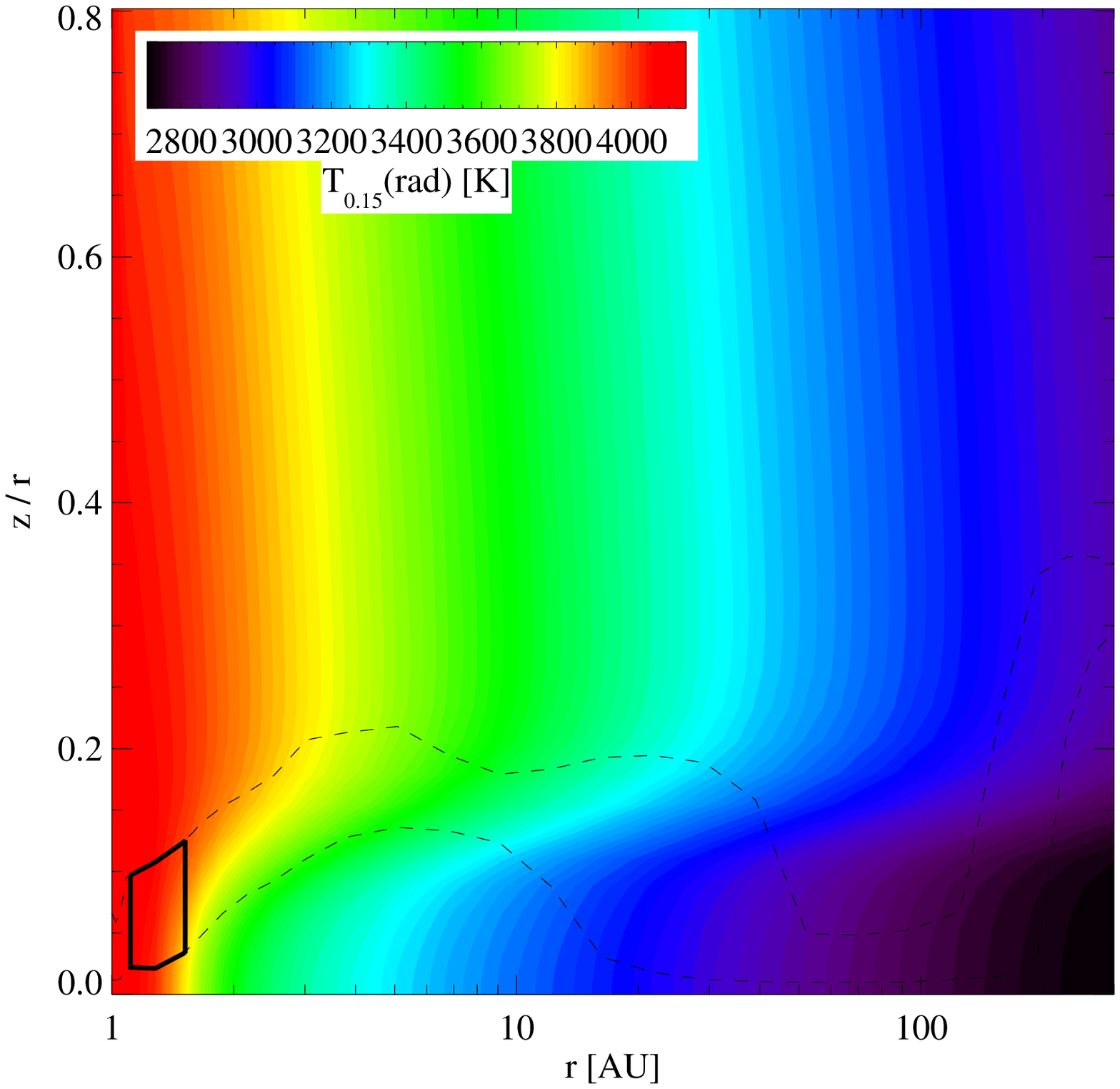}
  \includegraphics[scale=0.5,angle=0]{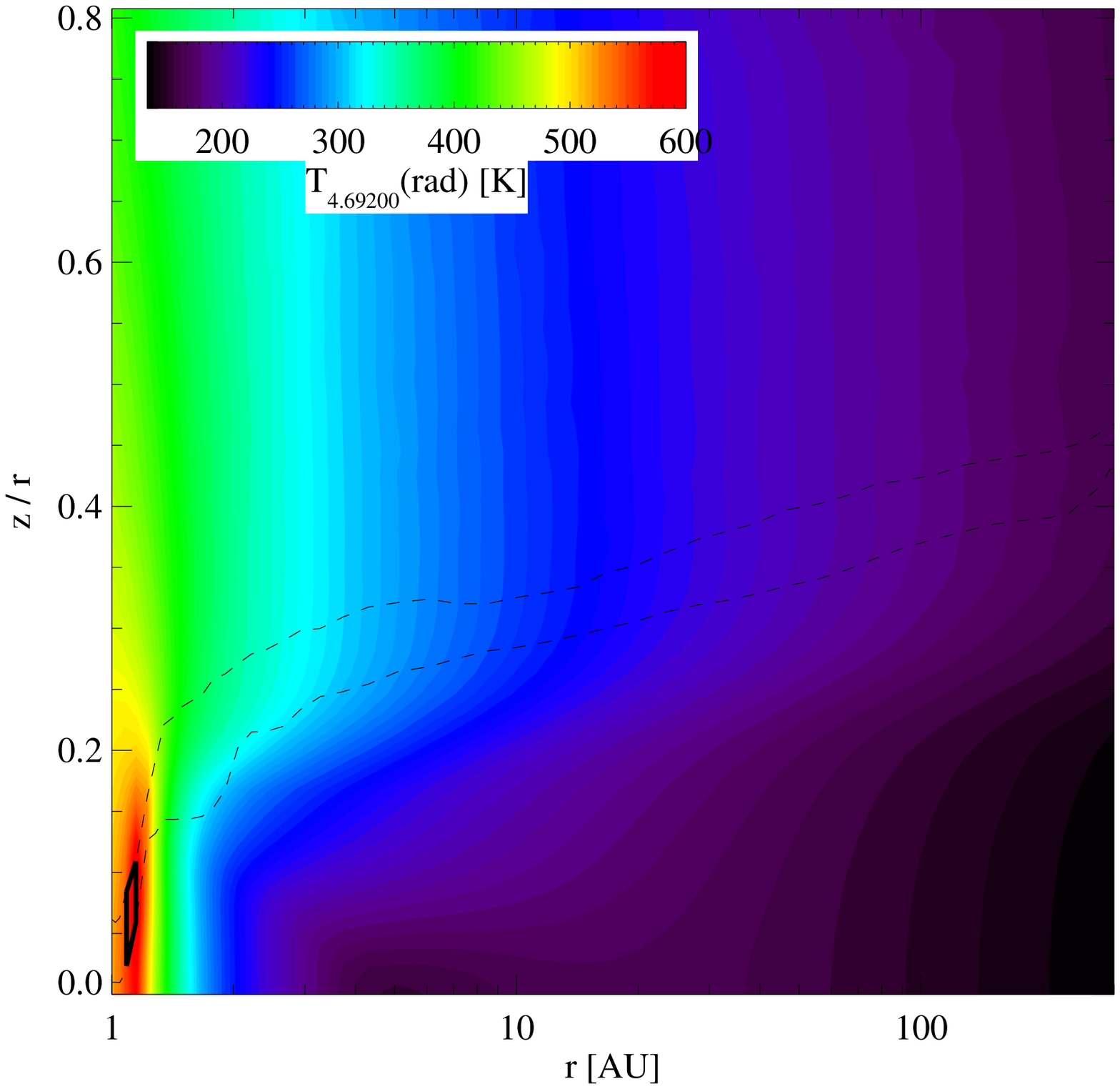}
  \includegraphics[scale=0.5,angle=0]{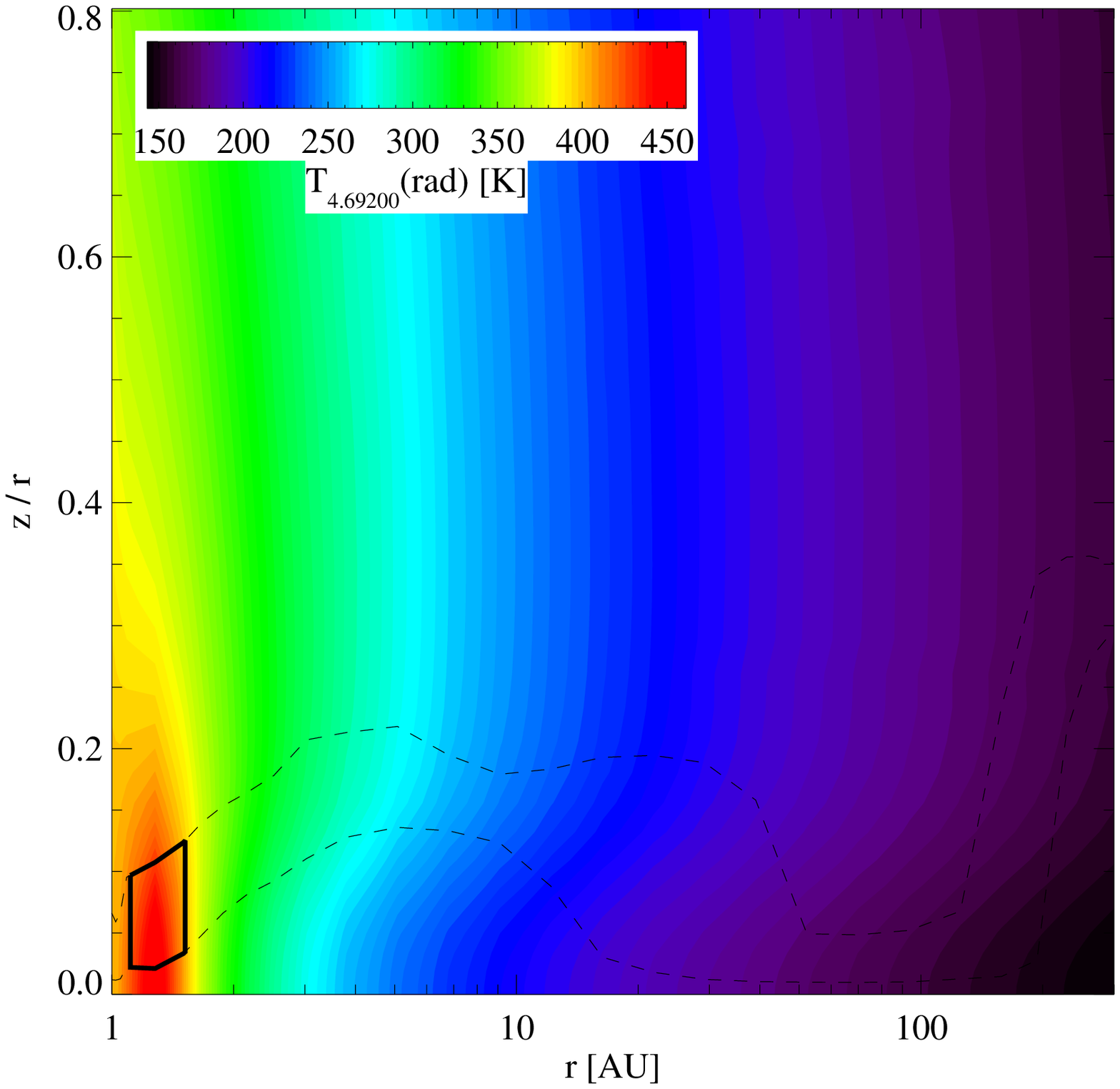}
  \caption{\label{Trad_hot} Radiation temperature $T_{\mathrm{rad}}$
    at 0.15 micron and 4.692 micron for the
    $M_{\mathrm{disc}}=10^{-2}$ M$_\odot$, $R_{\mathrm{in}}$=1~AU disc
    models (left panel) and for the $M_{\mathrm{disc}}=10^{-4}$
    M$_\odot$, $R_{\mathrm{in}}$=1~AU disc models (right panel). The
    black contours show the regions that emit 49\% of the fluxes at
    4.692 micron.  The black dashed-line contours contain 70\% of the
    fluxes in the vertical direction.}
\end{figure*}

\begin{figure*}
  \centering
  \includegraphics[scale=0.5,angle=0]{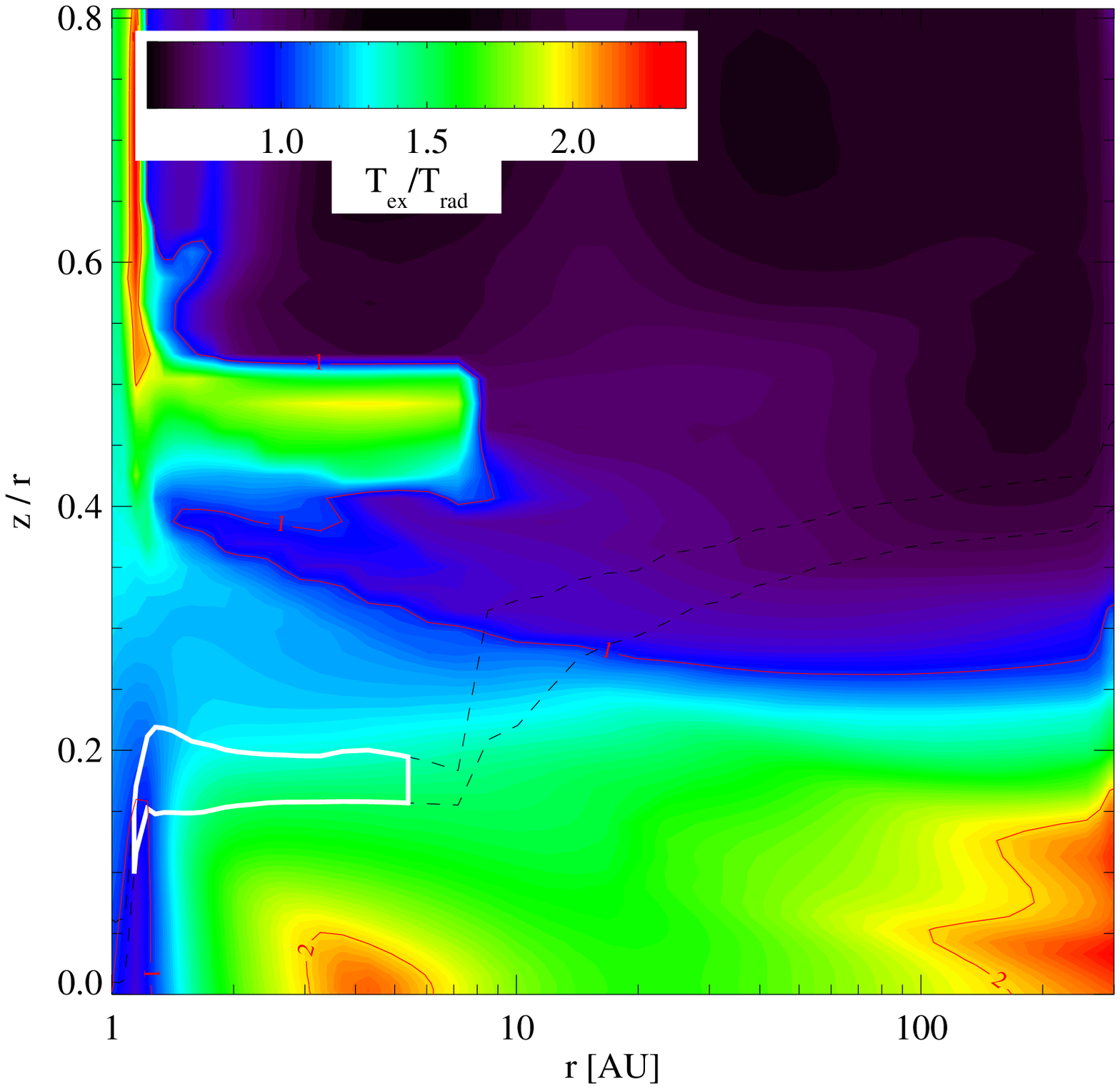}
  \includegraphics[scale=0.5,angle=0]{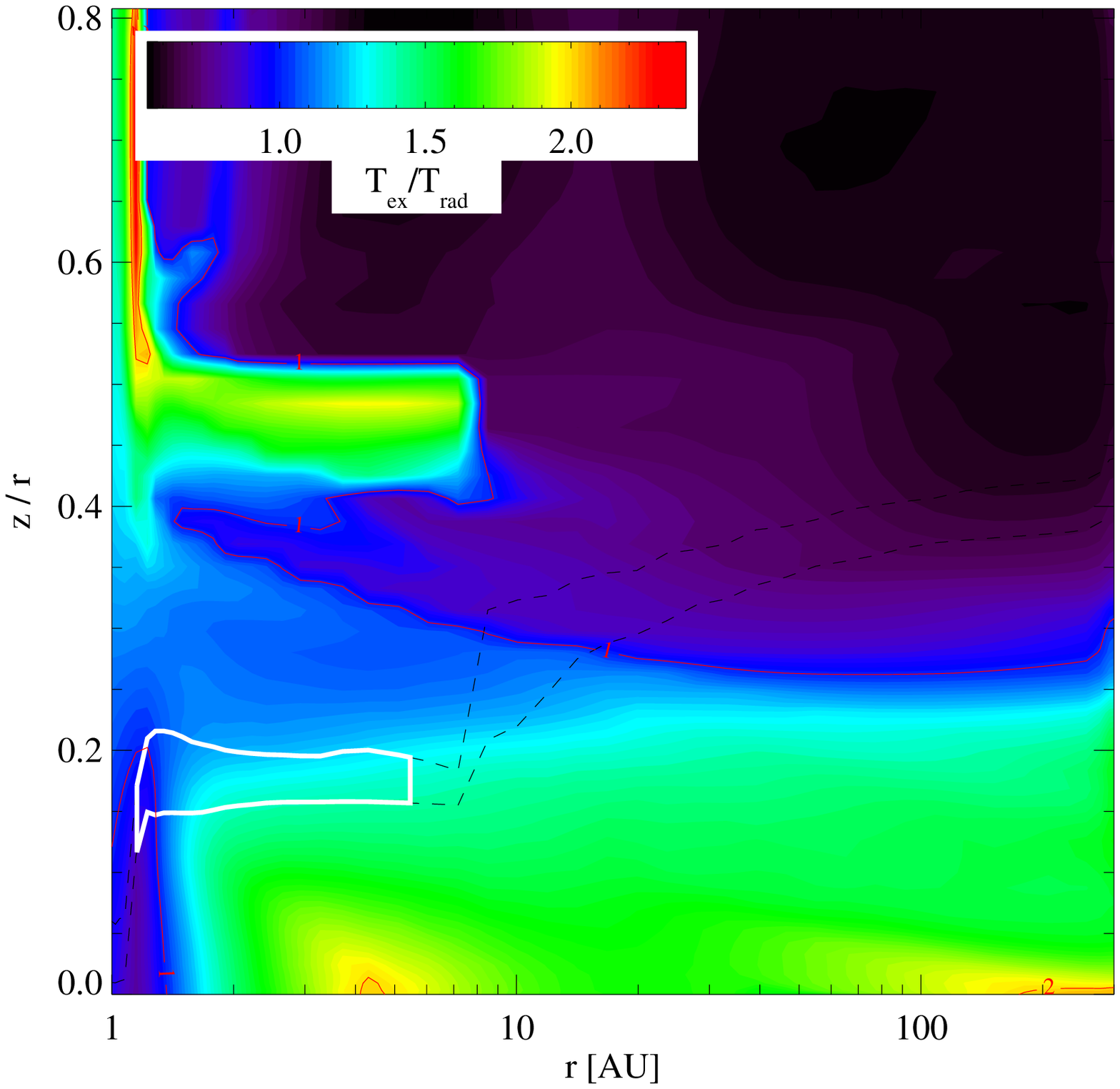}
  \includegraphics[scale=0.5,angle=0]{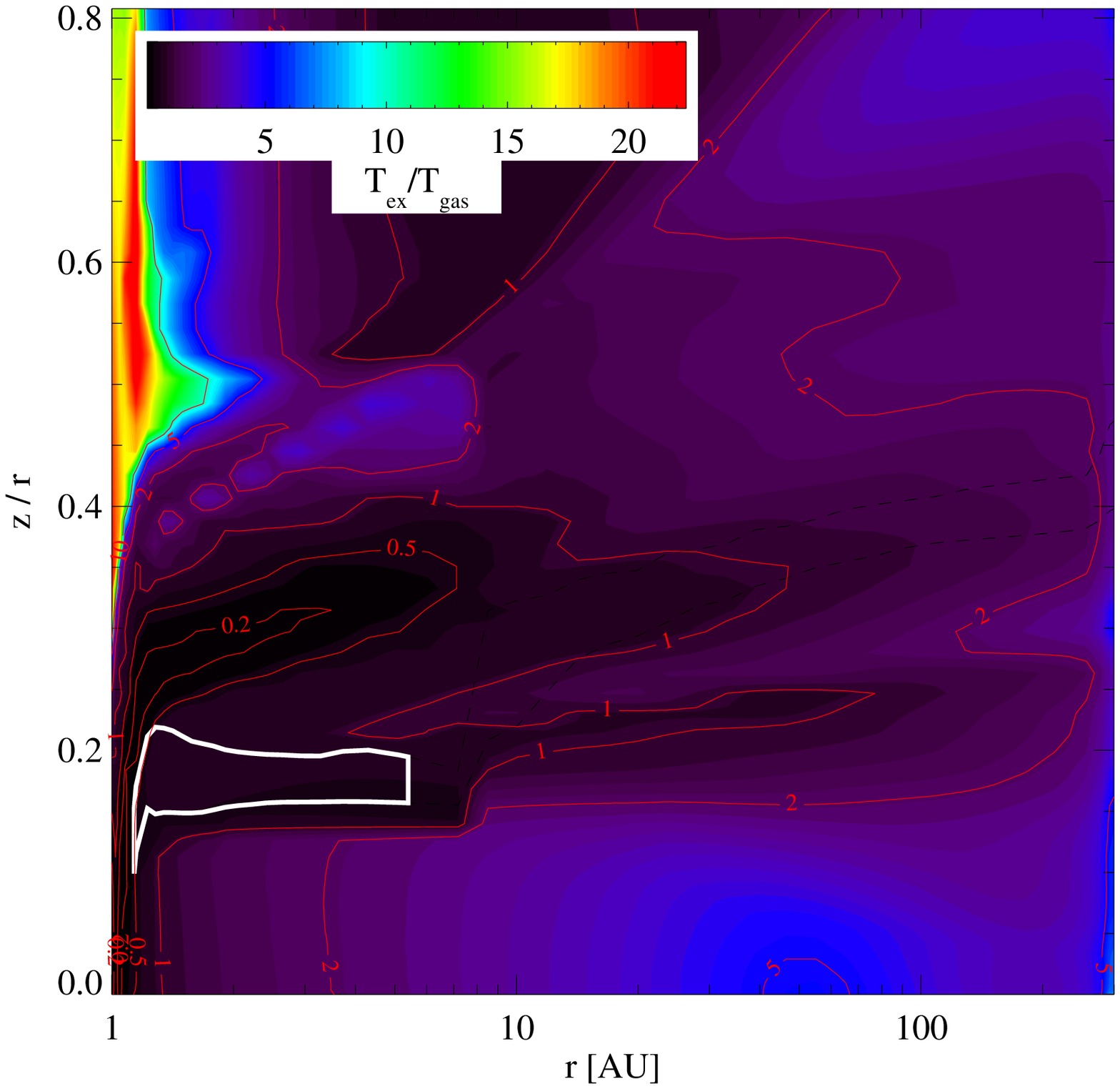}
  \includegraphics[scale=0.5,angle=0]{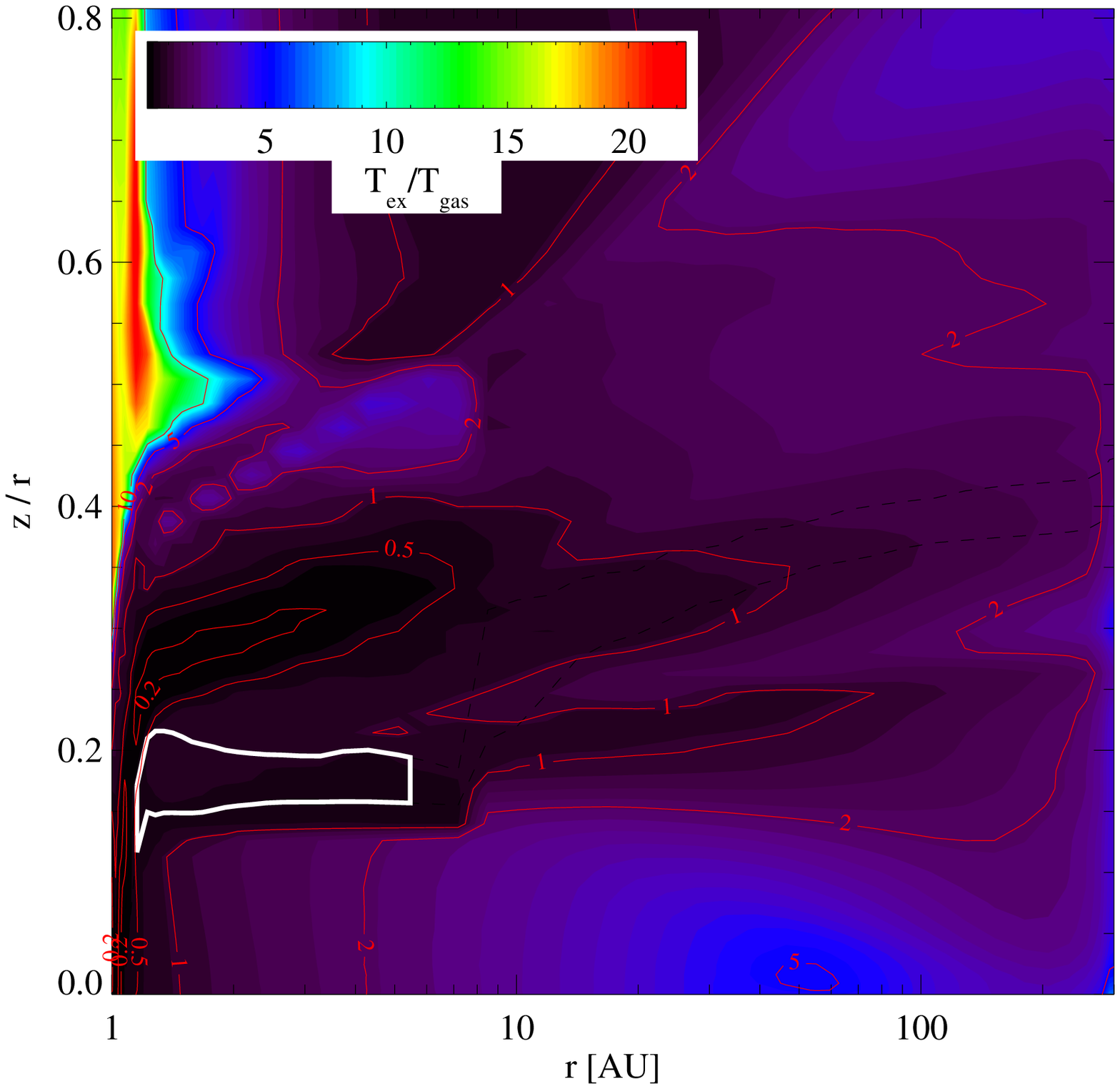}
\centering
\caption{\label{Tex_Trad_Tgas_massive_disc_ratio}
  $T_{\mathrm{ex}}(v=1,J=19)/T_{\mathrm{rad}}(4.844)$ (upper panels) and
  $T_{\mathrm{ex}}(v=1,J=19)/T_{\mathrm{gas}}$ (lower panels)
  structures for the $M_{\mathrm{disc}}$=10$^{-2}$ M$_\odot$,
  $R_{\mathrm{in}}$=1~AU disc models. The left panels correspond to
  the model with UV pumping and the right panels to models without UV
  pumping. The contours are labelled in red. The white contours
  encompass the regions that emit 49\% of the fluxes at 4.844 micron.
  The dashed-line contours contain 70\% of the fluxes in the vertical
  direction.}
\end{figure*}

\begin{figure*}[ht] 
  \centering
  \includegraphics[scale=0.5,angle=0]{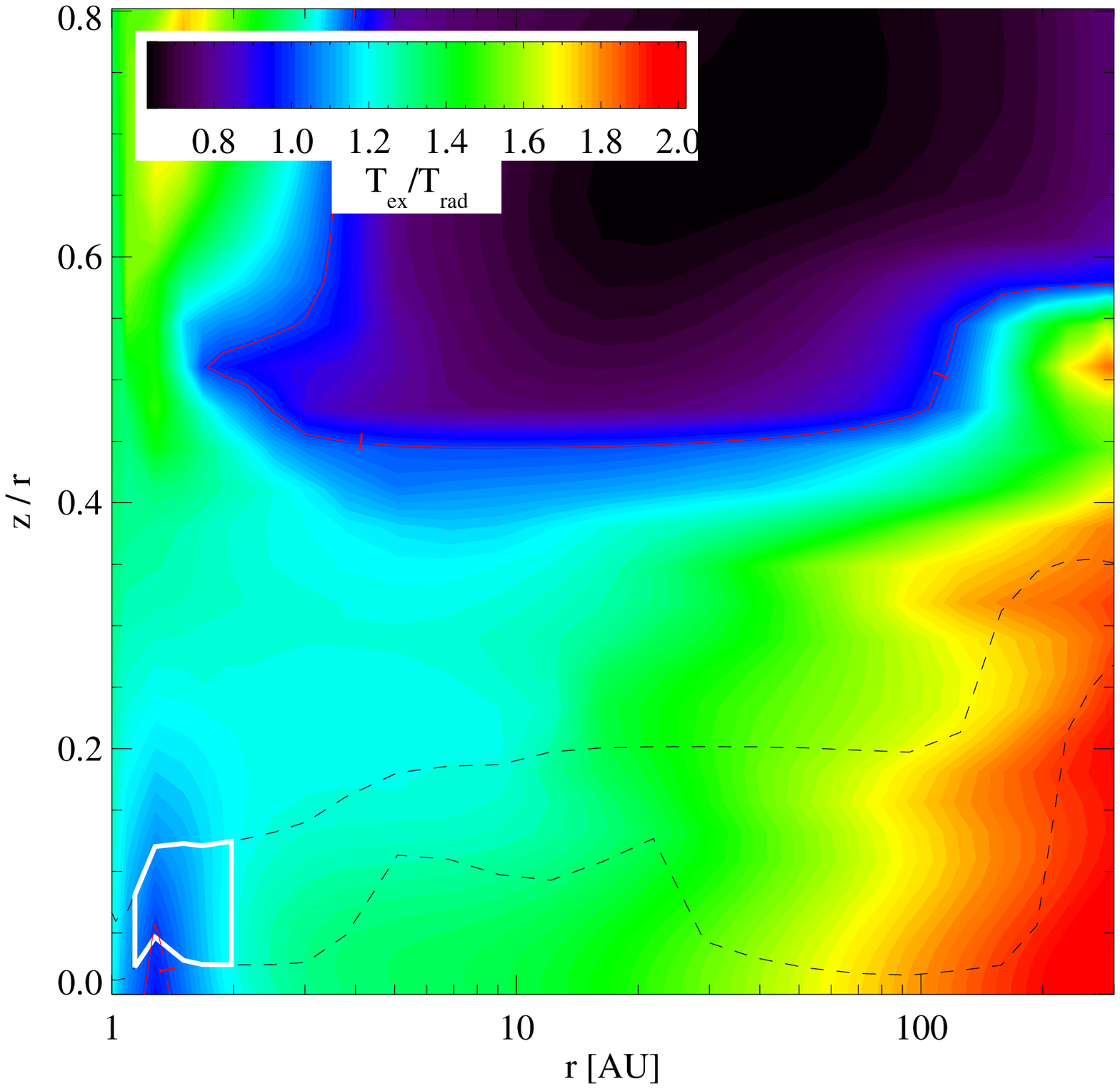}
  \includegraphics[scale=0.5,angle=0]{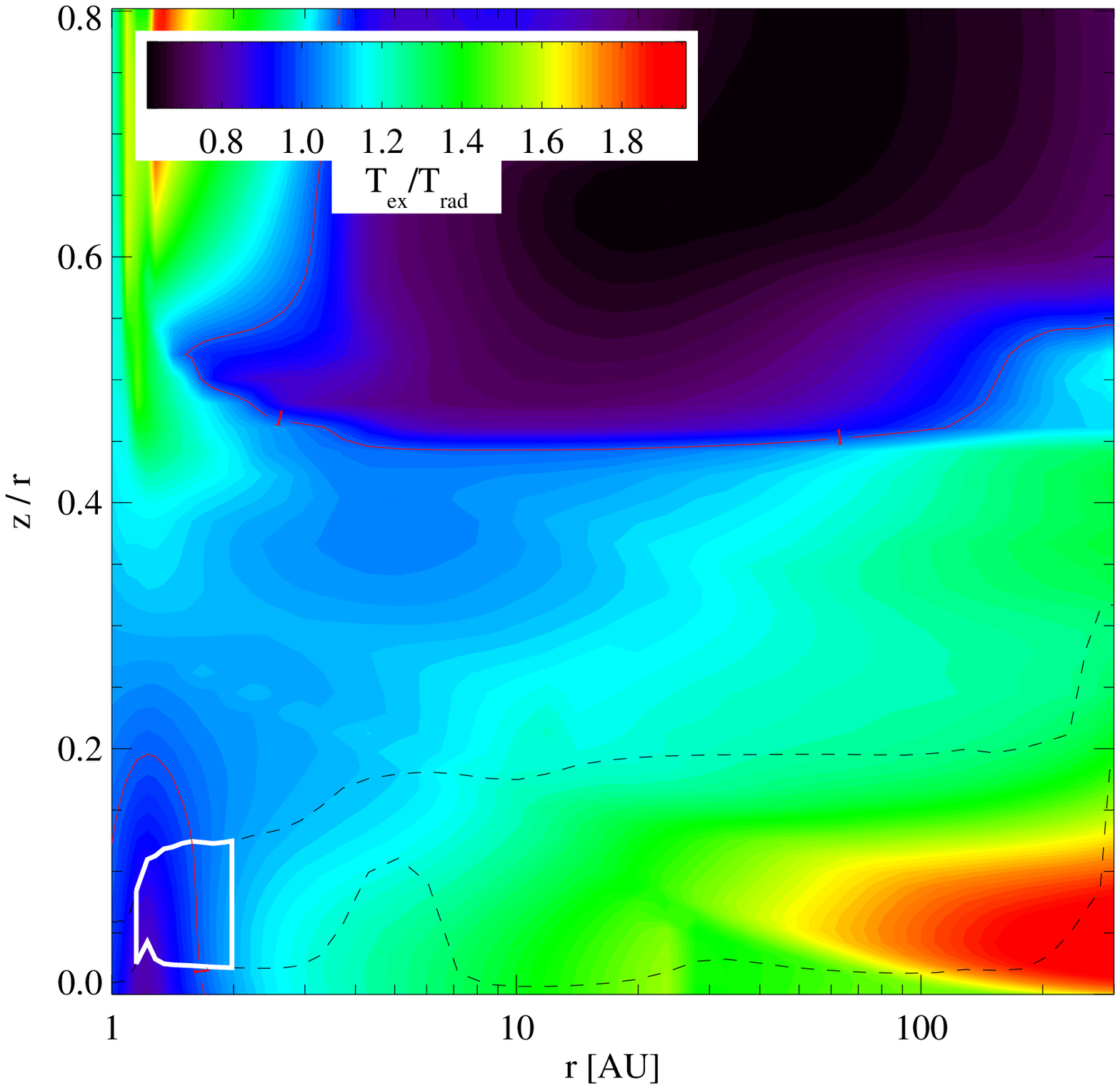}
  \includegraphics[scale=0.5,angle=0]{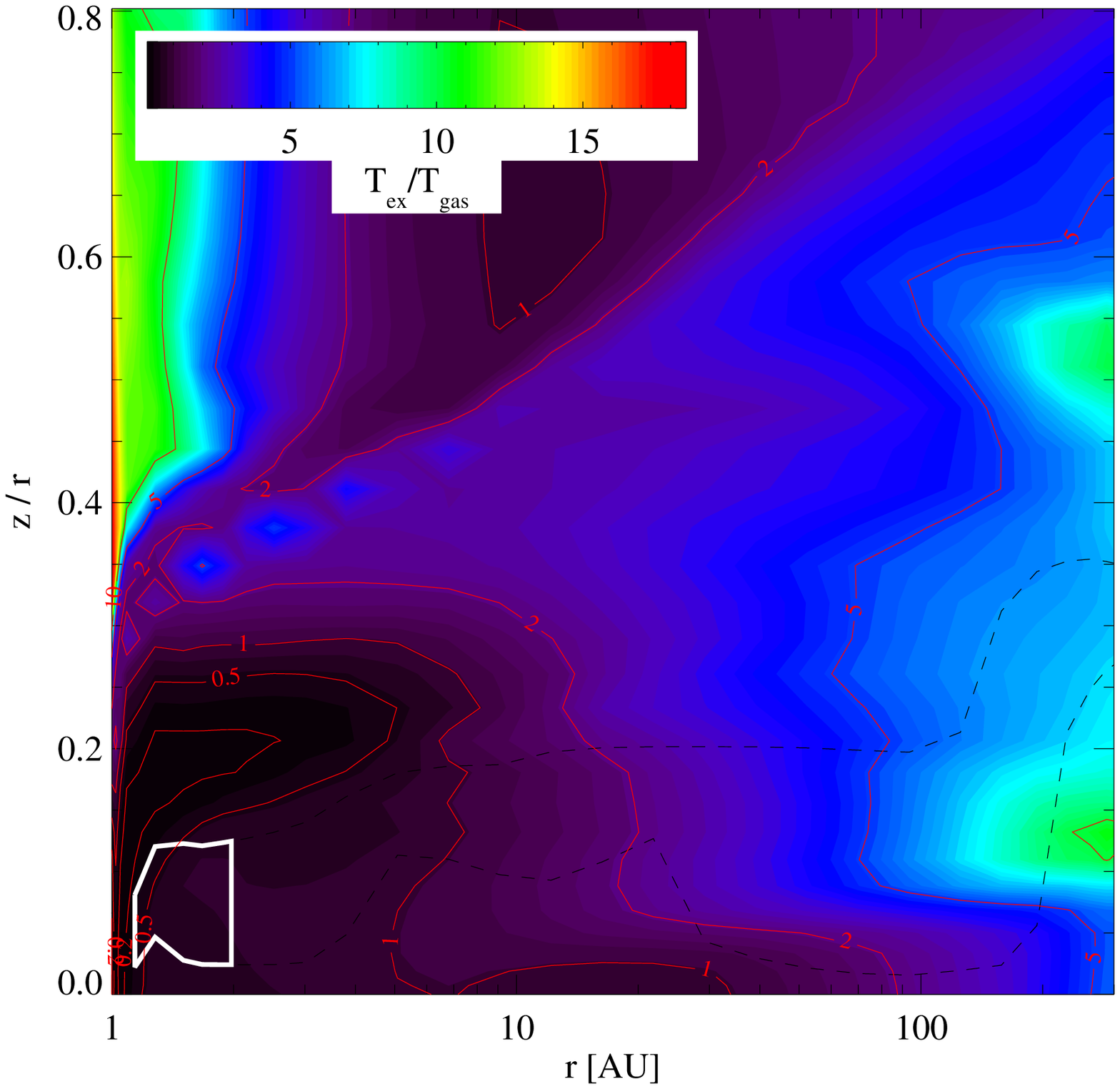}
  \includegraphics[scale=0.5,angle=0]{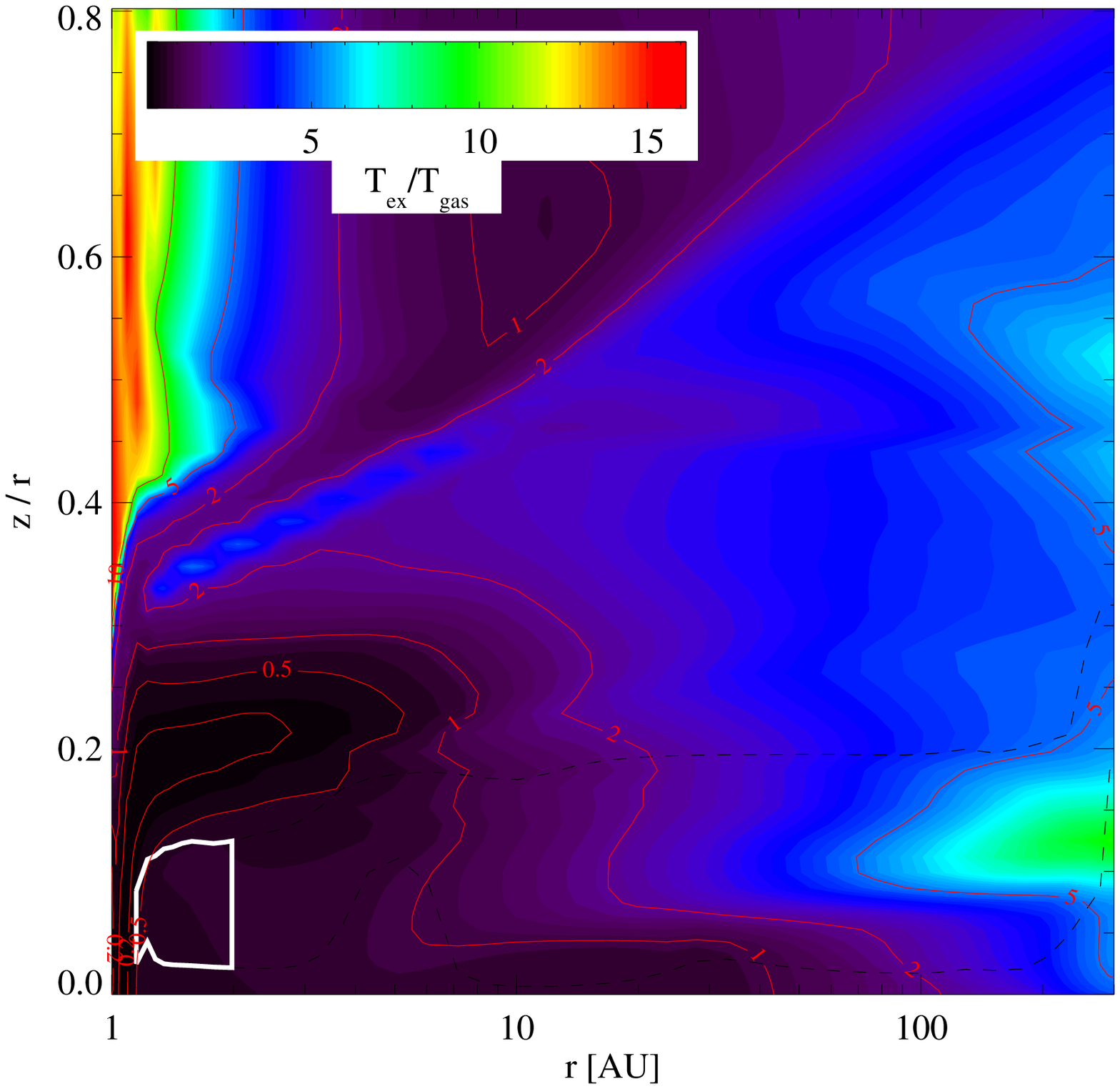}
\centering
\caption{\label{Tex_Trad_Tgas_low_mass_disc_ratio}
  $T_{\mathrm{ex}}(v=1,J=19)/T_{\mathrm{rad}}(4.844)$ (upper panels) and
  $T_{\mathrm{ex}}(v=1,J=19)/T_{\mathrm{gas}}$ (lower panels)
  structures for the $M_{\mathrm{disc}}$=10$^{-4}$ M$_\odot$,
  $R_{\mathrm{in}}$=1~AU disc models. The left panels correspond to
  the model with UV pumping and the right panels to models without UV
  pumping. The contours are labelled in red. The white contours
  encompass the regions that emit 49\% of the fluxes at 4.844 micron.
  The dashed-line contours contain 70\% of the fluxes in the vertical
  direction.}
\end{figure*}

We define the radiation (brightness) temperature $T_{\mathrm{rad}}$ at
a given wavelength as the equivalent blackbody temperature that will
match the specific intensity computed by the continuum radiative
transfer at a given location in the disc (Fig.~\ref{Trad_fundamental}
and \ref{Trad_hot}).  We also show temperature ratios in
Fig.~\ref{Tex_Trad_Tgas_massive_disc_ratio} and
\ref{Tex_Trad_Tgas_low_mass_disc_ratio}: the excitation over the
radiation and the excitation over the gas kinetic temperatures.



\end{appendix}

\end{document}